\documentclass{aa}

\usepackage{amssymb} 

\usepackage{graphicx}
\usepackage{amsmath}
\usepackage{latexsym}
\usepackage{amssymb}
\usepackage{longtable}
\usepackage{gensymb}
\usepackage{rotating}

\begin{document}

\title{Extremely Metal-Poor Galaxies: The \ion{H}{i} Content}


\author{M. E. Filho\inst{1}\thanks{E-mail: mfilho@astro.up.pt} \and B. Winkel\inst{2} \and J. S\'anchez Almeida\inst{3,4} \and J. A. Aguerri\inst{3,4} \and R. Amor\'\i n\inst{5,6} \and Y. Ascasibar\inst{7} \and B. G. Elmegreen\inst{8} \and D. M. Elmegreen\inst{9} \and J. M. Gomes\inst{1} \and A. Humphrey\inst{1} \and P. Lagos\inst{1} \and A. B. Morales-Luis\inst{3,4} \and C. Mu\~noz-Tu\~n\'on\inst{3,4} \and P. Papaderos\inst{1} \and J. M. V\'\i lchez\inst{5}}

\institute{Centro de Astrof\'isica da Universidade do Porto, 4150-762, Porto, Portugal 
\and Max-Planck-Institut f\"{u}r Radioastronomie (MPIfR), Auf dem H\"{u}gel 69, 53121 Bonn, Germany 
\and Instituto de Astrof\'\i sica de Canarias, E-38200 La Laguna (Tenerife), Spain 
\and Departamento de Astrof\'\i sica, Universidad La Laguna, E-38206 La Laguna (Tenerife), Spain 
\and Instituto de Astrof\'\i sica de Andaluc\'\i a, E-18008 Granada, Spain 
\and INAF - Osservatorio Astronomico di Roma, Via di Frascati 33, 00040 Monte Porzio Catone, Rome, Italy 
\and Universidad Aut\'onoma de Madrid, E-28049 Madrid, Spain 
\and IBM, T. J. Watson Research Center, Yorktown Heights, New York 10598, USA 
\and Vassar College, Department of Physics and Astronomy, Poughkeepsie, New York 12604, USA}

\date{}

\abstract{Extremely metal-poor (XMP) galaxies are chemically, and possibly dynamically, primordial objects in the local Universe.}
{Our objective is to characterize the \ion{H}{i} content of the XMP galaxies as a class, using as a reference the list of 140 known local XMPs compiled by Morales-Luis et al. (2011).}
{We have observed 29 XMPs, which had not been observed before at 21 cm, using the Effelsberg radio telescope. This information was complemented with \ion{H}{i} data published in literature for a further 53 XMPs. In addition, optical data from the literature provided morphologies, stellar masses, star-formation rates and metallicities.}
{Effelsberg \ion{H}{i} integrated flux densities are between 1 and 15 Jy km s$^{-1}$, while line widths are between 20 and 120 km s$^{-1}$. \ion{H}{i} integrated flux densities and line widths from literature are in the range 0.1 -- 200 Jy km s$^{-1}$ and 15 -- 150 km s$^{-1}$, respectively. Of the 10 new Effelsberg detections, two sources show an asymmetric double-horn profile, while the remaining sources show either asymmetric (seven sources) or symmetric (one source) single-peak 21 cm line profiles. An asymmetry in the \ion{H}{i} line profile is systematically accompanied by an asymmetry in the optical morphology.
Typically, the $g$-band stellar mass-to-light ratios are $\sim$0.1, whereas the \ion{H}{i} gas mass-to-light ratios may be up to two orders of magnitude larger. Moreover, \ion{H}{i} gas-to-stellar mass ratios fall typically between 10 and 20, denoting that XMPs are extremely gas-rich. 
We find an anti-correlation between the \ion{H}{i} gas mass-to-light ratio and the luminosity, whereby fainter XMPs are more gas-rich than brighter XMPs, suggesting that brighter sources have converted a larger fraction of their \ion{H}{i} gas into stars. (abridged)}
%
%
%
%
%
%
%
{XMP galaxies are among the most gas-rich objects in the local Universe. The observed \ion{H}{i} component suggests kinematical disruption and hints at a primordial composition.} 


\authorrunning{M. E. Filho et al.}
\titlerunning{Extremely Metal-Poor Galaxies: The \ion{H}{i} Content}

\maketitle

\section{Introduction}



According to the hierarchical paradigm of structure formation, massive galaxies assemble through mergers and cannibalism of smaller systems. Interactions between galaxies and secular processes induce episodes of star-formation. Stellar winds and the death of stars chemically enrich both the interstellar medium and the subsequent stellar generations. In this scenario, extremely metal-poor dwarf galaxies (XMPs) should be common in the early Universe, whereas they should be very rare at low redshift (York et al. 2000; Pustilnik et al. 2005; Guseva et al. 2007; Izotov, Thuan \& Guseva 2012; Mamon et al. 2012). Indeed, one of the most recent searches in the Sloan Digital Sky Survey (SDSS) Data Release 7 (DR7; Abazajian et al. 2009) and in literature, has yielded only 140 XMPs in the local Universe, corresponding to 0.1\% of the galaxies in the local volume (Morales-Luis, S\'anchez Almeida, Aguerri \& Mu\~noz-Tu\~n\'on 2011; hereinafter ML11). In this case, XMPs are defined as having ionized gas with an oxygen abundance smaller than a tenth of the solar value. These XMPs are thus the best local analogs of the first generation of low-mass galaxies formed early on, possessing chemical abundances as close as possible to that of the primordial Universe. 


It has been found that local XMP galaxies are commonly star-forming blue compact dwarfs (BCDs), which are characterized by strong emission lines, blue colors, high surface brightness, low luminosity, compactness, and blue and faint optical continuum (Sargent \& Searle 1970; Thuan \& Martin 1981; Papaderos et al. 1996a, b; Telles \& Terlevich 1997; Kunth \& \"Ostlin 2000; Cair\'os et al. 2001; Bergvall \& \"Ostlin 2002; Cair\'os et al. 2003; Noeske et al. 2003; Gil de Paz \& Madore 2005; Amor\'\i n et al. 2007, 2009; ML11; Lagos et al. 2011; Bergvall 2012; Micheva et al. 2013). 

Furthermore, in 75\% of the cases, XMPs exhibit cometary or multi-knotted asymmetric optical structures (Papaderos et al. 2008; ML11). As a comparison, only 0.2\% of the star-forming galaxies in the Kiso survey of UV-bright galaxies (Miyauchi-Isobe, Maehara \& Nakajima 2010) are cometary (Elmegreen et al. 2012).




There are various interpretations for the asymmetric optical morphology of these galaxies. They could be diffuse edge-on disks in the early stages of evolution, with massive star-forming regions viewed from the side (Elmegreen \& Elmegreen 2010), resulting from the spontaneous excitation of gravitational instabilities (Elmegreen et al. 2009). Alternatively, these structures may also arise from gravitational triggering due to a merger with a low-mass companion (Straughn et al. 2006) or it may be self-propagation of the star-formation activity within an already existing gas-rich galaxy or chemically pristine gas cloud (Papaderos et al. 1998, 2008). The large starburst that gives rise to the asymmetry could also be due to the infall of pristine external gas (S\'anchez Almeida et al. 2013).

In the few XMPs where the \ion{H}{i} has been investigated with interferometric observations, the \ion{H}{i} spatial distributions and velocity profiles are found to be distorted (Ekta, Chengalur \& Pustilnik 2008; Ekta \& Chengalur 2010a), indicating infall of external unenriched gas that may feed the starburst and drop the metallicity (Kewley, Geller \& Barton 2006) or gas stripping forced by an interaction with an external medium (Gavazzi et al. 2001; Elmegreen \& Elmegreen 2010). In the latter case, the asymmetric starburst results from the ram compression by the intergalactic medium: gas-rich disks with star-formation at the leading edge and the rest of the disk visible as the tail, or with star-formation at the leading edge and a tail of star-formation in the stripped gas. In any case, the study of the dynamical, stellar, ionized gas and neutral atomic gas structure is crucial to disentangle the nature of the XMPs and their association with a particular morphology.


Our team is involved in the full observational characterization of a representative sample of local XMP galaxies. Their \ion{H}{i} gas content is particularly important, since these galaxies are expected to have large gas reservoirs responsible for many of their observational properties, including sustaining the current star-formation episode and even diluting the interstellar gas to yield their low metallicity. This work aims at describing and quantifying, for the first time, the \ion{H}{i} content of the XMPs as a class.


In Sect.~2.1 we present a compilation of published \ion{H}{i} information for the reference list of XMPs in ML11. The existing data were completed with new observations obtained with the 100-meter single-dish Effelsberg radio telescope. The observations, data reduction (following a novel technique) and properties of the detected sources are described in Sect.~2.2. Because derived parameters, such as dynamical masses, rely on ancillary optical data, we have compiled this information from the SDSS optical images and spectra (Sect. 3). Global galaxy parameters are derived combining the \ion{H}{i} and optical data, as explained in Sect.~4. The results of our investigation, namely, the description of the \ion{H}{i} content of the XMP galaxies with respect to other physical properties, such as the Tully -- Fisher relation, are described in Sect.~5. In Sect.~6 we describe properties that rely exclusively on optical data, such as the morphology. Our conclusions are summarized in Sect.~7.

Throughout this paper, we adopt the cosmological parameters  $\Omega_m = 0.27$, $\Omega_\Lambda = 0.73$ and H$_0 = 73$ km s$^{-1}$ Mpc$^{-1}$. 

\section{\ion{H}{i} Radio Data}

The sources presented in this analysis were extracted from the work of ML11, which contains 140 extremely metal-poor galaxies selected from the SDSS DR7 (Abazajian et al. 2009), including 11 new XMP candidates, and completed with all the galaxies in literature having an oxygen metallicity less than a tenth of the solar value (explicitly, $12 \, + \, \log$~(O/H)~$\leq$~7.65). 


The \ion{H}{i} gas data comes partly from literature (Sect.~2.1) and is complemented by new observations (Sect.~2.2). 

\subsection{Radio Observations From Literature}

53 out of the 140 XMP galaxies possess published \ion{H}{i} line observations, 5 of which were non-detections. The \ion{H}{i} data were primarily gathered from the Arecibo Legacy Fast ALFA Survey (ALFALFA; Giovanelli et al. 2007), \ion{H}{i} Parkes All-Sky Survey (HIPASS; Barnes et al. 2001) and The \ion{H}{i} Nearby Galaxy Survey (THINGS; Walter et al. 2008), as well as sources observed with the Nan\c{c}ay Radio Telescope (NRT; Pustilnik \& Martin 2007), the Green Bank Telescope (GBT; Schneider et al. 1992; Hogg et al. 2007), the Australia Telescope Compact Array (ATCA; Warren, Jerjen \& Koribalski 2006; O'Brien et al. 2010), the Westerbork Synthesis Radio Telescope (WSRT; Kova\u{c}, Osterloo \& van der Hulst 2009), the Giant Metrewave Radio Telescope (GMRT; Begum et al. 2006, 2008), the Effelsberg Radio Telescope (Huchtmeier, Krishna \& Petrosian 2005) and the Very Large Array (VLA; Thuan, Hibbard \& L\'evrier 2004). 

These observations have yielded typical \ion{H}{i} integrated flux densities, S$_{\ion{H}{i}}$, in the 0.1 -- 200 Jy km s$^{-1}$ range and line widths, at 50\% of the peak flux density level, $w_{50}$, in the 15 -- 150 km s$^{-1}$ range.

In Table~1 we list the compilation of \ion{H}{i} line observations from literature. In addition to the \ion{H}{i} integrated flux density, which can be used to estimate the \ion{H}{i} mass (Eq.~4; Sect.~4), Table~1 includes the \ion{H}{i} diameter, d$_{\ion{H}{i}}$, the systemic heliocentric or local group-corrected radial velocity (optical convention; Eq.~2; Sect.~4), v$_{sys}$, and the line width, $w_{50}$. The systemic radial velocity is the midpoint of the \ion{H}{i} emission-line profile and can be used to estimate a redshift distance using the Hubble law. The line width provides a measure of the Doppler broadening and, together with the \ion{H}{i} diameter, can be used to estimate the dynamical mass (Eq.~3; Sect.~4). Original references and observational facilities or surveys are also listed in Table~1.



\setcounter{table}{0}

\begin{table*}

\footnotesize

\begin{center}

\begin{minipage}[c]{177mm}

\caption{Compilation of the \ion{H}{i} data from literature for 53 XMP galaxies in the local Universe.
Col.~1: Source name.
Col.~2: Right Ascension.
Col.~3: Declination.
Col.~4: \ion{H}{i} integrated flux density and error. ''$<$'' designates an upper limit. Data from references (a) are corrected for pointing, (b) are flux density-corrected for beam resolution and (h) are self-absorption-corrected line flux densities. 
Col.~5: \ion{H}{i} systemic heliocentric or local group-corrected ($\star$) radial velocity (optical convention) and error. 
Col.~6: \ion{H}{i} line width at 50\% of the peak flux density and error.
Col.~7: \ion{H}{i} diameter at a column density of $\sim$ 10$^{19}$ atoms cm$^{-2}$.
Col.~8: Telescope or survey.
Col.~9: Bibliographical reference. 
(a) - Green Bank Telescope (GBT) -- Schneider et al. 1992; (b) - Nan\c{c}ay Radio Telescope (NRT) -- Pustilnik \& Martin 2007; (c) - The \ion{H}{i} Parkes All Sky Survey (HIPASS) -- Koribalski et al. 2004; (d) - Australia Telescope Compact Array (ATCA) -- Warren, Jerjen \& Koribalski 2006; (e) - Giant Metrewave Radio Telescope (GMRT) -- Begum et al. 2008; (f) - Arecibo Radio Telescope -- Lu et al. 1993; (g) - Giant Metrewave Radio Telescope (GMRT) -- Ekta \& Chengalur 2010a; (h) - Arecibo Radio Telescope -- Springob et al. 2005; (i) - The \ion{H}{i} Parkes All Sky Survey (HIPASS) -- Doyle et al. 2005; (j) - Giant Metrewave Radio Telescope (GMRT) -- Ekta, Pustilnik \& Chengalur 2009; (k) - Effelsberg Radio Telescope -- Huchtmeier, Karachentsev \& Karachentsev 2003; (l) - The \ion{H}{i} Nearby Galaxy Survey (THINGS) performed with the Very Large Array (VLA) -- Walter et al. 2008; (m) - Giant Metrewave Radio Telescope (GMRT) -- Chengalur et al. 2006; (n) Very Large Array (VLA) - ACS Nearby Galaxy Survey Treasury (VLA-ANGST) -- Ott et al. 2012; (o) - \ion{H}{i} catalog \textit{A General Catalog of \ion{H}{i} Observations of Galaxies} -- Huchtmeier \& Richter 1989; (p) - Green Bank Telescope (GBT) --  Hogg et al. 2007; (q) - Effelsberg Radio Telescope -- Huchtmeier, Krishna \& Petrosian 2005; (r) - Giant Metrewave Radio Telescope (GMRT) -- Begum et al. 2006; (s) - Parkes Radio Telescope -- Barnes \& de Blok 2004; (t) - Very Large Array (VLA) -- Thuan, Hibbard \& L\'evrier 2004; (u) - Giant Metrewave Radio Telescope (GMRT) -- Ekta, Chengalur \& Pustilnik 2006; (v) - The Arecibo Legacy Fast ALFA Survey (ALFALFA) -- Giovanelli et al. 2007; (w) - Westerbork Synthesis Radio Telescope (WSRT) -- Kova\u{c}, Osterloo \& van der Hulst 2009; (x) - Green Bank Telescope (GBT) -- O'Neil 2004; (y) - Giant Metrewave Radio Telescope (GMRT) -- Ekta, Chengalur \& Pustilnik 2008; (z) - Australia Telescope Compact Array (ATCA) -- O'Brien et al. 2010.}

\begin{tabular}{l c c c c c c c c}

\hline

Name      &    RA(J2000)        &  DEC(J2000)                    & S$_{\ion{H}{i}}$       & v$_{sys}$       & $w_{50}$    & d$_{\ion{H}{i}}$   & Telescope   &  Ref. \\
          & $^h$ $^m$ $^s$      &  \degree \, \arcmin \, \arcsec & Jy km s$^{-1}$ 	  & km s$^{-1}$     & km s$^{-1}$ & \arcmin            & or Survey   &        \\
(1)	  &   (2)	        &  (3)		                 &  (4)   	    	  & (5) 	    &   (6)         &  (7)  & (8) & (9) \\

\hline
\hline			
 
UGC12894     &00 00 22    &+39 29 44  	& 6.75 $\pm$ 0.69 & 335 $\pm$ 4   & 34 $\pm$ 2  & 1.0   & GBT		& a	\\ 
HS0017+1055  &00 20 21    &+11 12 21   & $<$ 0.20        & 5630 $\pm$ 30 & 50          & \ldots& NRT		& b	\\ 
ESO473-G024  &00 31 22    &-22 45 57  	& 7.2 $\pm$ 1.8	  & 540 $\pm$ 4   & 45          & \ldots& HIPASS	& c	\\ 
	      &            &	   	& 5.7 $\pm$ 0.9	  & 542 $\pm$ 3   & 37 $\pm$ 2  & \ldots& ATCA		& d	\\ 
AndromedaIV  &00 42 32    &+40 34 19   & 19.5 $\pm$ 2.0  & 237           & 90          & 7.6  	& GMRT		& e	\\	
IC1613       &01 04 48    &+02 07 04   & 218 $\pm$ 21.8  & 234 $\pm$ 1   & 24 $\pm$ 1  & \ldots& Arecibo	& f	\\
J0119-0935    &01 19 14    &-09 35 46   & 0.95 $\pm$ 0.3  & 1932 $\pm$ 1.67&  \ldots    & \ldots& GMRT		& g  	\\ 
HS0122+0743  &01 25 34    &+07 59 24   & 5.6  $\pm$ 0.56 & 2926 $\pm$ 1 & 53 $\pm$ 2   & \ldots& Arecibo	& f	\\
	      &		   &		& 7.60 $\pm$ 0.30 & 2899 $\pm$ 5 & 50 $\pm$ 6   & \ldots& NRT		& b	\\
J0133+1342    &01 33 53    &+13 42 09   & 0.10 $\pm$ 0.05 & 2580 $\pm$ 4 & 33 $\pm$ 11  & \ldots& NRT		& b     \\
UGCA20        &01 43 15    &+19 58 32   & 11.43 $\pm$ 0.47& 498	         & 65           & \ldots& Arecibo	& h	\\
UM133         &01 44 42    &+04 53 42   & 3.65 $\pm$ 0.16 & 1621 $\pm$ 1.67&  \ldots    & \ldots& GMRT		& g	\\
J0205-0949    &02 05 49    &-09 49 18   & 11.8            & 1898.1       & 121.8        & \ldots& HIPASS	& i	\\
	      &		   &	   	& 12.9 $\pm$ 0.22 & 1885 $\pm$ 1 & 112 $\pm$ 2	& \ldots& NRT		& b	\\	
UGC2684      &03 20 24    &+17 17 45   & 10.54 $\pm$ 0.39& 350          & 78	        & \ldots& Arecibo	& h	\\ 
SBS0335-052W &03 37 38    &-05 02 37   & 0.86            & 4014.7 $\pm$ 1.7 & 47.4 $\pm$ 3.3 & \ldots & GMRT   & j	\\ 
SBS0335-052E &03 37 44    &-05 02 40   & 0.61	          & 4053.6 $\pm$ 1.7 & 50.8 $\pm$ 3.3 &	\ldots & GMRT   & j	\\
ESO358-G060  &03 45 12    &-35 34 15   & 10.6 $\pm$ 1.8  &  808 $\pm$ 4 & 73           & \ldots& HIPASS	& c	\\ 
ESO489-G56   &06 26 17    &-26 15 56    & 2.1             & 492.3 $\pm$ 0.5 & 23.6 $\pm$ 1.3& \ldots& Effelsberg & k \\
              &            &             & 2.4	          &  491.5       & 33.8         & \ldots& HIPASS	& i	\\      
	      &		   &    	& 2.79 $\pm$ 0.08 &  491 $\pm$ 1 & 27 $\pm$ 1	& \ldots& NRT		& b	\\             
UGC4305      &08 19 05    &+70 43 12   & 219             &  157.1       & 57.4         & \ldots& THINGS	& l	\\ 
HS0822+03542 &08 25 55    &+35 32 31   & 0.27 $\pm$ 0.06 & 726.6 $\pm$ 2& 21.7 $\pm$ 4 & \ldots& GMRT		& m	\\
DD053        &08 34 07    &+66 10 54   & 21.5 $\pm$ 2.2  & 19.2         & 29.6         & 4.5 	& GMRT		& e	\\ 
              &            &            & 20              & 17.7         & 28.3         & \ldots& THINGS	& l	\\   	
              &            &            & 13.8            & 20.1 $\pm $0.3 & 25.0 $\pm$ 0.8 & \ldots& Effelsberg   & k \\
UGC4483      &08 37 03    &+69 46 31   & 12.0            & 153.9        & 34.3         & \ldots & VLA-ANGST & n \\
             &            &            & 12.90           & \ldots       & \ldots       & \ldots & \ldots	& o	\\      
HS0846+3522  &08 49 40    &+35 11 39   & 0.10 $\pm$ 0.03 & 2169 $\pm$ 3 & 33 $\pm$ 6	& \ldots& NRT		& b	\\
IZw18        &09 34 02    &+55 14 25   & 2.87            & 745          & 31           & \ldots& GBT		& p	\\ 
J0940+2935    &09 40 13    &+29 35 30   & 2.05 $\pm$ 0.25 & 505 $\pm$ 2  & 77 $\pm$ 10  & \ldots& NRT		& b	\\
SBS940+544   &09 44 17    &+54 11 34   & $<$ 2.6         &  \ldots      &  \ldots      & \ldots& Effelsberg	& q	\\
LeoA         &09 59 26    &+30 44 47   & 42.0 $\pm$ 4.0  & 21.7 $\pm$ 0.2 & 18.8 $\pm$ 0.7 & \ldots& GMRT	& r	\\
              &            &            & 48.3            & 23.9$\pm$0.1 & 19.1 $\pm$ 0.2   & \ldots& Effelsberg & k \\
SextansB     &10 00 00    &+05 19 56    & 91.0              & 302.2       & 40.6            & \ldots & VLA-ANGST & n \\
             &            &             & 72.9              & 300.5 $\pm$ 0.1 & 38.0 $\pm$ 0.2 & \ldots& Effelsberg & k \\
              &            &            & 198.61 $\pm$ 49.71& 301        & 37           & \ldots& Arecibo	& h	\\ 
SextansA     &10 11 00    &-04 41 34    & 138.1           & 324.8        & 46.2         & \ldots & VLA-ANGST & n \\
             &            &             & 131.8           & 324          & 45           & \ldots& GBT		& p	\\
	      &		   &	   	& 168.5 $\pm$ 20.9& 324 $\pm$ 2  & 46      	& \ldots& HIPASS	& c  \\ 
	      &		   &	   	& 168 $\pm$ 12	  & 324 $\pm$ 1  & 46           & \ldots& Parkes	& s	\\           

\hline

\end{tabular}

\end{minipage}

\end{center}

\end{table*}



\setcounter{table}{0}

\begin{table*}

\footnotesize

\begin{center}

\begin{minipage}[c]{177mm}

\caption{Compilation of the \ion{H}{i} data from literature for 53 XMP galaxies in the local Universe. Continued.}

\begin{tabular}{l c c c c c c c c}

\hline

Name      &    RA(J2000)        &  DEC(J2000)                    & S$_{\ion{H}{i}}$       & v$_{sys}$       & $w_{50}$    & d$_{\ion{H}{i}}$   & Telescope   &  Ref. \\
          & $^h$ $^m$ $^s$      &  \degree \, \arcmin \, \arcsec & Jy km s$^{-1}$ 	  & km s$^{-1}$     & km s$^{-1}$ & \arcmin            & or Survey   &        \\
(1)	  &   (2)	        &  (3)		                 &  (4)   	    	  & (5) 	    &   (6)         &  (7)  & (8) & (9) \\

\hline
\hline	

KUG1013+381  &10 16 24    &+37 54 44   & 1.51 $\pm$ 0.39   & 1169 $\pm$ 4 & 86 $\pm$ 7	& \ldots& NRT		& b	\\ 
UGCA211      &10 27 02    &+56 16 14   & 3.00	            & \ldots       & \ldots     & \ldots& \ldots	& o	\\
HS1033+4757  &10 36 25    &+47 41 52   & 1.32 $\pm$ 0.15   & 1541 $\pm$ 9 & 86 $\pm$ 7	& \ldots& NRT		& b 	\\
HS1059+3934  &11 02 10    &+39 18 45   & 1.39 $\pm$ 0.06   & 3019 $\pm$ 3 & 59 $\pm$ 6	& \ldots& NRT		& b 	\\
J1105+6022    &11 05 54    &+60 22 29   & 2.48 $\pm$ 0.18   & 1333 $\pm$ 3 & 48 $\pm$ 6	& \ldots& NRT		& b	\\
J1121+0324    &11 21 53    &+03 24 21   & 2.67 $\pm$ 0.16   & 1171 $\pm$ 3 & 89 $\pm$ 6	& \ldots& NRT		& b 	\\		
UGC6456      &11 28 00    &+78 59 39   & 10.1 $\pm$ 1.0   & -93.69        & 37.4       & 3.7	& GMRT		& e	\\
              &            &            & 11.0	           & \ldots        & 34.2       & 3.6$^{a}$& VLA        & t	\\	   	
              &            &            & 14.1             & -103 $\pm$ 0.3 & 22.1 $\pm$ 0.8 & \ldots & Effelsberg & k \\
SBS1129+576  &11 32 02    &+57 22 46   & 3.9 	           & 1506 $\pm$ 3.3& 67 $\pm$ 3.3& \ldots& GMRT		& u	\\
              &            &            & 13.85	           & 1559 $\pm$ 2  & 90         & \ldots& Effelsberg	& q	\\	  	
J1201+0211    &12 01 22    &+02 11 08   & 0.96 $\pm$ 0.09  & 974 $\pm$ 3   & 29 $\pm$ 7	& \ldots& NRT		& b	\\
SBS1159+545  &12 02 02    &+54 15 50   & $<$ 0.10         & 3560 $\pm$ 15 & \ldots	& \ldots& NRT		& b	\\
SBS1211+540  &12 14 02    &+53 45 17   & 0.63             & 894 $\pm$ 7   & 47         & \ldots& Effelsberg	& q	\\
J1215+5223    &12 15 47    &+52 23 14   & 4.7 $\pm$ 0.5	   & 159           & 26.6       & 2.6   & GMRT		& e	\\
              &            &            & 5.24 $\pm$ 0.16  & 158 $\pm$ 1   & 27 $\pm$ 1	& \ldots& NRT		& b	\\		
Tol1214-277  &12 17 17    &-28 02 33   & $<$ 0.10         & 7785 $\pm$ 50 & \ldots	& \ldots& NRT		& b	\\
VCC0428      &12 20 40    &+13 53 22   & 0.65 $\pm$ 0.03  & 794           & 54         & \ldots& ALFALFA	& v	\\
Tol65        &12 25 47    &-36 14 01   & 2.13 $\pm$ 0.24  & 2790 $\pm$ 3  & 40 $\pm$ 6	& \ldots& NRT		& b 	\\
J1230+1202    &12 30 49    &+12 02 43   & 1.03 $\pm$ 0.06  & 1227          & 40 $\pm$ 6 & \ldots& ALFALFA	& v	\\
UGCA292      &12 38 40    &+32 46 01   & 12.9             & 308.3         & 25.2       & \ldots & VLA-ANGST & n \\      
             &            &            & 14.3             & 308.3 $\pm$ 0.1 & 26.9 $\pm$ 0.3 & \ldots & Effelsberg & k    \\
              &            &            & 14.36            & 295.8$\star$ & 25.4       & \ldots& WSRT		& w	\\                  
GR8           &12 58 40    &+14 13 03   & 9.0 $\pm$ 0.9    & 217 $\pm$ 2.2 & 26 $\pm$ 1.2 & 4.3 & GMRT		& e,r 	\\
              &            &            & 5.8              & 213.7         & 21.4         & \ldots & VLA-ANGST & n \\
              &            &            & 8.97 $\pm$ 0.03  & 213           & 26         & \ldots& ALFALFA	& v    \\
	      &		   &	   	& 8.8 $\pm$ 0.09   & 213 $\pm$ 8   & 41 $\pm$ 16& \ldots& GBT		& x	\\	
DD0167       &13 13 23    &+46 19 22   & 3.7 $\pm$ 0.4	   & 150.24       & 18.6        & 2.0   & GMRT		& e	\\
SBS1415+437  &14 17 01    &+43 30 05   & 5.40             & 605 $\pm$ 2   & 49         & \ldots& Effelsberg	& q	\\	
HS1442+4250  &14 44 13    &+42 37 44   & 7.05 $\pm$ 0.16  & 647 $\pm$ 1   & 85 $\pm$ 2	& \ldots& NRT		& b 	\\
HS1704+4332  &17 05 45    &+43 28 49  	& 0.24 $\pm$ 0.05  & 2082 $\pm$ 8  & 33 $\pm$ 15& \ldots& NRT		& b 	\\ 
SagDIG        &19 29 59    &-17 40 41   & 23.0 $\pm$ 1.0   & -78.5 $\pm$ 1 & 19.4 $\pm$ 0.8& \ldots & GMRT	& r	\\
              &            &            & 29.2 $\pm$ 4.9   & -79 $\pm$ 1   & 28         & \ldots& HIPASS	& c	\\  	  	
J2104-0035    &21 04 55    &-00 35 2    & 2.0              & 1401          & 64           & \ldots & GMRT & y \\
HS2134+0400  &21 36 59    &+04 14 04   & 0.12 $\pm$ 0.05  & 5090 $\pm$ 4  & 25 $\pm$ 9	& \ldots& NRT		& b 	\\
ESO146-G14   &22 13 00    &-62 04 03   & 16.5 $\pm$ 3.1   & 1693 $\pm$ 5  & 130      	& \ldots & HIPASS	& c	\\   
	      &		   &	   	& 16.3	           & 1692.1        & 129.4      & \ldots& HIPASS	& i	\\
	      &		   &	 	& 8.4 	           & 1691.1        & 140.4      & \ldots& ATCA		& z	\\
J2238+1400    &22 38 31    &+14 00 30   & $<$ 0.15         & 6160 $\pm$ 20 & 50  	& \ldots& NRT		& b	\\
	
\hline

\end{tabular}

$^{a}$ Quoted value is for the major axis diameter only.

\end{minipage}

\end{center}

\end{table*}


\subsection{Effelsberg \ion{H}{i} Observations}

Of the 87 sources with no published \ion{H}{i} line observations, we chose the subsample of 29 targets, with declinations above $-25\,^{\circ}$, suitable for the latitude of Effelsberg, and observable during the period July through November 2012.

\subsubsection{Observations and Data Reduction}

To obtain measurements of the \ion{H}{i} flux density, S$_{\nu}$, at frequency $\nu$, we performed pointed observations with the 100-m single-dish Effelsberg radio telescope\footnote{http://www.mpifr-bonn.mpg.de/8964/effelsberg} using the central feed of the 7-beam L-band receiver and the AFFTS backend. The latter is a Field Programmable Gate Array (FPGA)-based Fast Fourier Transform (FFT) spectrometer (Klein et al. 2006), providing 16k spectral channels over a bandwidth of 100 MHz (spectral resolution of 7.1 kHz). This results in a velocity resolution at 21 cm of 1.5 km s$^{-1}$. 

Each of the candidate sources was observed for about one hour in \textit{on-off} mode (effective observing time is $\sim$30 minutes per source). This position switching allows to remove the effect of the bandpass (i.e., frequency-dependent gain).

The L-band unfortunately features a lot of narrow-band radio frequency interference (RFI) at the telescope site. However, due to the high spectral resolution, the total fraction of each spectrum which is polluted is relatively low. To improve the data reduction further, we applied a simple, but in this case, effective RFI mitigation strategy, where we median-filtered the spectra (5 channel width) and searched for 8$\sigma$ outliers in the residual (peaks in excess of 8 times the noise standard deviation or {\it rms}). The associated spectral channels in the spectrum were subsequently replaced with the median-filtered values.

The spectra have been flux-calibrated using a novel technique (Winkel, Kraus \& Bach 2012), which accounts for the frequency dependence of the system temperature and, as such, provides an unbiased calibration over the full bandwidth. This new method requires good knowledge of the temperature of the calibration diode, $T_\mathrm{cal}$, which we have measured repeatedly using the calibration sources NGC\,7027, 3C\,48 and 3C\,147 during the observing sessions in July and November 2012. For the most accurate absolute flux calibration, total-power measurements of the (Galactic) target S7 (Kalberla, Mebold \& Reif 1982) were used to fix $T_\mathrm{cal}$ at the frequency of 1420.1 MHz. The latter usually leads to an uncertainty of less than 3\% for the 7-beam system (Winkel et al. 2010), while we find a scatter of the individually determined $T_\mathrm{cal}$ spectra of about 5\%, with respect to the average of all $T_\mathrm{cal}$ spectra.

Using the SIMBAD Astronomical Database, we obtained the radial velocity (heliocentric, converted from the optical to the radio convention; Eq.~2; Sect.~4) value for the sources, in order to directly extract the relevant part of each spectrum. Residual baselines were removed using a polynomial fit. For each resulting profile, the residual {\it rms} and integrated flux density of the source, if detected, was determined. The baseline {\it rms} can also be used to calculate a theoretical flux limit ($\sigma$) for a hypothetical source with a Gaussian-like \ion{H}{i} profile of width\footnote{For different widths the flux density limit scales with $\sqrt{w_{50}}$.} $w_{50}$ = 50 km s$^{-1}$ (equal to the full width at half maximum, FWHM, for a Gaussian).


Of the 29 observed sources, there were seven (excluding J0014-0044; Sect.~2.2.2) detections ($\gg$ 5$\sigma$), three marginal detections ($>$ 5$\sigma$), six uncertain detections ($\sim$ 5$\sigma$) and 12 non-detections ($\ll$ 5$\sigma$).

\subsubsection{Effelsberg \ion{H}{i} Properties}

Figure~1 displays \ion{H}{i} spectra and SDSS Data Release 9 (DR9; Ahn et al. 2012) composite images of the 10 detected sources (including marginal detections) ordered by right ascension, and also of  J0014-0044 in the field of UCG 139 (see below). 

Table~2 contains the Effelsberg \ion{H}{i} parameters. In addition to the \ion{H}{i} integrated flux density and line width, Table~2 contains the effective observing time on-source, $t$, the systemic local standard-of-rest radial velocity (radio convention; Eq.~2; Sect.~4), v$_{sys}^{lsr}$, the 5$\sigma$ detection limit and a characterization of the \ion{H}{i} line profile shape.   

The \ion{H}{i} integrated flux density, S$_{\ion{H}{i}} = \int S_{\nu } \, d\nu$, systemic radial velocities and line widths were derived from the 21 cm baseline-subtracted profiles obtained with the single-dish observations. Typically, the Effelsberg flux density errors are in the range 5 -- 10\%. In the cases where the \ion{H}{i} profile is double-horn (J0014-0044 in the field of UGC 139; Fig.~1 and see below; J0113+0052 and J0204-1009), we marked the positions of 50\% of the peak on each side of the double-horn profile, and estimated the line width at these points (e.g., Springob et al. 2005).



\begin{figure*} 
\begin{center} 

\includegraphics[width=9cm]{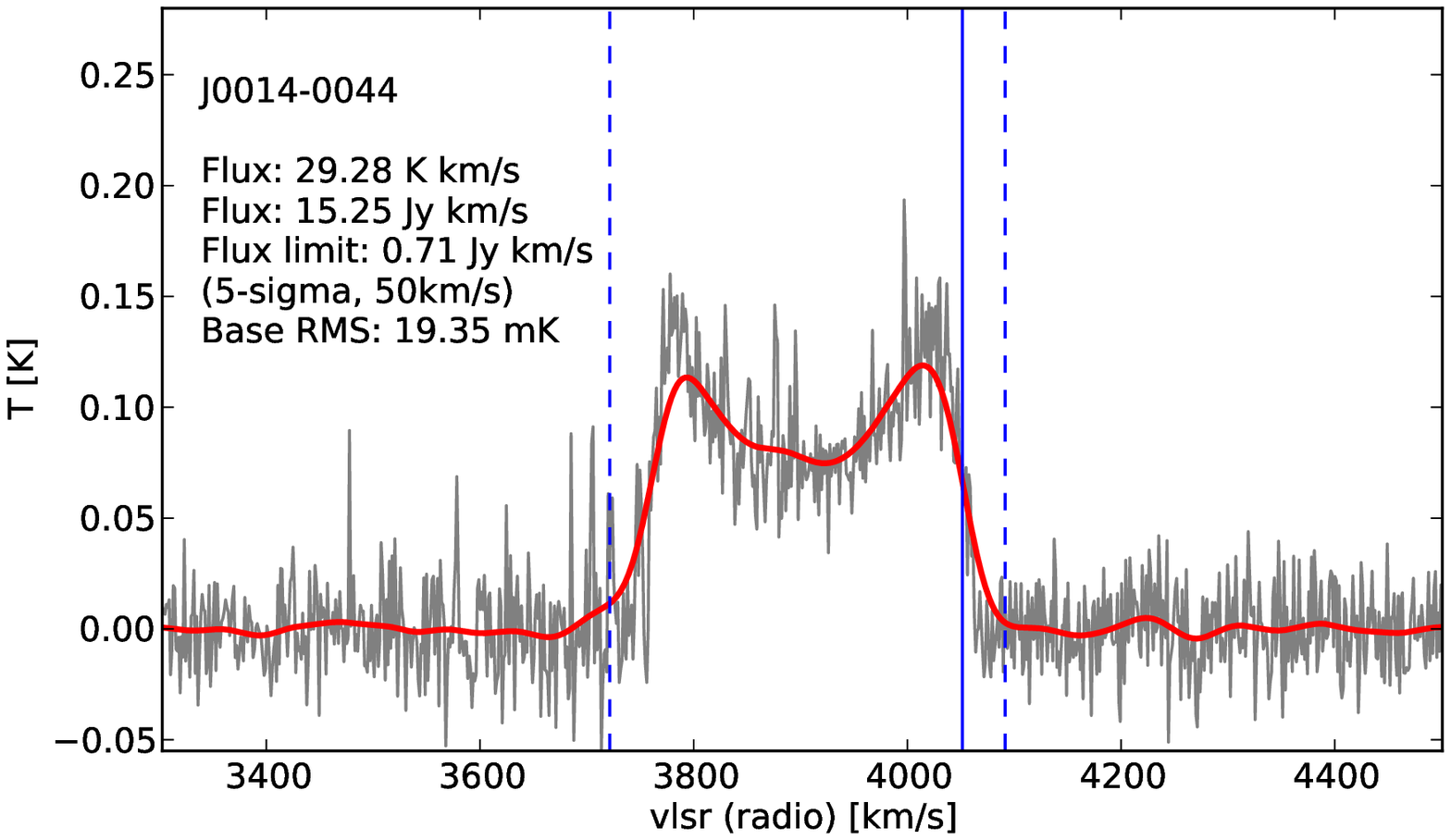} 
\includegraphics[width=7cm]{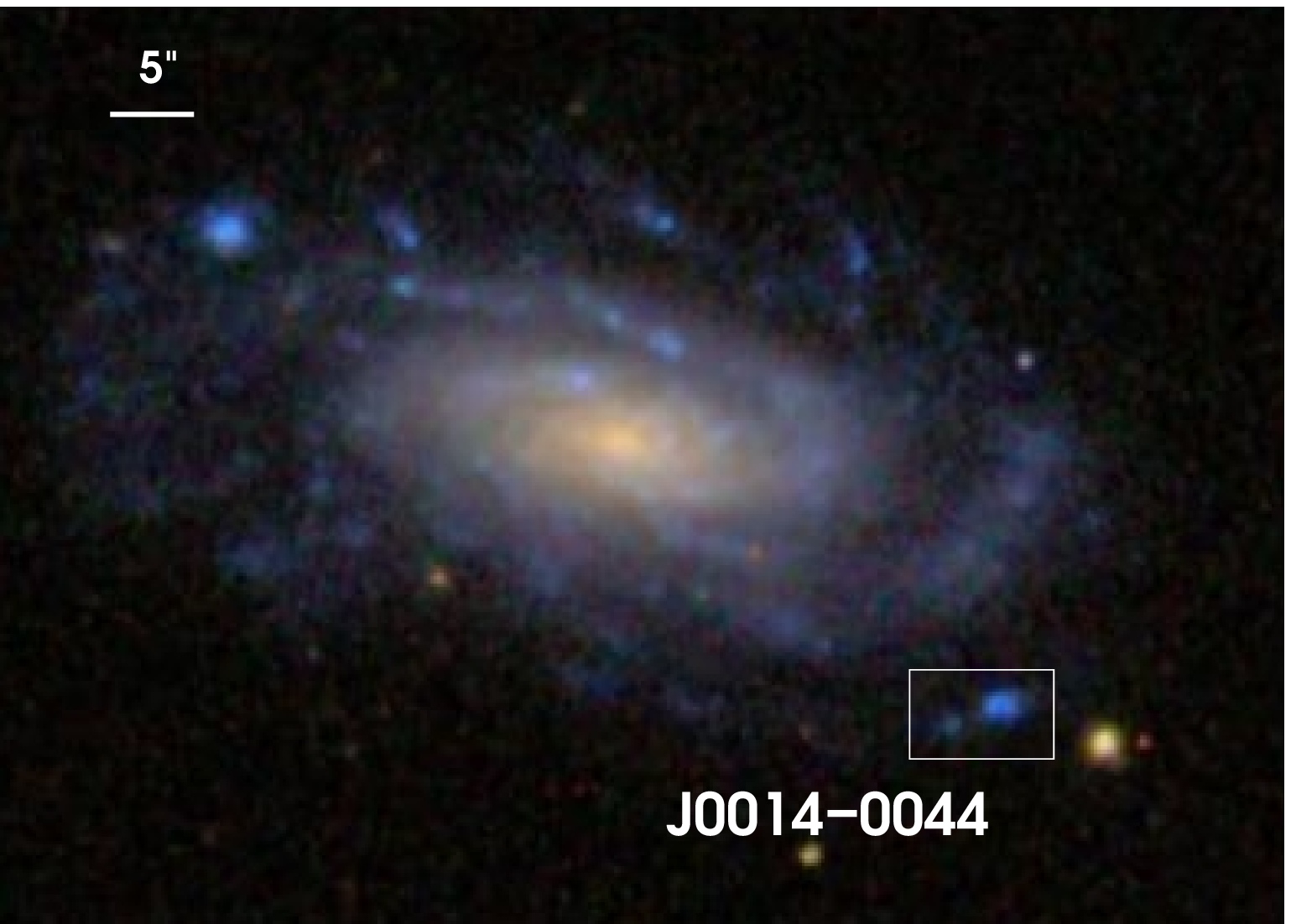} 

\includegraphics[width=9cm]{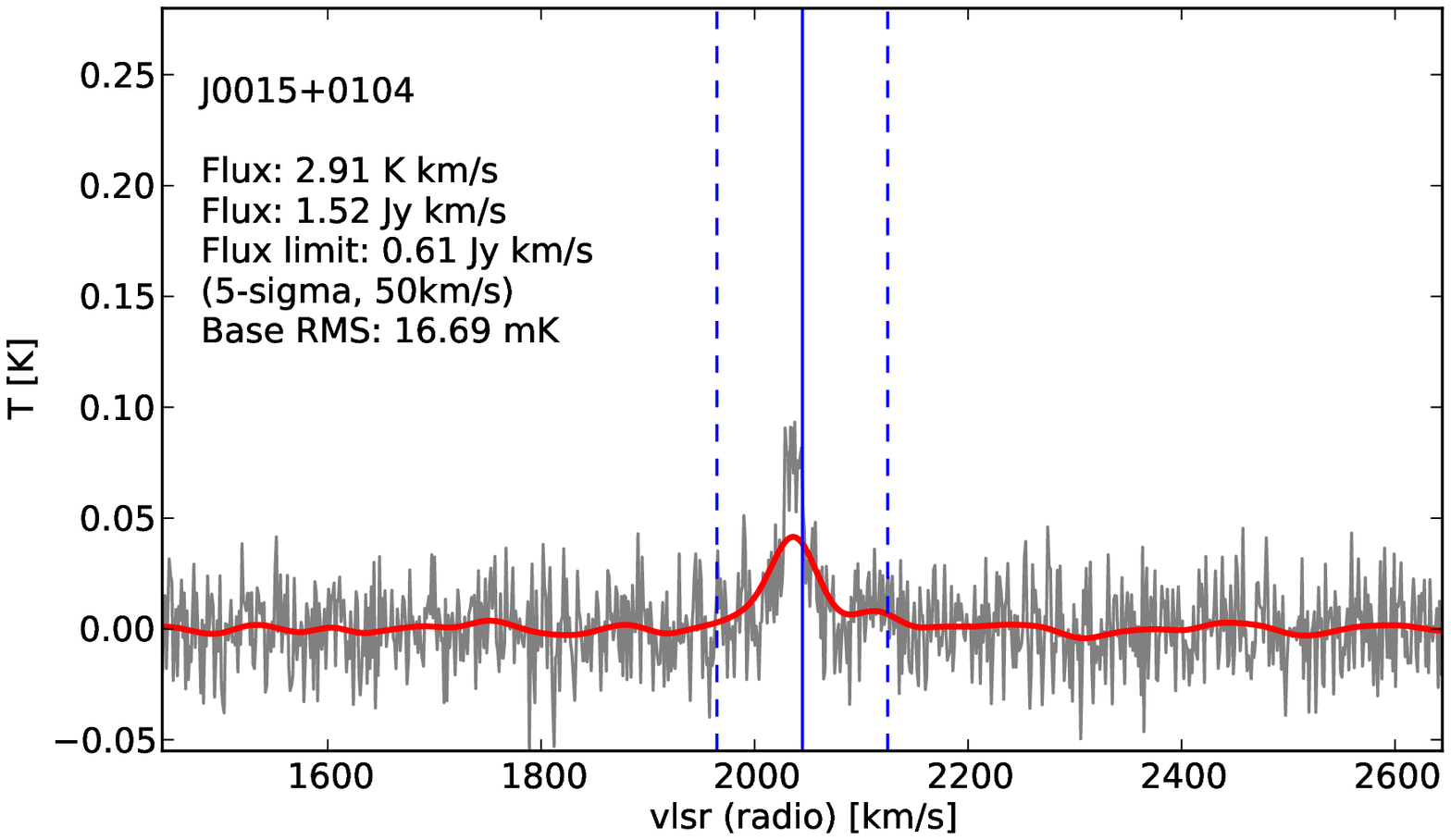} 
\includegraphics[width=5cm]{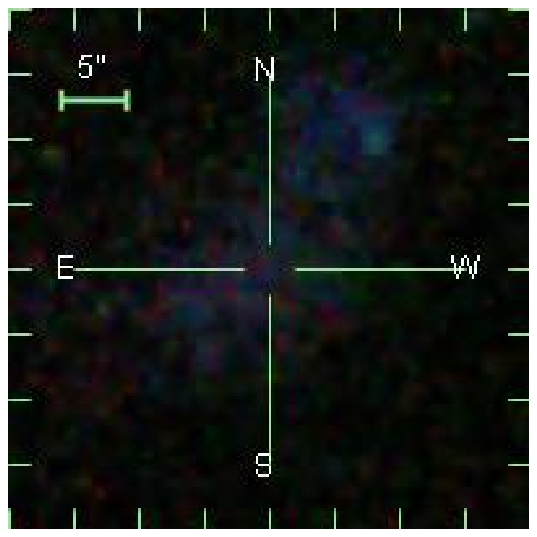}

\includegraphics[width=9cm]{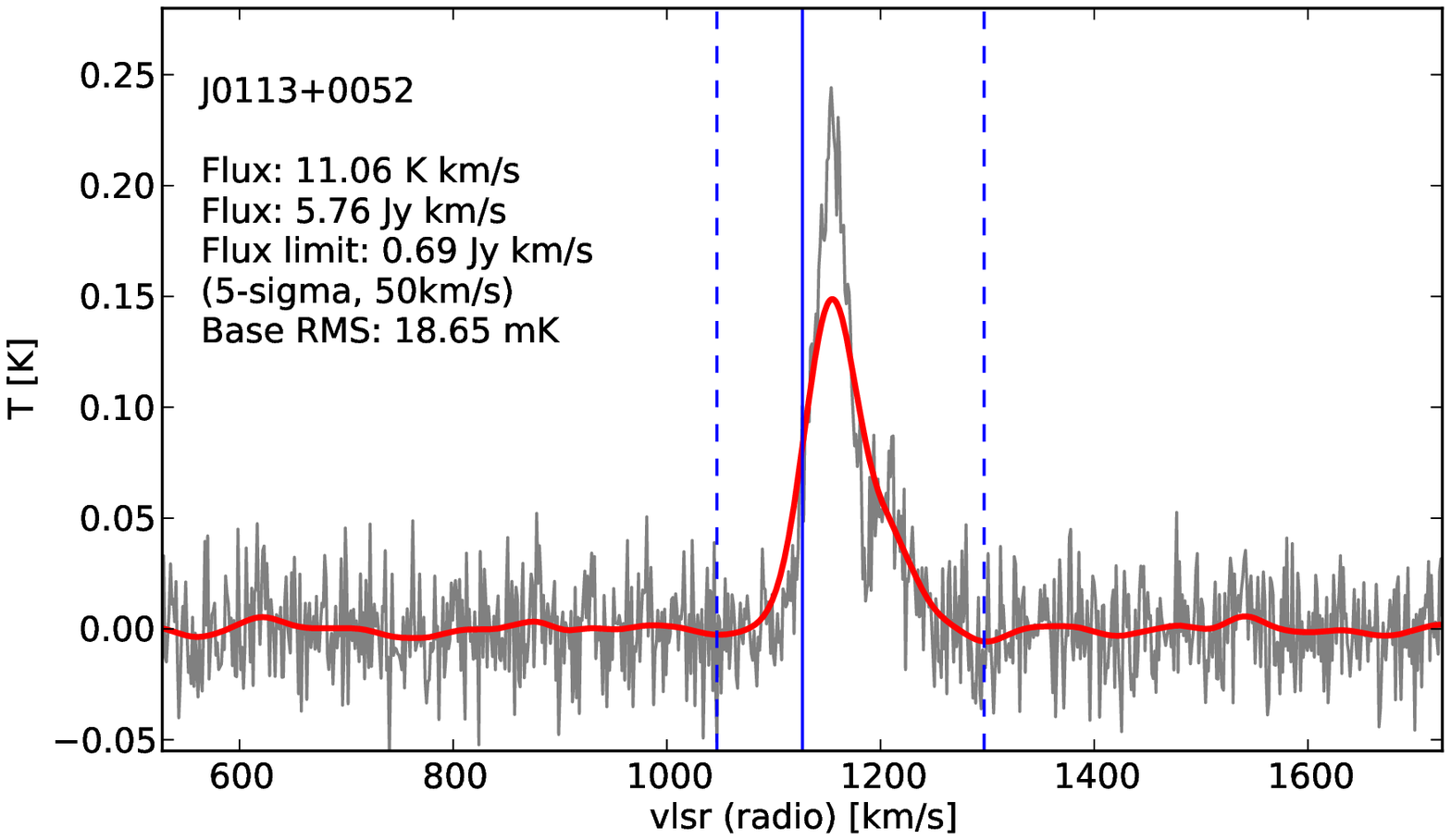} 
\includegraphics[width=5cm]{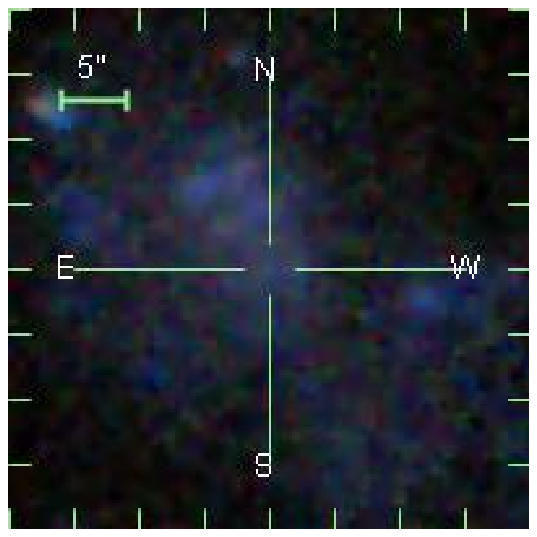}

\caption{\ion{H}{i} Effelsberg spectra (left) and SDSS fields (right) of the detected (including marginal) XMP galaxies. The \ion{H}{i} spectra of J0014-0044 is dominated by the \ion{H}{i} gas of the nearby spiral galaxy UGC 139 (Sect.~2.2.2). In the spectra, the blue vertical solid lines mark the radial velocity from SIMBAD (heliocentric; converted from the optical to the radio convention), the blue vertical dashed lines mark the regions for the baseline computation and flux density integration, and the red lines represent a smoothed
version of the spectrum, using a Gaussian kernel of width $\sigma_\mathrm{sm}=21.22~\mathrm{km\,s}^{-1}$. The smoothing kernel was chosen such that the resulting spectral resolution becomes
$\sigma_\mathrm{res}=21.25~\mathrm{km\,s}^{-1}$, which is optimally suited to detect a (Gaussian) feature of line width $\sqrt{8\ln2}\,\sigma_\mathrm{res}=50~\mathrm{km\,s}^{-1}$ (FWHM). Note that the integrated flux densities and limits were derived from the full-resolution spectra/{\it rms}.}


\end{center}
\end{figure*}

\setcounter{figure}{0}

\begin{figure*}
\begin{center}

\includegraphics[width=9cm]{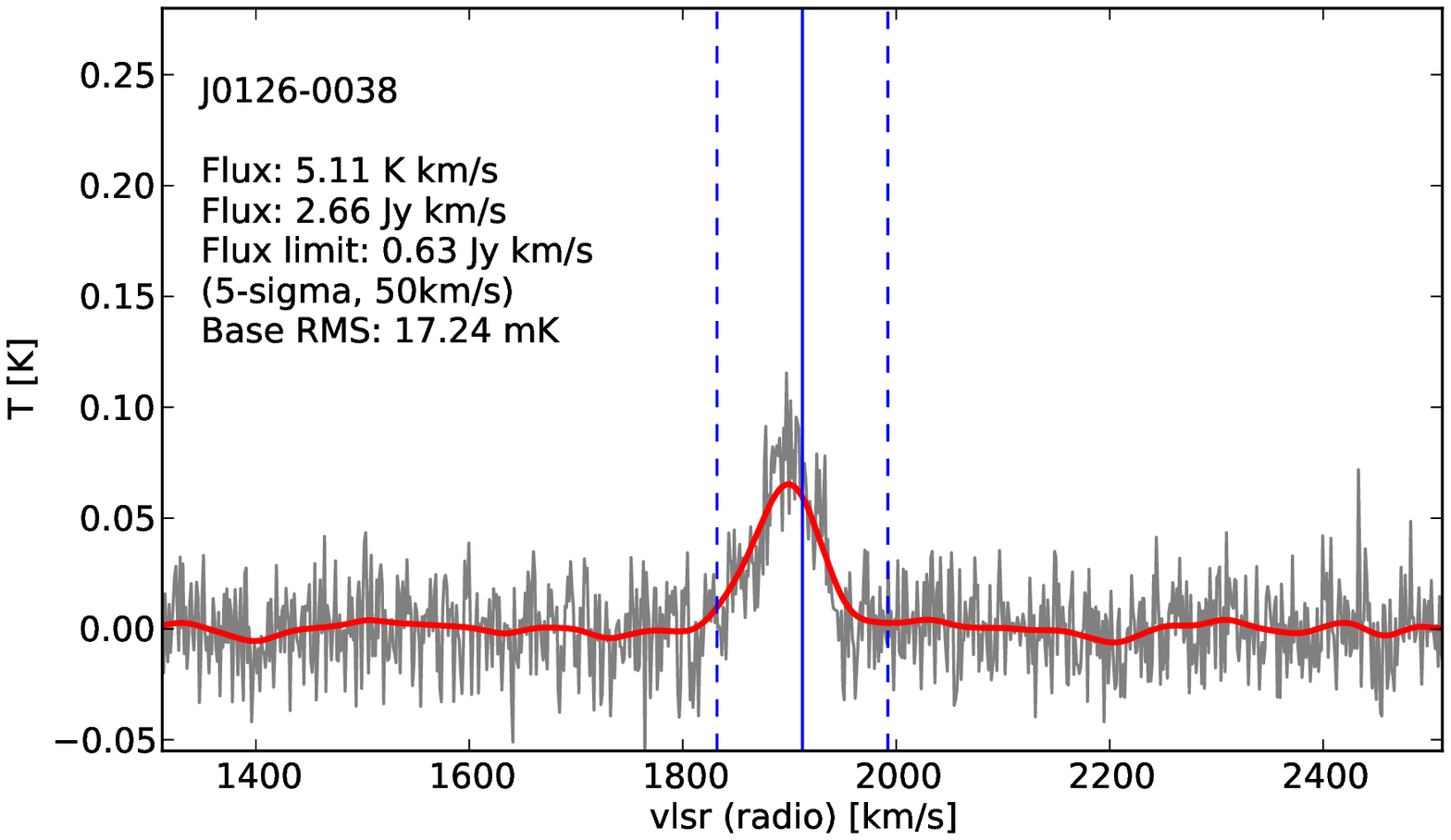} 
\includegraphics[width=5cm]{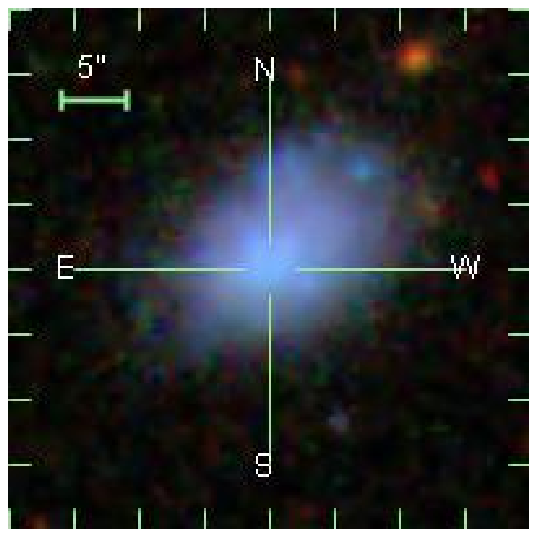}

\includegraphics[width=9cm]{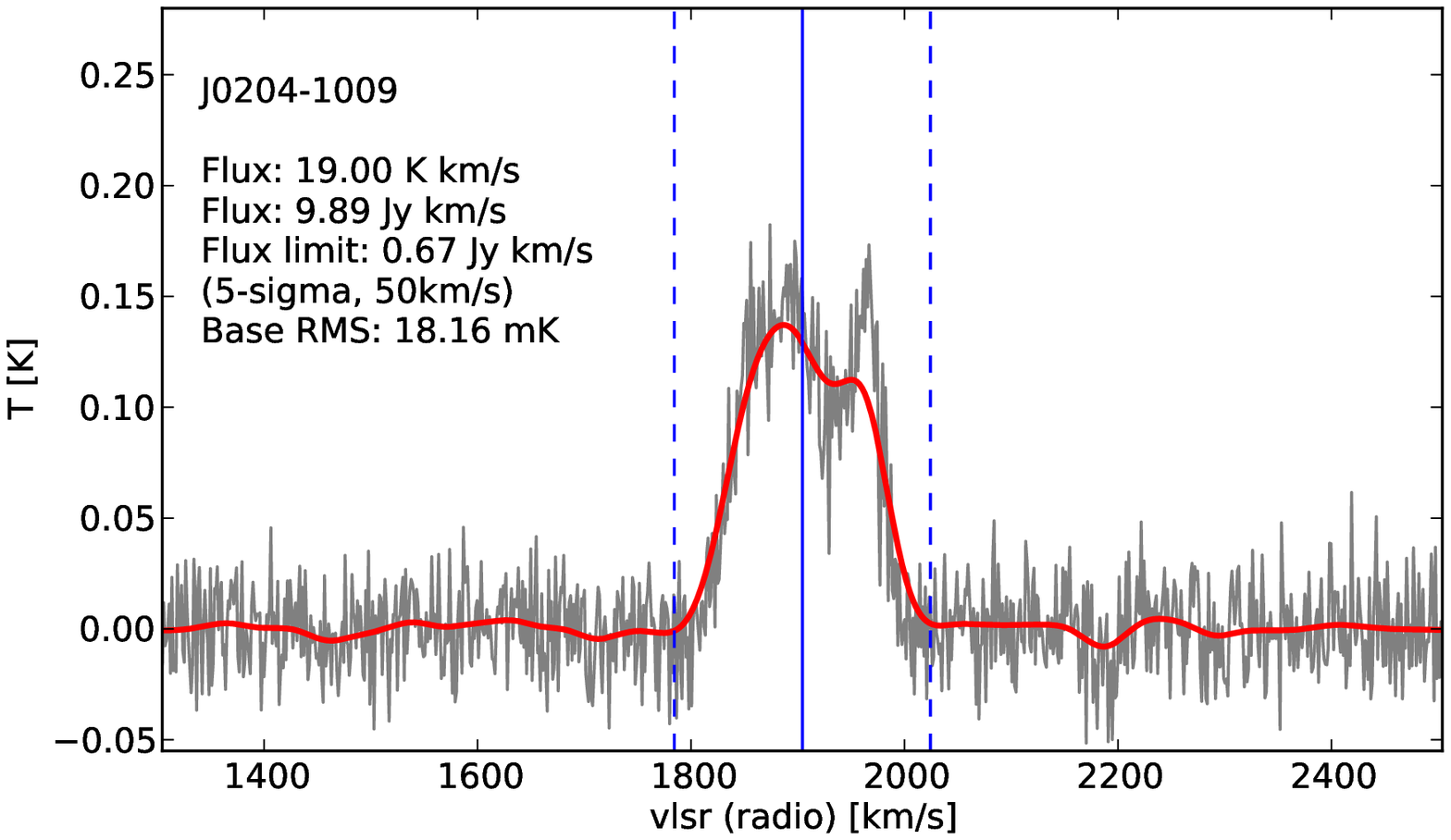} 
\includegraphics[width=5cm]{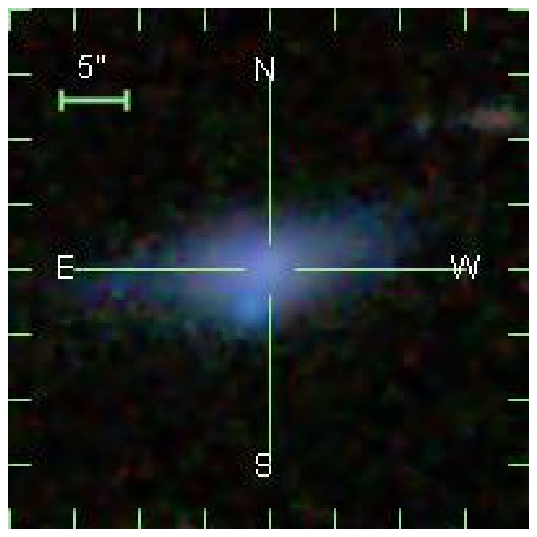}

\includegraphics[width=9cm]{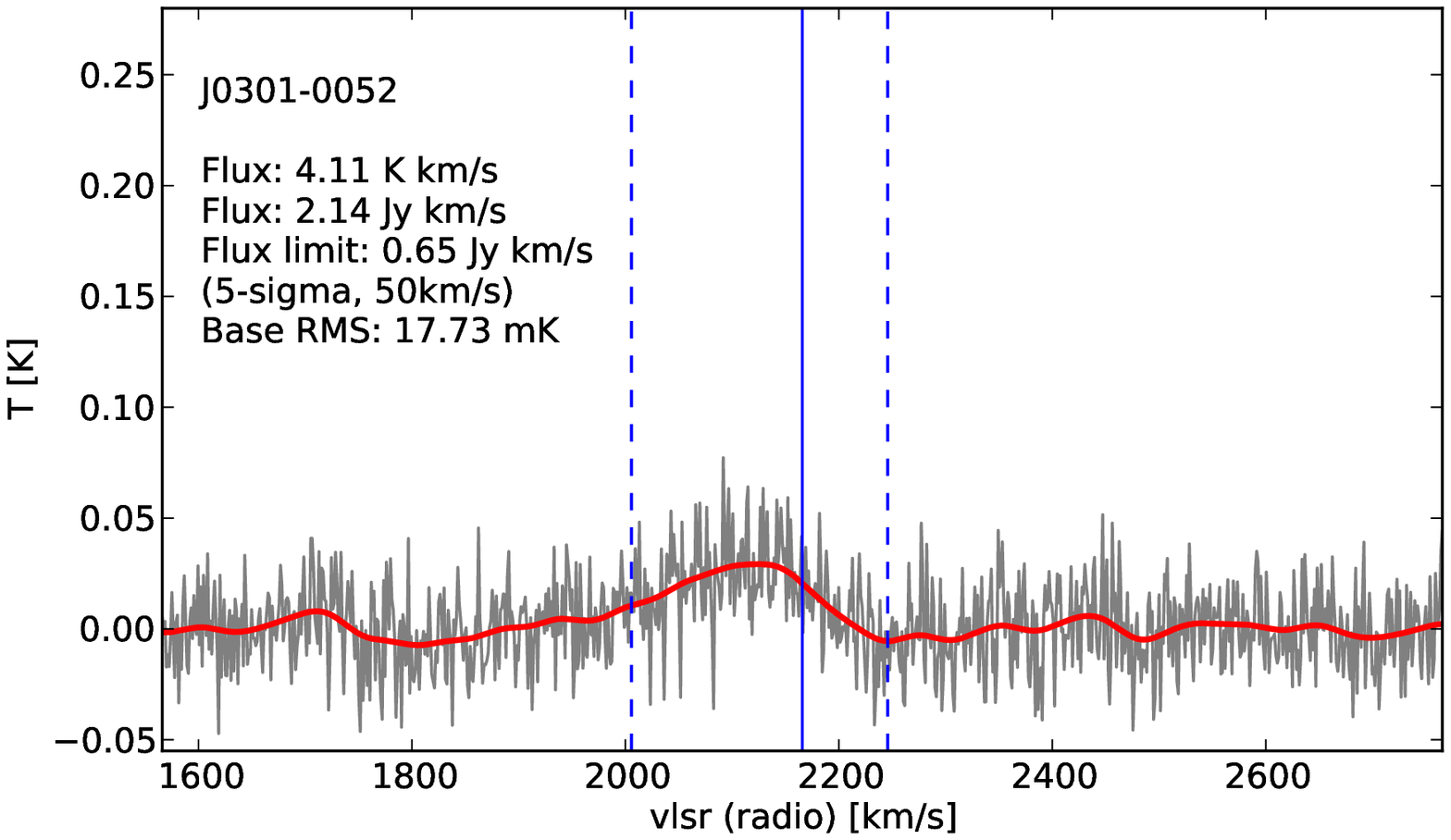} 
\includegraphics[width=5cm]{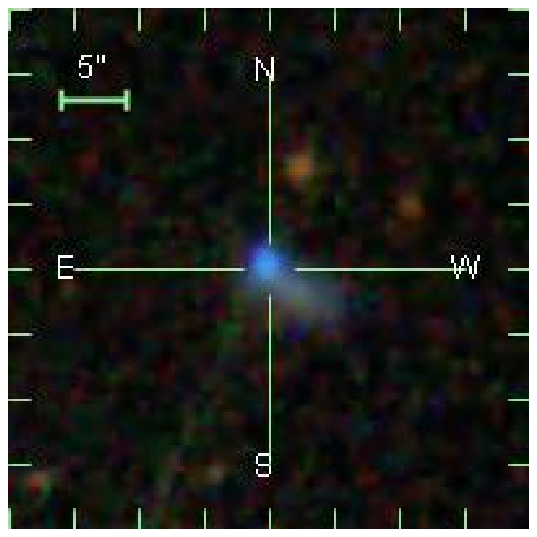}

\caption{Continued from Fig.~1.}

\end{center}
\end{figure*}

\setcounter{figure}{0}

\begin{figure*} 
\begin{center}

\includegraphics[width=9cm]{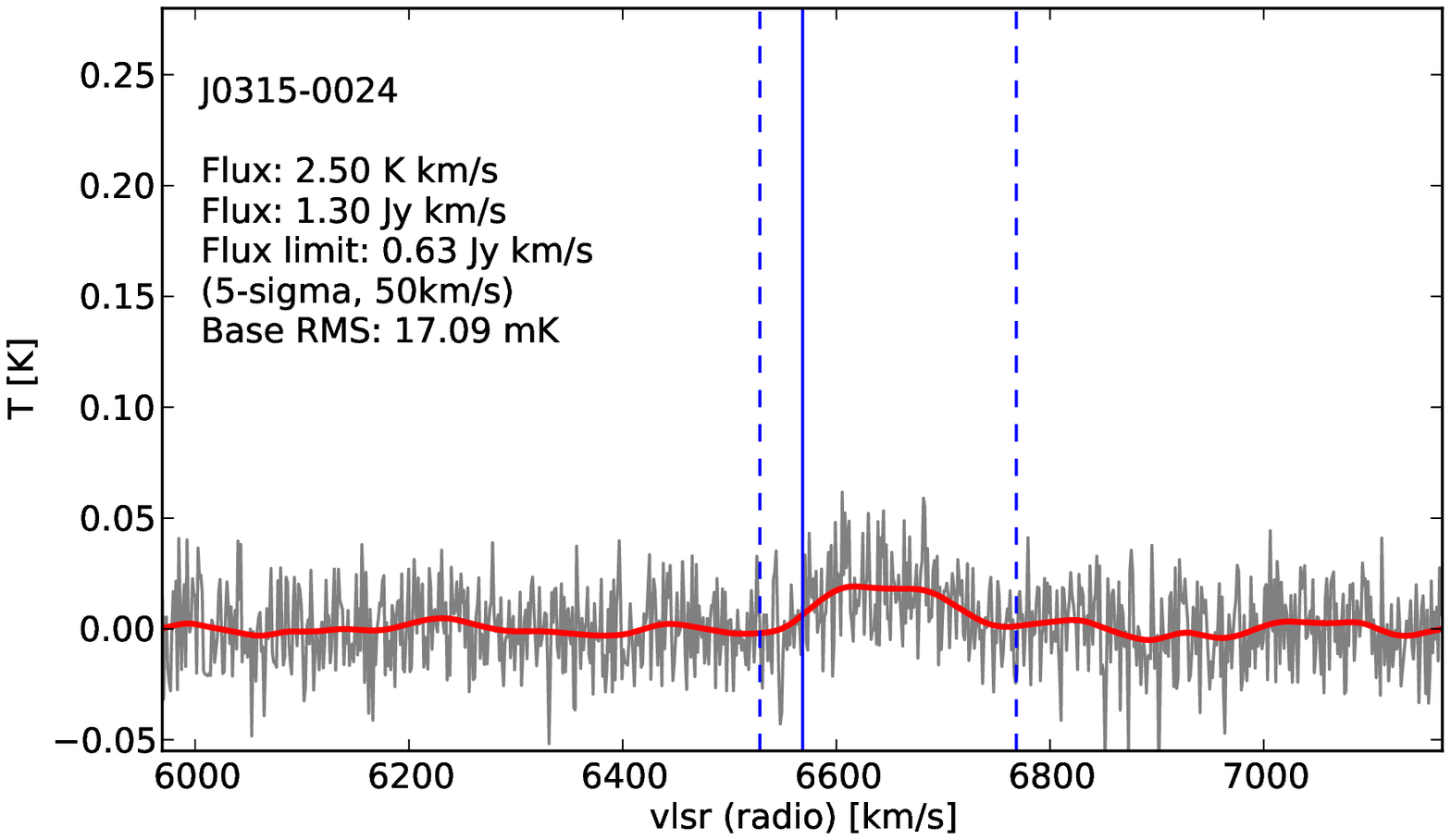} 
\includegraphics[width=5cm]{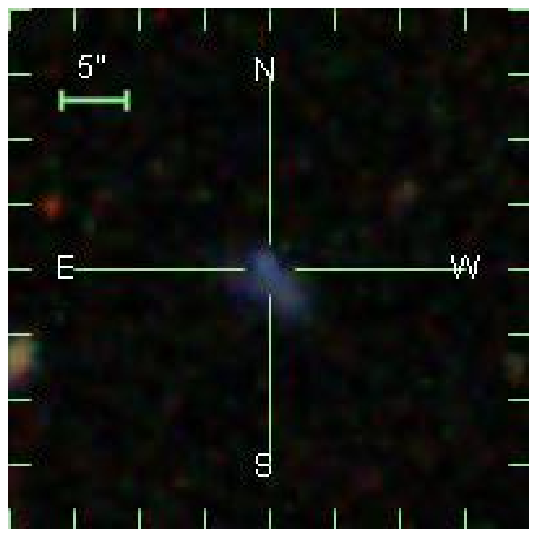}

\includegraphics[width=9cm]{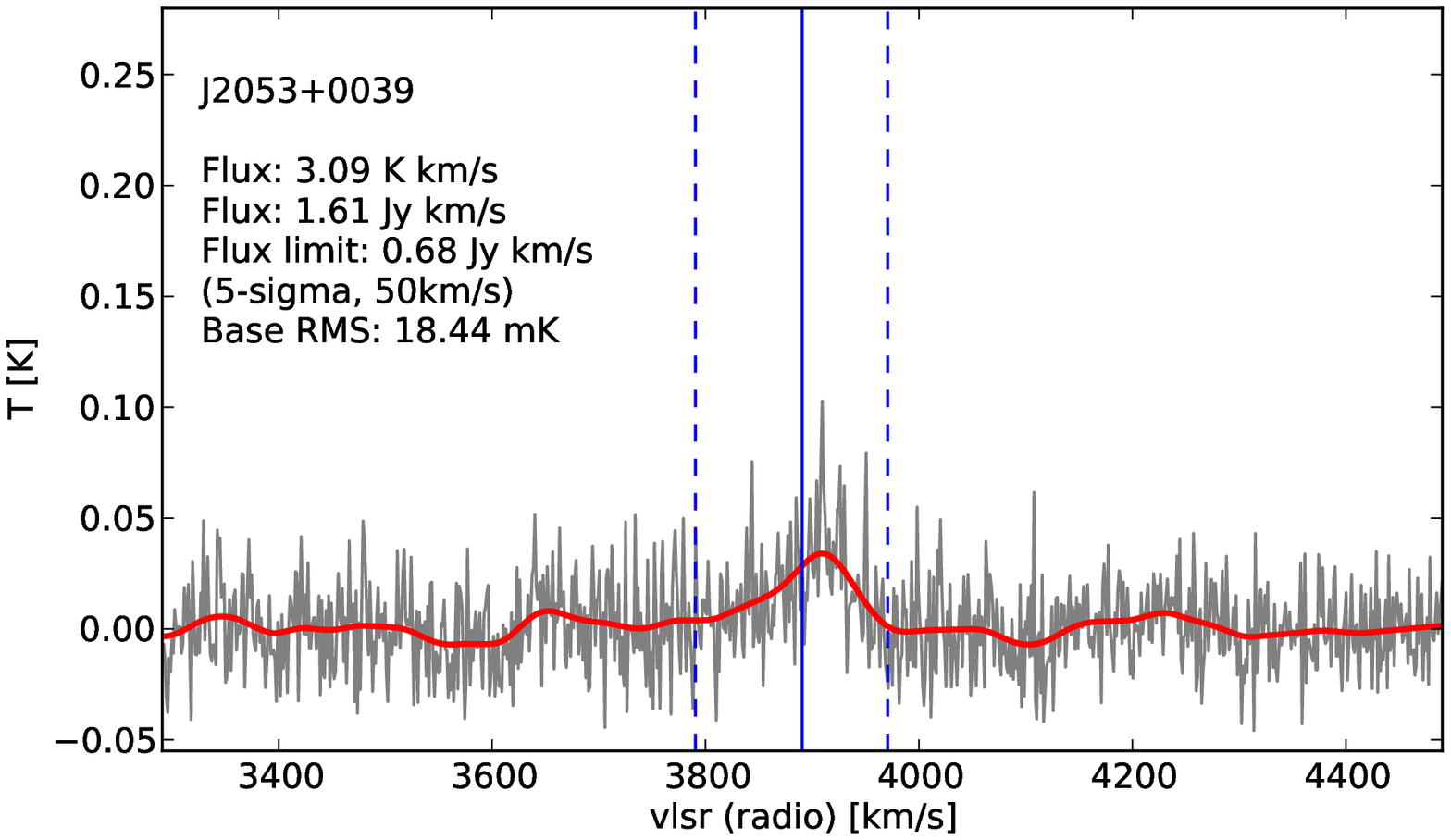}  
\includegraphics[width=5cm]{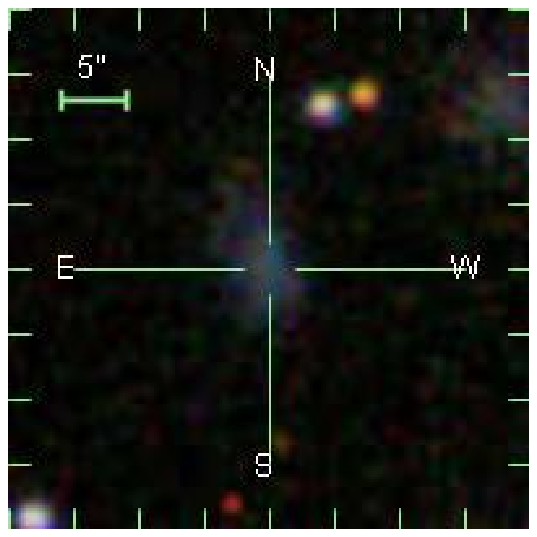}

\includegraphics[width=9cm]{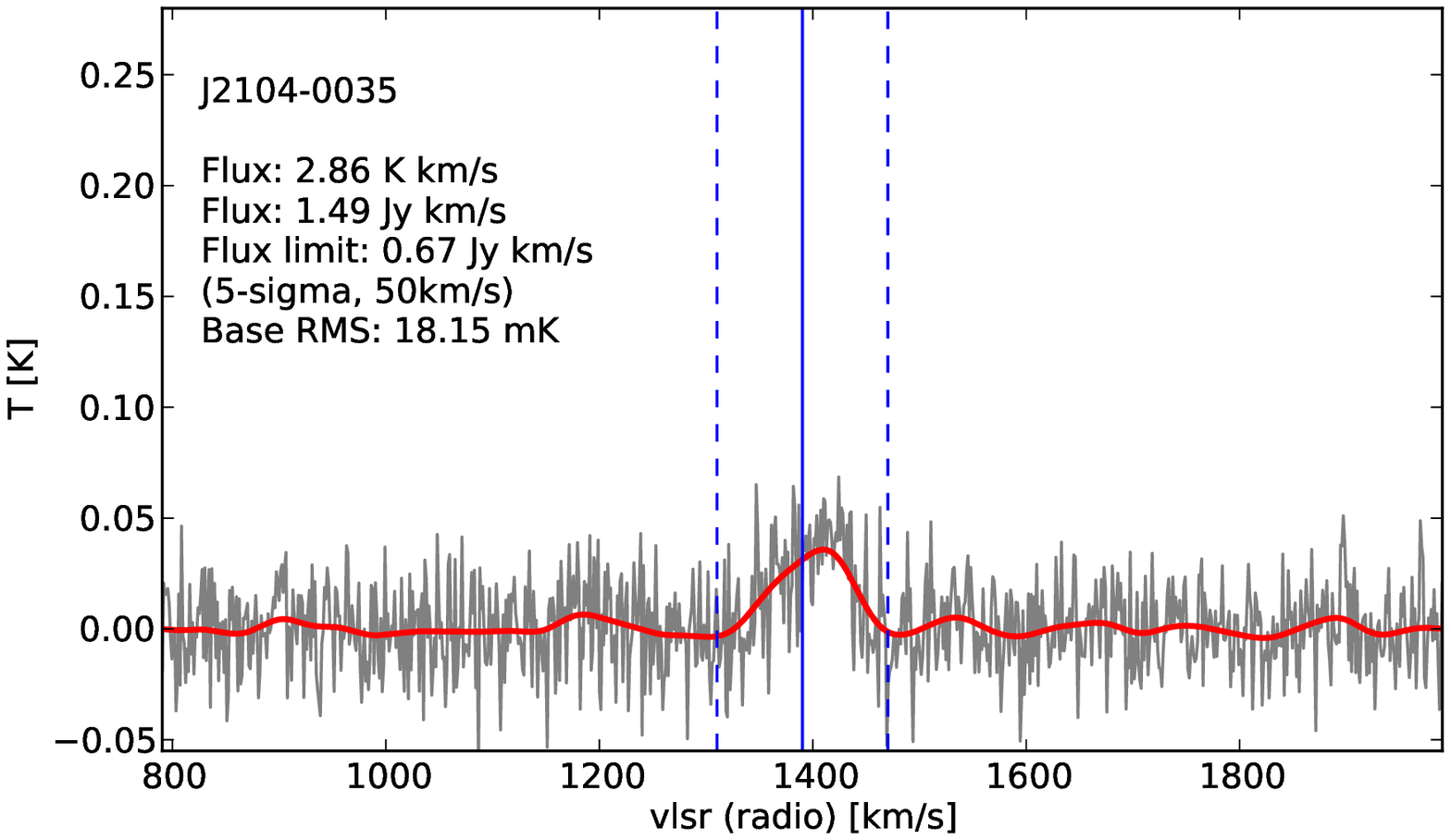} 
\includegraphics[width=5cm]{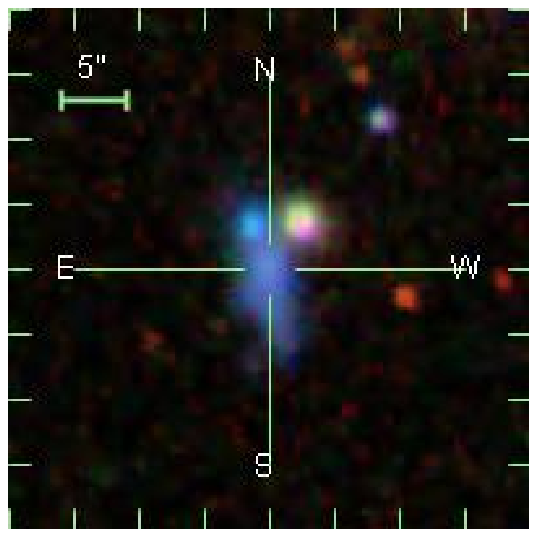}

\caption{Continued from Fig.~1.}

\end{center}
\end{figure*}

\setcounter{figure}{0}

\begin{figure*} 
\begin{center}

\includegraphics[width=9cm]{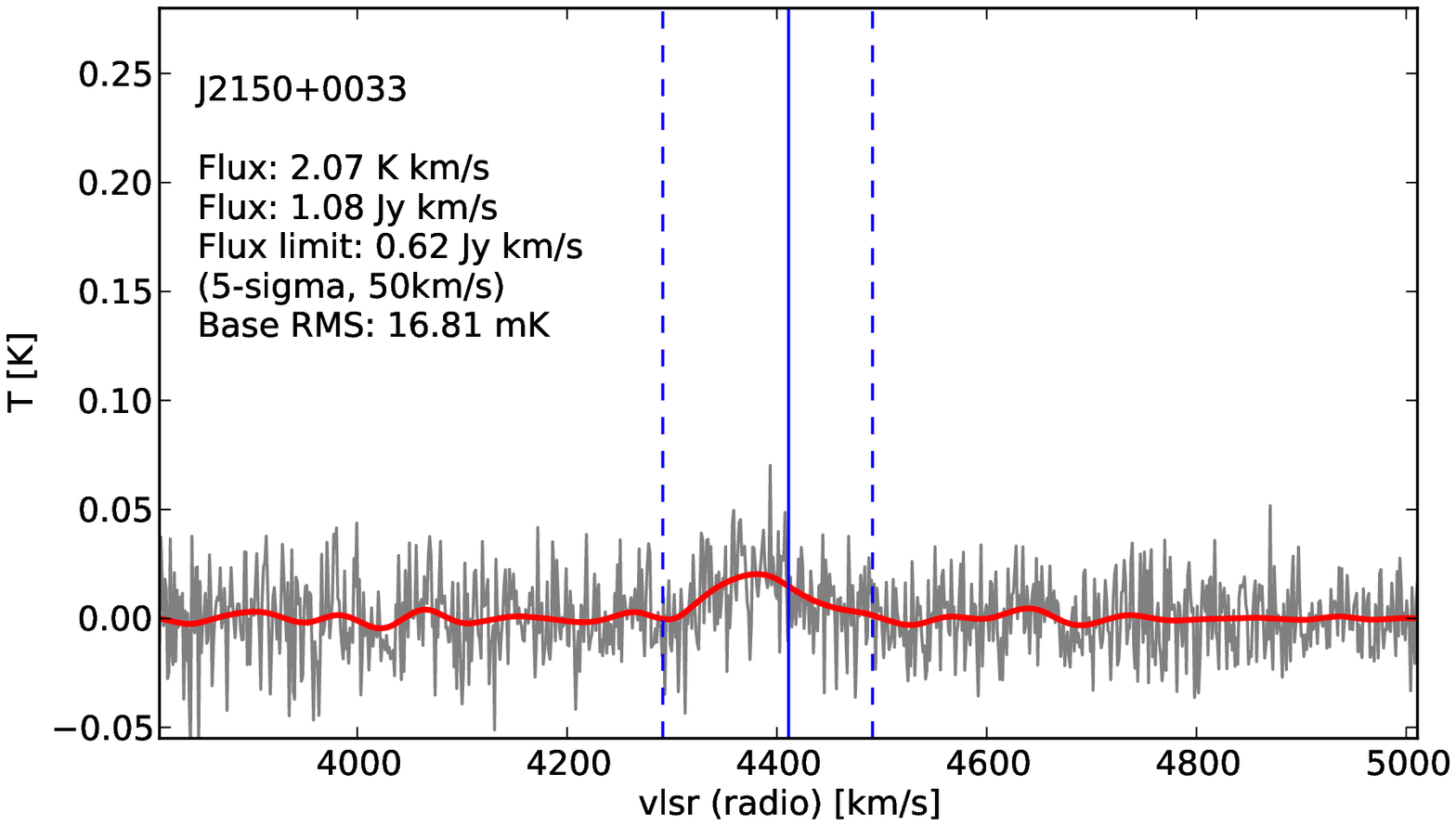} 
\includegraphics[width=5cm]{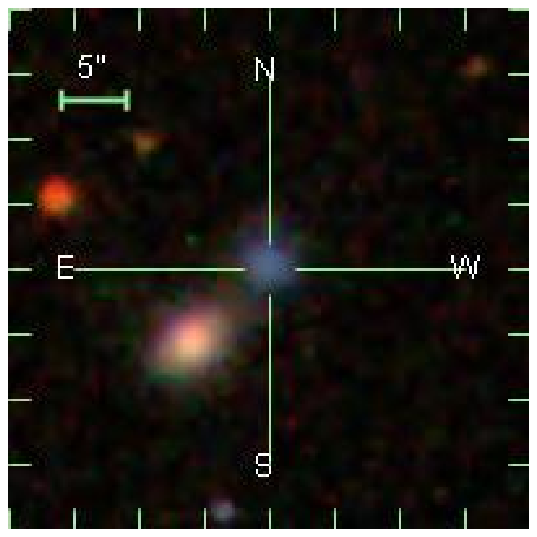}

\includegraphics[width=9cm]{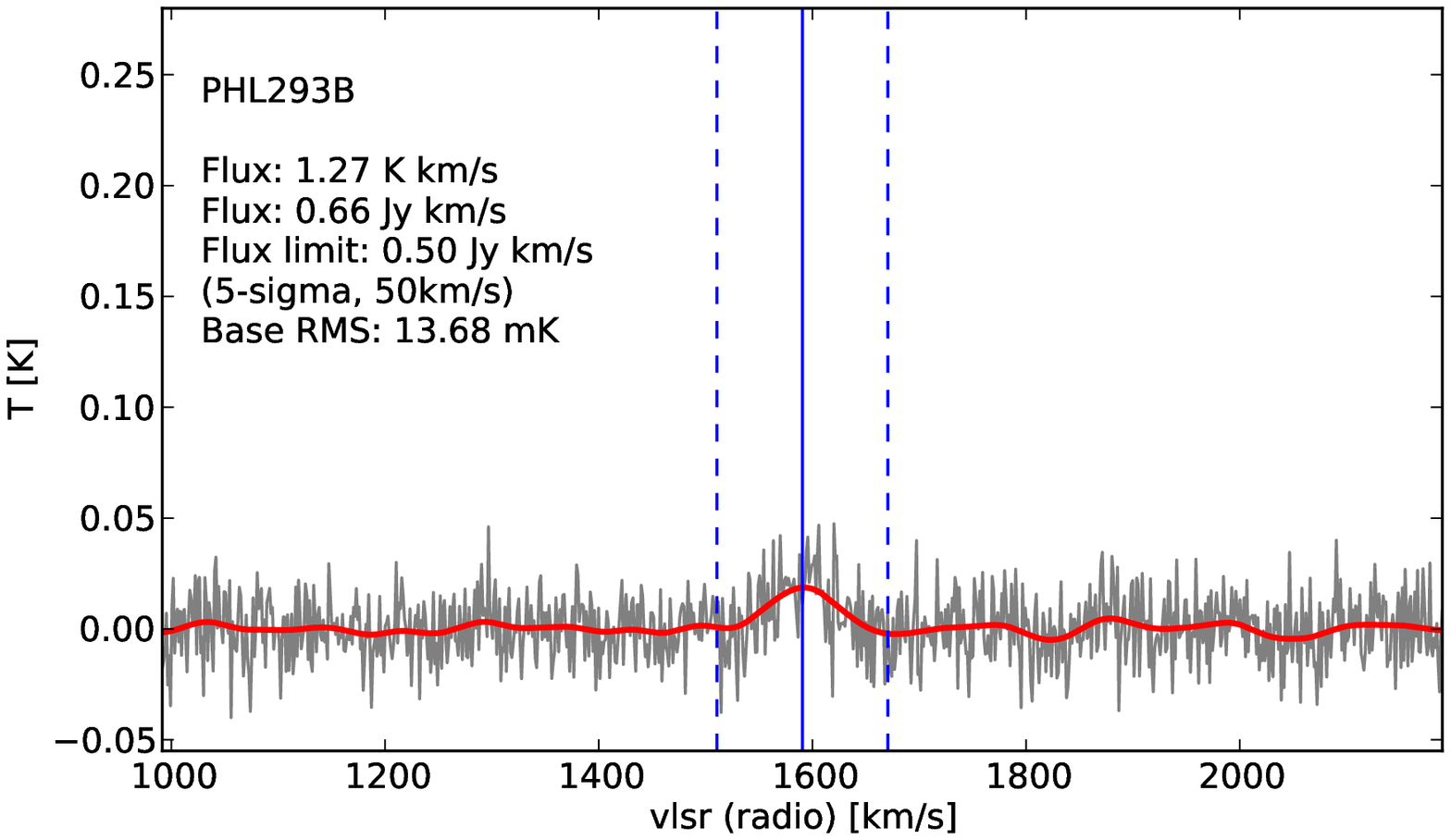} 
\includegraphics[width=5cm]{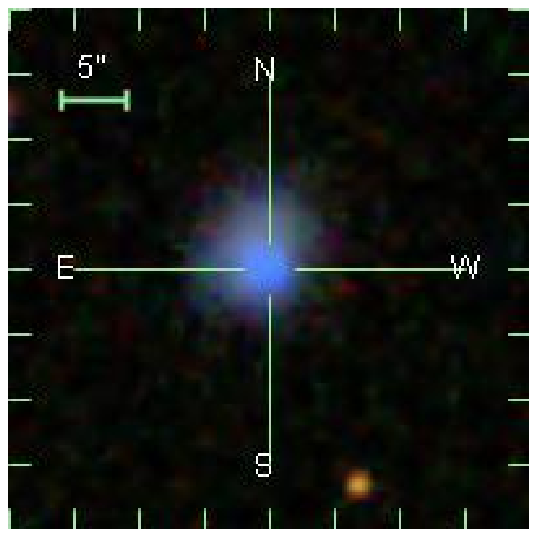}

\caption{Continued from Fig.~1.}

\end{center} 
\end{figure*} 



\setcounter{table}{1}

\begin{table*}

\begin{center}

\begin{minipage}[c]{175mm}

\caption{The \ion{H}{i} parameters from our new Effelsberg observations for 29 XMP galaxies in the local Universe.
Col.~1: Source name.
Col.~2: Right Ascension.
Col.~3: Declination.
Col.~4: Effective integration time on-source.    
Col.~5: Systemic local standard-of-rest radial velocity (radio covention).   
Col.~6: \ion{H}{i} line width at 50\% of the peak flux density. 
Col.~7: \ion{H}{i} integrated flux density.
Col.~8: 5$\sigma$ error in integrated flux density.
Col.~9: Note on detection. Detections are $\gg$ 5$\sigma$, non-detections are $\ll$ 5$\sigma$, marginal detections are $>$ 5$\sigma$ and uncertain detections are $\sim$ 5$\sigma$.
Col.~10: Profile of the \ion{H}{i} line, single (s)- or double (d)-peaked and symmetric (sym) or asymmetric (asym).}

\begin{tabular}{l c c c c c c c c c c }

\hline

Name & RA(J2000)           &  DEC(J2000)                    & $t$  & v$_{sys}^{lsr}$    & $w_{50}$     & S$_{\ion{H}{i}}$ & 5$\sigma$       & Detection & Profile \\
     &  $^h$ $^m$ $^s$     &  \degree \, \arcmin \, \arcsec & min    & km s$^{-1}$        & km s$^{-1}$  & Jy km s$^{-1}$   & Jy km s$^{-1}$  &      &         \\
(1)  & (2)      & (3)                       & (4)           & (5)         &(6)          & (7)          & (8)      & (9)       & (10)  \\

\hline
\hline

J0004+0025	& 00 04 21.6	& +00 25 36	& 30 & \ldots	 	& \ldots	& \ldots& 0.7	& uncertain	& \ldots \\
J0014-0044$^a$	& 00 14 28.8	& -00 44 44	& 30 & 3915.17		& 320.07        & 15.3	& 0.7	& yes	        & d-sym \\
J0015+0104	& 00 15 20.679	& +01 04 36.99	& 30 & 2036.79		& 27.87	        & 1.5	& 0.6	& yes		& s-asym \\
J0016+0108	& 00 16 28.254	& +01 08 01.92	& 60 & \ldots	 	& \ldots	& \ldots& 0.5	& no		& \ldots \\
J0029-0108	& 00 29 04.73	& -01 08 26.3 	& 30 & \ldots		& \ldots	& \ldots& 0.7	& uncertain	& \ldots \\
J0029-0025	& 00 29 49.497	& -00 25 39.89	& 30 & \ldots		&\ldots	        & \ldots& 0.8	& no		& \ldots \\
J0057-0022	& 00 57 12.603	& -00 21 57.67	& 30 & \ldots		&\ldots	        & \ldots& 0.6	& uncertain   	& \ldots \\
J0107+0001	& 01 07 50.817	& +00 01 28.42	& 30 & \ldots		&\ldots	        & \ldots& 0.6	& no		& \ldots \\
J0113+0052	& 01 13 40.45	& +00 52 39.2	& 30 & 1156.22		& 34.58	        & 5.8	& 0.7	& yes		& d-asym  \\
J0126-0038	& 01 26 46.4	& -00 38 39 	& 30 & 1898.32		& 49.91	        & 2.7	& 0.6	& yes		& s-asym  \\
J0135-0023	& 01 35 44.037	& -00 23 16.89	& 30 & \ldots 	 	& \ldots	& \ldots& 0.6	& uncertain  	& \ldots \\
HKK97L14	& 02 00 09.9	& +28 49 57	& 30 & \ldots		& \ldots	& \ldots& 0.6	& uncertain	&\ldots  \\
J0204-1009	& 02 04 25.55	& -10 09 36.0 	& 30 & 1906  		& 112.04        & 9.9	& 0.7	& yes		& d-asym  \\
J0254+0035	& 02 54 28.94	& +00 35 50.4	& 30 & \ldots	 	&\ldots	        & \ldots& 0.7	& no		& \ldots \\
J0301-0059	& 03 01 26.3	& -00 59 26	& 30 & \ldots		&\ldots	        & \ldots& 0.6	& no		& \ldots \\
J0301-0052$^b$	& 03 01 49.03	& -00 52 57.4	& 30 & 2107.85		& 110.2	        & 2.1	& 0.7	& yes		& s-asym   \\
J0303-0109	& 03 03 31.3	& -01 09 47	& 30 & \ldots		& \ldots        & \ldots& 0.8	& no		& \ldots \\
J0313+0010	& 03 13 01.57	& +00 10 40.3	& 30 & \ldots		&\ldots	        & \ldots& 0.6	& no		& \ldots \\
J0315-0024	& 03 15 59.9	& -00 24 26	& 30 & 6649.89		& 91.71	        & 1.3	& 0.6	& marginal  	& s-asym  \\	
J0341-0026	& 03 41 18.1	&-00 26 28	& 30 & \ldots		&\ldots	        & \ldots& 0.6	& no		& \ldots \\
J2053+0039	& 20 53 12.597	& +00 39 14.25	& 30 & 3906.98	  	& 52.79         & 1.6	& 0.7	& yes		& s-asym  \\
J2104-0035	& 21 04 55.3	& -00 35 22	& 30 & 1404.00		& 52.04	        & 1.5	& 0.7	& yes		& s-asym  \\
J2105+0032	& 21 05 08.6	& +00 32 23	& 30 & \ldots		& \ldots	& \ldots& 0.7	& no		& \ldots        \\
J2120-0058	& 21 20 25.937	& -00 58 26.53	& 30 & \ldots		& \ldots	& \ldots& 0.6	& uncertain	& \ldots        \\
J2150+0033	& 21 50 31.957	& +00 33 05.07	& 30 & 4382.42		& 64.45	        & 1.1	& 0.6	& marginal 	& s-sym  \\
PHL293B		& 22 30 36.8	& -00 06 37	& 60 & 1590.99		& 44.84         & 0.7	& 0.5	& marginal 	& s-asym  \\	
J2238+1400	& 22 38 31.1	& +14 00 29	& 30 & \ldots		& \ldots	& \ldots& 0.5	& no		& \ldots \\
J2259+1413	& 22 59 00.86	& +14 13 43.5	& 30 & \ldots		& \ldots	& \ldots& 0.6	& no		& \ldots \\
J2302+0049	& 23 02 10.0	& +00 49 39	&30 & \ldots		&  \ldots	& \ldots& 0.6	& no		& \ldots \\
		
\hline

\end{tabular}

$^a$ The \ion{H}{i} line is dominated by the \ion{H}{i} gas in the nearby spiral galaxy UGC 139 (Sect.~2.2.2).

$^b$ Baseline problems.

\end{minipage}

\end{center}

\end{table*}


Usually \ion{H}{i} sources are relatively isolated. However, given the large Effelsberg beam ($\sim$9\arcmin), we have checked the fields in which the XMPs reside, in order to assess possible \ion{H}{i} contamination. For this, we have visually analyzed the SDSS DR9 images (Fig.~1) in a field of view of about 6\arcmin. The results of this inspection are detailed in the list below.

$\bullet$ {\bf J0014-0044} -- The SDSS image (Fig.~1) shows a two-knot source next to a large spiral (SBc) galaxy (UGC 139). The NASA/IPAC Extragalactic Database (NED\footnote{http://ned.ipac.caltech.edu/}) has flagged the brightest SDSS XMP knot as a Western \ion{H}{ii} region of UGC 139. Indeed, SDSS spectroscopy of the bright knot and UGC 139 put them at a similar redshift of $\sim$0.01. Furthermore, the \ion{H}{i} profile is double-horn symmetric, typical of ordered disk rotation, with a large line width ($\sim$320 km s$^{-1}$; Table 2). Therefore, the \ion{H}{i} line detected with Effelsberg is primarily associated with the spiral galaxy UGC 139 and the source has been excluded from the following analysis.

$\bullet$ {\bf J0015+0104, J0113+0052, J0204-1009, J0301-0052, J0315-0024 and PHL293B} -- The SDSS images (Fig.~1) show knotted or cometary structures in relatively clean fields. Therefore, we are confident that the detected \ion{H}{i} flux density is associated with the XMP galaxies.

$\bullet$ {\bf J0126-0038 and J2104-0035} -- These sources show cometary structures (Fig.~1) and are found in fields with nearby stars and small red objects, the latter likely high redshift galaxies. Therefore, it is unlikely that there is contamination of the \ion{H}{i} data.

$\bullet$ {\bf J2053+0039} -- This source (Fig.~1) is found in a SDSS field with nearby stars and small red objects, likely to be high redshift galaxies. However, there is a small blue cloud to the Northwest of the original XMP position (top right-hand corner; Fig.~1), with a SDSS photometric redshift of 0.17$\pm$0.12. The main XMP source has a spectroscopic redshift of 0.01. The large error in the photometric redshift, the close proximity, the diffuse structures and similar colours, has led us to associate these two sources with the XMP, whose morphology is classified as two-knot. The \ion{H}{i} data will likely contain information from both these components. 

$\bullet$ {\bf J2150+0033} -- In the SDSS image (Fig.~1), this symmetric XMP shows a nearby galaxy in projection. The SDSS redshifts for the XMP source and the galaxy are 0.02 and 0.06, respectively. Therefore, the \ion{H}{i} data is not likely to be affected by contamination. 

\section{Auxiliary Optical Data}

In order to estimate physical parameters like dynamical and \ion{H}{i} gas masses, we have compiled from literature optical sizes, inclinations, distances, $g$-band magnitudes, \ion{H}{ii} gas-phase metallicities and other relevant observables. These are included in Tables 3, 4 and 5.

\subsection{Distance, Size, Inclination, Magnitude and Metallicity}

Hubble flow distances (Table~5), D, are obtained from NED and are corrected for Virgocentric infall. The optical radii (Table~3), r$_{\rm opt}$,  have been obtained from the SDSS data, as the radii containing 90\% of the galaxy light in the $g$-band. When these values are not available, we use NED SDSS $r$-band Petrosian radii, or the average between the semi-major axis, $\theta _M$, and the semi-minor axis, $\theta _m$, at the 25 mag arcsec$^{-2}$ isophote. The inclination angle of the source (Table~3), $i$, is computed as:

\begin{equation}
\sin \, i = \sqrt{ \frac{1 - \left(\frac{\theta _m}{\theta _M}\right)^2}{1 - q_0^2}},
\end{equation}

\noindent where $q_0 = 0.25$, which implicitly assumes galaxies to be disks of intrinsic thickness $q_0$ (e.g., S\'anchez-Janssen et al. 2010). Absolute $g$-band magnitudes, M$_{\rm g}$, have been obtained from the Hubble flow Virgocentric infall-corrected distances (Table~5) and SDSS DR7 Petrosian $g$-band magnitudes (Table~3), included in Table~1 and 2 of ML11. The $g$-band luminosity has been obtained from the absolute magnitude, assuming a solar absolute $g$-band magnitude of 5.12\footnote{http://www.sdss.org/dr7/algorithms/sdssUBVRI\\Transform.html}. Metallicity values are \ion{H}{ii} gas-phase metallicities (Table~3) taken from Table~1 and 2 of ML11 (and references therein). They have been derived using the direct (T$_{\rm e}$) method or strong-line methods, based on empirical calibrations consistent with the direct method.


\setcounter{table}{2}


\begin{table*}

\footnotesize

\begin{center}

\begin{minipage}[c]{165mm}

\caption{The optical data for the 140 XMP galaxies in the local Universe.
Col.~1: Source name.
Col.~2: Right Ascension.
Col.~3: Declination.
Col.~4: SDSS $g$-band Petrosian magnitude, from ML11.
Col.~5: Optical radius containing 90\% of the $g$-band galaxy light, from the SDSS. Otherwise, NED SDSS $r$-band Petrosian radius ($\star$) or the average source size at isophote 25 mag arcsec$^{-2}$ ($\dag$), is used.
Col.~6: Inclination angle of the source from the SDSS $g$-band data, NED SDSS $r$-band Petrosian data ($\star$) or estimated from the NED source size ($\dag$).
Col.~7: Metallicity, $12 \, + \, \log$ (O/H), from ML11. Tabulated values are based on the direct method or strong-line method ($\ddag$; see ML11 for details).
Col.~8: SDSS optical morphology adapted from ML11 or obtained from SDSS DR9 images ($\ast$). Symmetric (spherical, elliptical or disk-like symmetric structure), two-knot (two-knot structure), multi-knot (multiple knot structure) and cometary (head-tail structure), are used.
Col.~9: Spatial offset, $\Delta$R/r, where $\Delta$R is the difference between the position of the brightest star-formation region and the position of the center of the galaxy and r is the radius of the outer isophote.}

\begin{tabular}{l c c c c c c c c}

\hline

Name &   RA(J2000)        &   DEC(J2000)  	         & m$_{\rm g}$    &   r$_{\rm opt}$ & $\sin \, i$       & $12 \, + \, \log$ (O/H)   & Optical  & Spatial \\
     &   $^h$ $^m$ $^s$   & \degree \, \arcmin \, \arcsec& mag            &   \arcsec       &                   &       & Morphology & Offset \\
(1)  &  (2)               & (3)                          & (4)            &   (5)           &  (6)              & (7)   &   (8)   & (9) \\

\hline
\hline

UGC12894   &  00 00 22   & +39 29 44    		& \ldots     &  27.0$\star$	& \ldots		   & 7.64	&   \ldots & \ldots \\		
J0004+0025  &  00 04 22   & +00 25 36    		& 19.4       &  5.23    & 0.70             & 7.37   	& symmetric$\ast$  & \ldots \\
J0014-0044$^a$  &  00 14 29   & -00 44 44    		& 18.7       &  1.43$\star$   & 0.89$\star$		   & 7.63   	& \ldots  & \ldots \\
J0015+0104  &  00 15 21   & +01 04 37    		& 18.3       &  15.25   & 0.62             & 7.07  	& multi-knot$\ast$ & \ldots \\
J0016+0108  &  00 16 28   & +01 08 02    		& 18.9       &  4.54 	& 0.57	           & 7.53  	& symmetric$\ast$ & \ldots \\
HS0017+1055&  00 20 21   & +11 12 21    		& \ldots     & \ldots  & \ldots         & 7.63   	&  cometary$\ast$ & 0.2 \\
J0029-0108  &  00 29 05   & -01 08 26    		& 19.2       & 5.3	& 0.88		   & 7.35       & cometary$\ast$ & 0.4 \\
J0029-0025  &  00 29 49   & -00 25 40    		& 20.4       & 17.34$\star$ 	& 0.50$\star$		   & 7.29       &  symmetric$\ast$  & \ldots \\
ESO473-G024&  00 31 22   & -22 45 57    		& \ldots     & 25.5$\dag$ 	& 0.89$\dag$    	   & 7.45       & \ldots  & \ldots \\		
J0036+0052  &  00 36 30   & +00 52 34    	        & 18.8	     & 2.32     & 0.48             & 7.64$\ddag$      & symmetric  & \ldots \\		
AndromedaIV&  00 42 32   & +40 34 19    		& \ldots     & 34.5$\dag$ 	& 0.68$\dag$           & 7.49       &  \ldots & \ldots \\ 		
J0057-0022  &  00 57 13   & -00 21 58    		& 19.1       & 1.51$\star$ 	& 0.42$\star$		   & 7.60	&  symmetric$\ast$ & \ldots \\
IC1613     &  01 04 48   & +02 07 04    		& \ldots     & 460.5$\dag$	& 0.48$\dag$           & 7.64       &  multi-knot$\ast$ & \ldots \\		
J0107+0001  &  01 07 51   & +00 01 28    		& 19.4       & 1.30     & 0.63             & 7.23       & cometary$\ast$  & 0.4 \\
AM0106-382 &  01 08 22   & -38 12 34    		& \ldots     & 17.1$\dag$	& 0.88$\dag$           & 7.56       &  \ldots  & \ldots \\
J0113+0052  &  01 13 40   & +00 52 39    		& 20.1       & 3.23	& 0.42	 	   & 7.24  	&  multi-knot$\ast$   &  0.5 \\
J0119-0935  &  01 19 14   & -09 35 46    		& 19.5       & 2.04 	& 0.61		   & 7.31 	& cometary$\ast$ & 0.4 \\
HS0122+0743&  01 25 34   & +07 59 24    		& 15.7       & 10.14	& 0.67	 	   & 7.60      &  multi-knot & 0.5   \\		 
J0126-0038  &  01 26 46   & -00 38 39    		& 18.4       & 3.36$\star$	& 0.78$\star$	   & 7.51       &  cometary$\ast$ & 0.2 \\ 
J0133+1342  &  01 33 53   & +13 42 09    		& 18.1       & 5.64	& 0.81	  	   & 7.56       & symmetric$\ast$ & \ldots \\
J0135-0023  &  01 35 44   & -00 23 17    		& 18.9       & 2.23$\star$ 	& 0.94$\star$		   & 7.38       &  symmetric$\ast$ & \ldots \\		 
UGCA20     &  01 43 15   & +19 58 32    		& 18.0       & 58.5$\dag$ 	& 1.03$\dag$           & 7.60       &  cometary$\ast$ & 0.7 \\		
UM133      &  01 44 42   & +04 53 42   			& 15.4       & \ldots  & \ldots         & 7.63       &  cometary$\ast$ & 0.8 \\		
J0158+0006  &  01 58 09   & +00 06 37                   & 18.1       & 5.16     & 0.78             & 7.75$\ddag$      & cometary & \ldots \\ 
HKK97L14    &  02 00 10   & +28 49 53    		& \ldots     & 13.5$\dag$ 	& 0.92$\dag$           & 7.56       & cometary$\ast$ & 0.5 \\  
J0204-1009  &  02 04 26   & -10 09 35   		& 17.1       & 10.09 	& 0.94		   & 7.36       & cometary$\ast$  & \ldots  \\
J0205-0949  &  02 05 49   & -09 49 18    		& 15.3       & 21.39 	& 0.98		   & 7.61       & multi-knot$\ast$  & \ldots \\		
J0216+0115  &  02 16 29   & +01 15 21    		& 17.4       & 9.22	& 0.73	 	   & 7.44       & cometary$\ast$  &  0.6 \\
096632      &  02 51 47   & -30 06 32    		& 16.3       & 17.55$\dag$	& 0.74$\dag$           & 7.51       &   \ldots & \ldots \\
J0254+0035  &  02 54 29   & +00 35 50    		& 19.8       & 3.29	& 0.83	 	   & 7.28       & cometary$\ast$  & 0.3 \\
J0301-0059  &  03 01 26   & -00 59 26    		& 21.5       & 3.18$\star$ 	& 0.82$\star$		   & 7.64       &   cometary$\ast$ & 0.1\\
J0301-0052  &  03 01 49   & -00 52 57    		& 18.8       & 2.48$\star$	& 0.87$\star$		   & 7.52       &   cometary$\ast$ & 0.8 \\
J0303-0109  &  03 03 31   & -01 09 47    		& 19.8       & 4.69 	& 0.69		   & 7.22       & two-knot & 0.7\\
J0313+0006  &  03 13 00   & +00 06 12			& 19.2       & 3.80     & 0.82             & 7.82$\ddag$     & cometary & 0.5 \\
J0313+0010  &  03 13 02   & +00 10 40    		& 18.9       & 4.35	& 0.43	  	   & 7.44       & symmetric$\ast$   &  \ldots \\
J0315-0024  &  03 16 00   & -00 24 26    		& 20.2       & 1.48$\star$ 	& 0.59$\star$		   & 7.41       & cometary$\ast$  & 0.7 \\		
UGC2684    &  03 20 24   & +17 17 45    		& 22.8       & 2.47	& 0.77	           & 7.60       & cometary$\ast$  & 0.0 \\		
SBS0335-052W& 03 37 38   & -05 02 37    		& 19.0       & 3.45$\dag$	& 0.68$\dag$       	   & 7.11       &  \ldots & \ldots \\		
SBS0335-052E& 03 37 44   & -05 02 40    		& 16.3       & 6.45$\dag$	& 0.53$\dag$       	   & 7.31       &  \ldots & \ldots \\		
J0338+0013  &  03 38 12   & +00 13 13    		& 24.4       & 4.31	& 0.84	 	   & 7.64       & cometary$\ast$   & 0.7 \\
J0341-0026  &  03 41 18   & -00 26 28    		& 18.8       & 1.72$\star$ 	& 0.97$\star$		   & 7.26       &  cometary$\ast$  & 0.7 \\
ESO358-G060&  03 45 12   & -35 34 15    		& \ldots     & 31.5$\dag$ 	& 1.04$\dag$       	   & 7.26       &   \ldots  & \ldots \\	
G0405-3648  &  04 05 19   & -36 48 49   		& \ldots     & 12.45$\dag$	& 0.73$\dag$      	   & 7.25       &   \ldots  & \ldots \\
J0519+0007  &  05 19 03   & +00 07 29    		& 18.4       & 2.16	& 0.47	           & 7.44       & symmetric$\ast$  & \ldots  \\
Tol0618-402 &  06 20 02   & -40 18 09    		& \ldots     & 10.5$\dag$ 	& 0.71$\dag$       	   & 7.56        &  \ldots &  \ldots  \\	
	
\hline
 
\end{tabular}

$^a$ Although this source has been selected as an XMP galaxy in ML11, NED has flagged this as a Western \ion{H}{ii} region of UGC 139 (Sect.~3.2.2).

\end{minipage}

\end{center}

\end{table*}



\setcounter{table}{2}

\begin{table*}

\footnotesize

\begin{center}

\begin{minipage}[c]{165mm}

\caption{The optical data for the 140 XMP galaxies in the local Universe. Continued.}

\begin{tabular}{l c c c c c c c c}

\hline

Name &   RA(J2000)        &   DEC(J2000)  	         & m$_{\rm g}$ &  r$_{\rm opt}$    & $\sin \, i$   & $12 \, + \, \log$ (O/H)  & Optical & Spatial \\
     &    $^h$ $^m$ $^s$  & \degree \, \arcmin \, \arcsec& mag         &   \arcsec         &               &      &    Morphology    & Offset \\
(1)  &  (2)               & (3)                          & (4)   &   (5)     &  (6)          & (7)  & (8)    & (9) \\

\hline
\hline

ESO489-G56 &  06 26 17   & -26 15 56    		& 15.6       & 22.5$\dag$ 	& 0.80$\dag$       	   & 7.49        &  \ldots  & \ldots  \\		
J0808+1728  &  08 08 41   & +17 28 56    		& 19.2       & 0.56$\star$    & 0.75$\star$ 	   & 7.48        & symmetric  &  \ldots  \\
J0812+4836  &  08 12 39   & +48 36 46    		& 16.0       & 13.10	& 0.89	 	   & 7.28        & cometary$\ast$  & 0.2 \\
UGC4305    &  08 19 05   & +70 43 12    		& \ldots     & 213.0$\dag$	& 0.64$\dag$       	   & 7.65        &   \ldots  & \ldots \\
J0825+1846  &  08 25 40   & +18 46 17                   & 19.0       & 2.16     & 0.58             & 7.75$\ddag$        & symmetric & \ldots \\	
HS0822+03542& 08 25 55   & +35 32 31    		& 17.8       & 4.45	& 0.92	           & 7.35        & cometary$\ast$ & 0.3 \\	 	
DD053      &  08 34 07   & +66 10 54    		& 20.3       & 42.0$\dag$ 	& 0.53$\dag$      	   & 7.62        &   multi-knot$\ast$ & \ldots \\		
UGC4483    &  08 37 03   & +69 46 31    		& 15.1       & 31.5$\dag$ 	& 0.92$\dag$       	   & 7.58        &   \ldots  & \ldots \\		
HS0837+4717&  08 40 30   & +47 07 10    		& 17.6       & 2.31	& 0.69	           & 7.64        & cometary$\ast$  & 0.3 \\
J0842+1033  &  08 42 36   & +10 33 13 			& 17.7       & 5.65     & 0.87             & 7.58$\ddag$        & cometary  &  0.4 \\
HS0846+3522&  08 49 40   & +35 11 39    		& 18.2       & 5.80	& 0.79	           & 7.65        & cometary$\ast$ & 0.2 \\		
J0859+3923  &  08 59 47   & +39 23 06    		& 17.2       & 10.62	& 0.74	           & 7.57        & two-knot$\ast$   & \ldots \\
J0910+0711  &  09 10 29   & +07 11 18    		& 16.9       & 11.61	& 0.94             & 7.63        & cometary$\ast$    & 0.5 \\
J0911+3135  &  09 11 59   & +31 35 36    		& 17.8       & 6.04	& 0.71	 	   & 7.51        & cometary$\ast$  &  0.3 \\
J0926+3343  &  09 26 09   & +33 43 04    		& 17.8       & 15.58	& 1.00	           & 7.12        & cometary$\ast$  & 0.9 \\	
IZw18      &  09 34 02   & +55 14 25    		& 16.4       & 4.47	& 0.51	 	   & 7.17        & two-knot  & \ldots \\		
J0940+2935  &  09 40 13   & +29 35 30    		& 16.5       & 16.44	& 0.95	 	   & 7.65        & cometary$\ast$  & 0.3 \\
J0942+3404  &  09 42 54   & +34 04 11                   & 19.1       & 2.29     & 0.68             & 7.67$\ddag$       & cometary & 0.4 \\	
CGCG007-025&  09 44 02   & -00 38 32    		& 16.0       & 9.47  	& 0.82 		   & 7.65        & multi-knot$\ast$  & \ldots \\
SBS940+544 &  09 44 17   & +54 11 34    		& 19.1       & 0.63$\star$    & 0.36$\star$            & 7.46        &  cometary & 0.7 \\		
CS0953-174 &  09 55 00   & -17 00 00                   & \ldots     &  \ldots   & \ldots   	    & 7.58        &  \ldots  & \ldots \\ 
J0956+2849  &  09 56 46   & +28 49 44    & 15.9       & 32.05	 & 0.97		    & 7.13        &  multi-knot$\ast$  & \ldots \\
LeoA       &  09 59 26   & +30 44 47    & 19.0       & 124.5$\dag$  & 0.86$\dag$      	    & 7.30        &  multi-knot$\ast$   & \ldots\\		
SextansB   &  10 00 00   & +05 19 56    & 20.5       & 129.0$\dag$  & 0.78$\dag$      	    & 7.53        & multi-knot$\ast$   & \ldots\\
J1003+4504  &  10 03 48   & +45 04 57    & 17.5       & 3.40     & 0.59             & 7.65$\ddag$       & symmetric & \ldots \\
SextansA   &  10 11 00   & -04 41 34    & \ldots     & 162.0$\dag$  & 0.59$\dag$    	    & 7.54        &  \ldots   &  \ldots \\			
KUG1013+381&  10 16 24   & +37 54 44    & 15.9       & 1.62	 & \ldots		    & 7.58        & cometary    & 0.2 \\	
SDSSJ1025+1402& 10 25 30 & +14 02 07    & 20.4       & 0.60$\star$ 	 & 0.54$\star$	    & 7.36        &  symmetric$\ast$ &  \ldots \\  
UGCA211      &10 27 02   & +56 16 14    & 16.2       & 2.11	 & 0.38	            & 7.56        & cometary$\ast$   &  0.4 \\		
J1031+0434    &10 31 37   & +04 34 22    & 16.2       & 6.08     & 0.70	            & 7.70        & cometary   &  0.3 \\
HS1033+4757  &10 36 25   & +47 41 52    & 17.5       & 6.92	 & 0.67		    & 7.65        & symmetric$\ast$ & \ldots \\		 
J1044+0353    &10 44 58   & +03 53 13    & 17.5       & 4.16	 & 0.86		    & 7.44        & cometary   & 0.4 \\
HS1059+3934  &11 02 10   & +39 18 45    & 17.9       & 7.89	 & 0.71	            & 7.62        & multi-knot$\ast$   & \ldots \\
J1105+6022    &11 05 54   & +60 22 29    & 16.4       & 17.08	 & 0.91	 	    & 7.64        & cometary$\ast$  & 0.4  \\
J1119+5130    &11 19 34   & +51 30 12    & 16.9       & 4.52	 & 0.73	            & 7.51        & cometary$\ast$  & 0.2  \\
J1121+0324    &11 21 53   & +03 24 21    & 18.1       & 9.62 	 & 0.80        	    & 7.64        & cometary$\ast$ & 0.5 \\
UGC6456      &11 28 00   & +78 59 39    & \ldots     & 33.0$\dag$   & 0.88$\dag$     	    & 7.35        & \ldots  & \ldots   \\
SBS1129+576  &11 32 02   & +57 22 46    & 16.7       & 6.59$\star$    & 0.97$\star$       	    & 7.36        &  cometary$\ast$  & 0.5 \\
J1145+5018    &11 45 06   & +50 18 02    & 17.8	      & 5.50     & 0.89             & 7.71$\ddag$       & cometary & 0.4 \\
J1151-0222    &11 51 32   & -02 22 22    & 16.8       & 11.30    & 0.52		    & 7.78        & two-knot    & 0.6 \\
J1157+5638    &11 57 54   & +56 38 16    & 16.9       & 10.39    & 0.61             & 7.83$\ddag$       & cometary & 0.5 \\
J1201+0211    &12 01 22   & +02 11 08    & 17.6       & 10.28	 & 0.88	 	    & 7.49        & cometary   &  0.6 \\	
SBS1159+545  &12 02 02   & +54 15 50    & 18.7       & 2.33	 & 0.40	            & 7.41        & two-knot$\ast$   & 0.8 \\
SBS1211+540  &12 14 02   & +53 45 17    & 17.4       & 5.72	 & 0.79		    & 7.64        & cometary   & 0.1 \\		 
J1215+5223    &12 15 47   & +52 23 14    & 15.2       & 24.67    & 0.85		    & 7.43        & multi-knot$\ast$  &  \ldots  \\	
Tol1214-277  &12 17 17   & -28 02 33    & \ldots     & 2.4$\dag$ & \ldots 	    & 7.55        & \ldots  &  \ldots \\			
VCC0428      &12 20 40   & +13 53 22    & 17.0       & 11.73	 & 0.91	 	    & 7.64        & cometary$\ast$  &  0.1 \\
HS1222+3741  &12 24 37   & +37 24 37    & 17.9       & 2.19	 & 0.47	 	    & 7.64        & symmetric$\ast$ & \ldots  \\
Tol65        &12 25 47   & -36 14 01    & 17.5       & 6.0$\dag$ & \ldots	    & 7.54        &  \ldots &  \ldots \\
J1230+1202    &12 30 49   & +12 02 43    & 16.7       & 4.88	 & 0.86 	    & 7.73        & cometary & 0.3  \\	
KISSR85      &12 37 18   & +29 14 55    & 19.9       & 3.16 	 & 0.56             & 7.61        & cometary$\ast$  & 0.3 \\
UGCA292      &12 38 40   & +32 46 01    & 18.9       & 1.51	 & 0.14 	    & 7.28        & multi-knot$\ast$  & \ldots  \\
HS1236+3937  &12 39 20   & +39 21 05    & 18.5       & 8.54     & 1.00       	    & 7.47        & two-knot$\ast$  & 0.4 \\
J1239+1456    &12 39 45   & +14 56 13    & 19.8       & 1.47     & 0.56             & 7.65        & symmetric$\ast$  & \ldots  \\
SBS1249+493  &12 51 52   & +49 03 28    & 18.0       & 3.00     & 0.84     	    & 7.64        & cometary$\ast$  & 0.7  \\
J1255-0213    &12 55 26   & -02 13 34    & 19.1       & 0.83$\star$  	 & 0.64$\star$	    & 7.83        & symmetric  &  \ldots \\
GR8           &12 58 40   & +14 13 03    & 17.9       & 23.54    & 0.88  	    & 7.65        & multi-knot$\ast$ &  \ldots \\	
KISSR1490    &13 13 16   & +44 02 30    & 19.0       & 5.60     & 0.85      	    & 7.56        & cometary$\ast$  & 0.4 \\	
DD0167       &13 13 23   & +46 19 22    & \ldots     & 19.64    & 0.82      	    & 7.20        & multi-knot$\ast$  &  0.8 \\
HS1319+3224  &13 21 20   & +32 08 25    & 18.6       & 4.57     & 0.66      	    & 7.59        & cometary$\ast$  & 0.2 \\
J1323-0132    &13 23 47   & -01 32 52    & 18.2       & 0.66$\star$ 	 & 0.39$\star$	    & 7.78        & symmetric  &  \ldots \\
J1327+4022    &13 27 23   & +40 22 04    & 19.0       & 2.83     & 0.46             & 7.67$\ddag$       & symmetric & \ldots \\

\hline
 
\end{tabular}

\end{minipage}

\end{center}

\end{table*}



\setcounter{table}{2}

\begin{table*}

\footnotesize

\begin{center}

\begin{minipage}[c]{165mm}

\caption{The optical data for the 140 XMP galaxies in the local Universe. Continued.}

\begin{tabular}{l c c c c c c c c}

\hline
  
Name &   RA(J2000)       &   DEC(J2000)  	        & m$_{\rm g}$ & r$_{\rm opt}$ &  $\sin \, i$    & $12 \, + \, \log$ (O/H)        & Optical & Spatial \\
     &    $^h$ $^m$ $^s$ & \degree \, \arcmin , \arcsec & mag         & \arcsec       &                 &             &  Morphology      & Offset \\
(1)  &  (2)              & (3)                          & (4)   &   (5)  &  (6)            & (7)         & (8)    & (9) \\

\hline
\hline

J1331+4151    &13 31 27   & +41 51 48    & 17.1       & 3.80 	 & 0.74		    & 7.75        & cometary  & 0.3 \\
ESO577-G27   &13 42 47   & -19 34 54    & \ldots     & 28.5$\dag$ 	 & 0.46$\dag$           & 7.57        &  \ldots  & \ldots \\
J1355+4651    &13 55 26   & +46 51 51    & 19.3       & 2.20 	 & 0.79		    & 7.63        & cometary$\ast$  &  0.4 \\
J1414-0208    &14 14 54   & -02 08 23    & 18.0       & 8.47 	 & 0.72		    & 7.28        & cometary$\ast$  & 0.5 \\
SBS1415+437  &14 17 01   & +43 30 05    & 17.8       & 1.40     & 0.57		    & 7.43        & multi-knot$\ast$   & 0.8 \\ 
J1418+2102    &14 18 51   & +21 02 39    & 17.6       & 2.87     & 0.85             & 7.64$\ddag$       & cometary & 0.4 \\
J1422+5145    &14 22 51   & +51 45 16    & 20.2       & 2.01     & 0.73     	    & 7.41        & symmetric$\ast$ & \ldots \\
J1423+2257    &14 23 43   & +22 57 29    & 17.9       & 2.36 	 & 0.66		    & 7.72        & symmetric   &  \ldots \\
J1441+2914    &14 41 58   & +29 14 34    & 20.1       & 2.25     & 0.55             & 7.47        & symmetric$\ast$  & \ldots \\
HS1442+4250  &14 44 13   & +42 37 44    & 15.9       & 20.04    & 1.00             & 7.54        & cometary$\ast$  & 0.3\\
J1509+3731    &15 09 34   & +37 31 46    & 17.3       & 2.87 	 & 0.75		    & 7.85        & cometary$\ast$   & 0.6 \\
KISSR666     &15 15 42   & +29 01 40    & 19.1       & 4.24     & 0.87      	    & 7.53        & cometary$\ast$ &  0.4 \\
KISSR1013    &16 16 39   & +29 03 33    & 18.2       & 1.28 	 & 0.37		    & 7.63        & two-knot$\ast$  & \ldots \\
J1644+2734    &16 44 03   & +27 34 05    & 17.7       & 10.05 	 & 1.00		    & 7.48        & symmetric$\ast$ & \ldots \\
J1647+2105    &16 47 11   & +21 05 15    & 17.3       & 9.16 	 & 0.82		    & 7.75        & multi-knot  &  \ldots \\
W1702+18     &17 02 33   & +18 03 06    & 18.4       & 1.57     & 0.56     	    & 7.63        & symmetric$\ast$  &  \ldots \\
HS1704+4332  &17 05 45   & +43 28 49    & 18.4       & 3.49     & 0.75             & 7.55        & cometary$\ast$  & 0.5 \\	
SagDIG        &19 29 59   & -17 40 41    & \ldots     & \ldots   & \ldots     	    & 7.44        &   \ldots & \ldots \\
J2053+0039    &20 53 13   & +00 39 15    & 19.4       & 6.48 	 & 0.83		    & 7.33        & two-knot$\ast$  & 0.4 \\
J2104-0035    &21 04 55   & -00 35 22    & 17.9       & \ldots	 &\ldots 	    & 7.05        &  cometary$\ast$ & 0.8 \\
J2105+0032    &21 05 09   & +00 32 23    & 19.0       & 0.43$\star$ 	 & 0.47$\star$	    & 7.42        &  cometary$\ast$  & 0.3 \\
J2120-0058    &21 20 26   & -00 58 27    & 18.8       & 1.02$\star$ 	 & 0.74$\star$	    & 7.65        & symmetric$\ast$  &  \ldots \\
HS2134+0400  &21 36 59   & +04 14 04    & \ldots     & \ldots   & \ldots          & 7.44        &  cometary$\ast$ & 0.2 \\
J2150+0033    &21 50 32   & +00 33 05    & 19.3       & 6.12	 & 0.71		    & 7.60        & symmetric$\ast$  & \ldots \\
ESO146-G14   &22 13 00   & -62 04 03    & \ldots     & 52.5$\dag$   & 1.06$\dag$      	    & 7.59        &  \ldots  &  \ldots \\
2dF171716    &22 13 26   & -25 26 43    & \ldots     & 5.55$\dag$   & 0.56$\dag$      	    & 7.54        &  \ldots  & \ldots \\
PHL293B      &22 30 37   & -00 06 37    & 17.2       & 2.13$\star$   & 0.53$\star$      	    & 7.62        &  cometary$\ast$ & 0.4  \\
2dF115901    &22 37 02   & -28 52 41    & \ldots     & 3.6$\dag$    & 0.75$\dag$           & 7.57        &  \ldots & \ldots \\
J2238+1400  &22 38 31   & +14 00 30    & 19.0       &  3.00	& 0.80	            & 7.45        & two-knot & 0.1 \\	
J2250+0000    &22 50 59   & +00 00 33    & 19.8       &  1.71	& 0.57		    & 7.61        & symmetric$\ast$ & \ldots \\
J2259+1413    &22 59 01   & +14 13 43    & 19.1       &  4.03   & 0.85              & 7.37        & cometary$\ast$  &  0.5 \\
J2302+0049    &23 02 10   & +00 49 39    & 18.8       &  2.19	& 0.66		    & 7.71        & two-knot & \ldots \\
J2354-0005    &23 54 37   & -00 05 02    & 18.7       &  2.68$\star$	& 0.75$\star$		    & 7.35        & two-knot$\ast$ & 0.8  \\

\hline
 
\end{tabular}


\end{minipage}

\end{center}

\end{table*}


\subsection{Morphology}

We have adapted, completed and/or revised, using the SDSS DR9 composite images, the optical morphological classification, based on the simple scheme presented in ML11 -- symmetric for a spherical, elliptical or disk-like symmetric structure, cometary for a head-tail structure, with or without a clear knot at the head, and a diffuse tail, two-knot for a structure with two knots, with or without a head-tail morphology, and multi-knot for a diffuse structure with multiple star-formation knots. In Fig.~2 we supply SDSS DR9 postage stamp images illustrating this scheme, while Table~3 includes the individual morphological classification for the XMPs. 



\setcounter{figure}{1}

\begin{figure*} 
\begin{center}
\includegraphics[width=6cm]{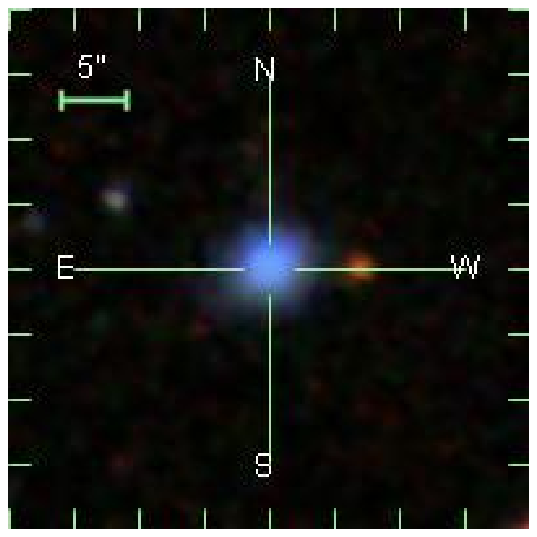}
\includegraphics[width=6cm]{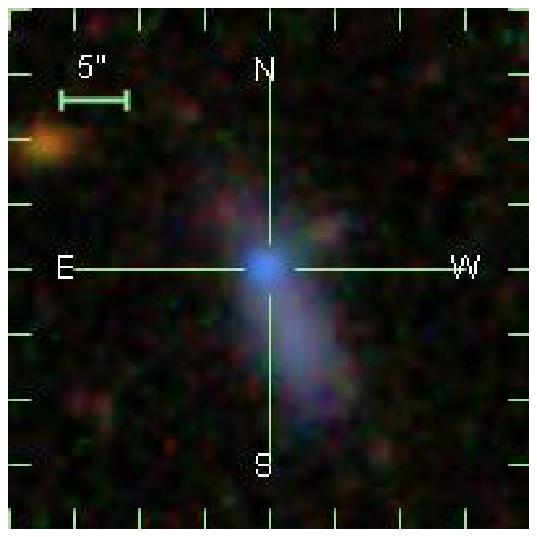}

\includegraphics[width=6cm]{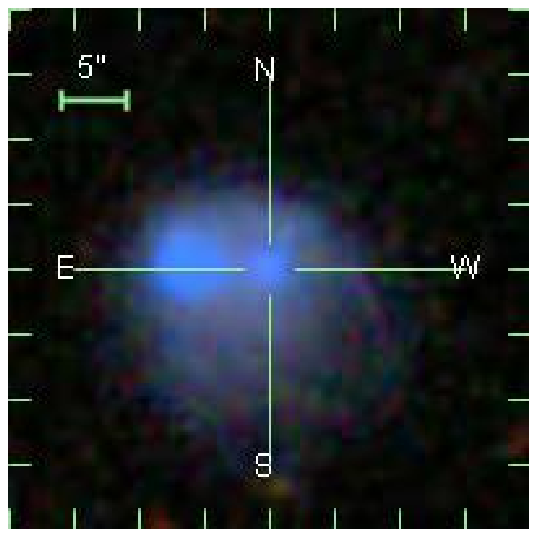}
\includegraphics[width=6cm]{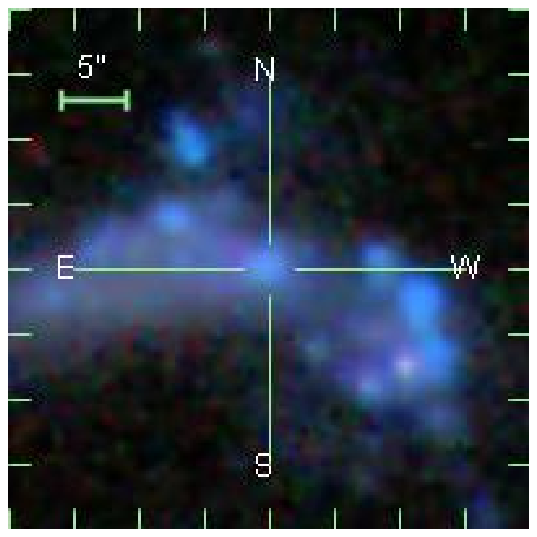}
\caption{SDSS images illustrating our optical morphology classification scheme: J1003+4504 (top left) showing symmetric, J0119-0935 (top right) showing cometary, J1151-0222 (also known as UM 461; bottom left) showing two-knot, and HS 0122+0743 (bottom right) showing multi-knot optical morphology.}
\end{center} 
\end{figure*} 


We have also parametrized the asymmetric optical morphology, by measuring the degree of optical asymmetry or spatial offset, $\Delta$R/r, in the cometary galaxies (and in some cases, also in two-knot or multi-knot sources with cometary-like morphology) from the SDSS images. $\Delta$R is the difference between the position of the brightest star-formation region and the position of the center of the galaxy (estimated from the outer isophote, measured at 25 mag arcsec$^{-2}$) and r is the radius of the same outer isophote. The results are included in Table 3.

\subsection{Stellar and Nebular Parameters}

The Max-Planck Institute for Astrophysics--Johns Hopkins University (MPA--JHU) group provides a number of derived physical parameters\footnote{http://www.sdss3.org/dr9/algorithms/galaxy$\_$mpa$\_$jhu.php} for all galaxies in the SDSS (Kauffmann et al. 2003; Brinchmann et al. 2004). We include in Table~4 the relevant parameters from this database: best-fit spectroscopic redshift, $z$ (heliocentric, optical convention; Eq.~2; Sect.~4), velocity offset of the Balmer, $\Delta$v$_{\rm Balmer}$, and forbidden, $\Delta$v$_{\rm forbidden}$, emission lines, and the total stellar mass and star-formation rate (SFR). 1$\sigma$ errors for the Balmer and fobidden emission-line velocity offsets, and the stellar masses are also provided in Table~4.


The best-fit spectroscopic redshift is measured using several nebular and stellar lines, via a code developed by David Schlegel\footnote{http://spectro.princeton.edu/$\#$dm$\_$spzbest}. The resulting radial velocity, v$_{\rm z}$ = c$z$ (heliocentric, optical convention; Eq.; Sect.~4; Table~5), is a measurement of the average nebular/stellar component radial velocity and shall be designated hereinafter as the best-fit radial velocity.

The MPA-JHU group also provides measurements of the Balmer and forbidden emission-line shifts relative to the best-fit radial velocity, after removing the contribution of the stellar component from the spectra, via stellar population synthesis models. Balmer and forbidden emission-line velocities (Table~5) have been calculated from the Balmer and forbidden line velocity offsets relative to the best-fit radial velocity. 


The stellar masses are the median of the distribution of the total mass estimates obtained using SDSS model photometry and include a correction for aperture effects and nebular emission\footnote{http://www.mpa-garching.mpg.de/SDSS/DR7/Data/\\stellarmass.html}. However, we stress that the stellar mass values are uncertain, as they are obtained by converting light into stellar mass via a mass-to-light (M/L) ratio that assumes a constant or exponentially decreasing star-formation history. This assumption is, however, not entirely appropriate for BCDs (and therefore also not for XMPs), since the young starburst component in these systems contributes a significant amount to the total $B$-band luminosity (Papaderos et al. 1996b; also Cair\'os et al. 2001; Gil de Paz \& Madore 2005; Amor\'\i n et al. 2007, 2009). Consequently, stellar mass determinations based on the integrated luminosity and a standard M/L ratio are overestimated by, typically, a factor of 2 (0.3 dex). This, and the strong contribution of nebular (line and continuum) emission in most XMPs (e.g., Papaderos et al. 2008; also Papaderos \& \"Ostlin 2012) conspires, along with aperture effects, to produce large uncertainties in stellar mass determinations.


In order to have an idea of the potential bias affecting the stellar mass estimates, we have also computed the stellar masses in an alternative way, using the correlations from Bell \& de Jong (2001). We have adopted the calibration obtained by assuming a mass-dependent formation epoch model with bursts, and a scaled-down Salpeter initial mass function (IMF; see their Table~1), using the SDSS $g-r$ colours to estimate the stellar M/L values, and therefore, the stellar masses. We note, however, that this method may also suffer from similar uncertainties as those described above, as well as those related to model assumptions.




\begin{figure} 
\begin{center}
\includegraphics[width=6cm]{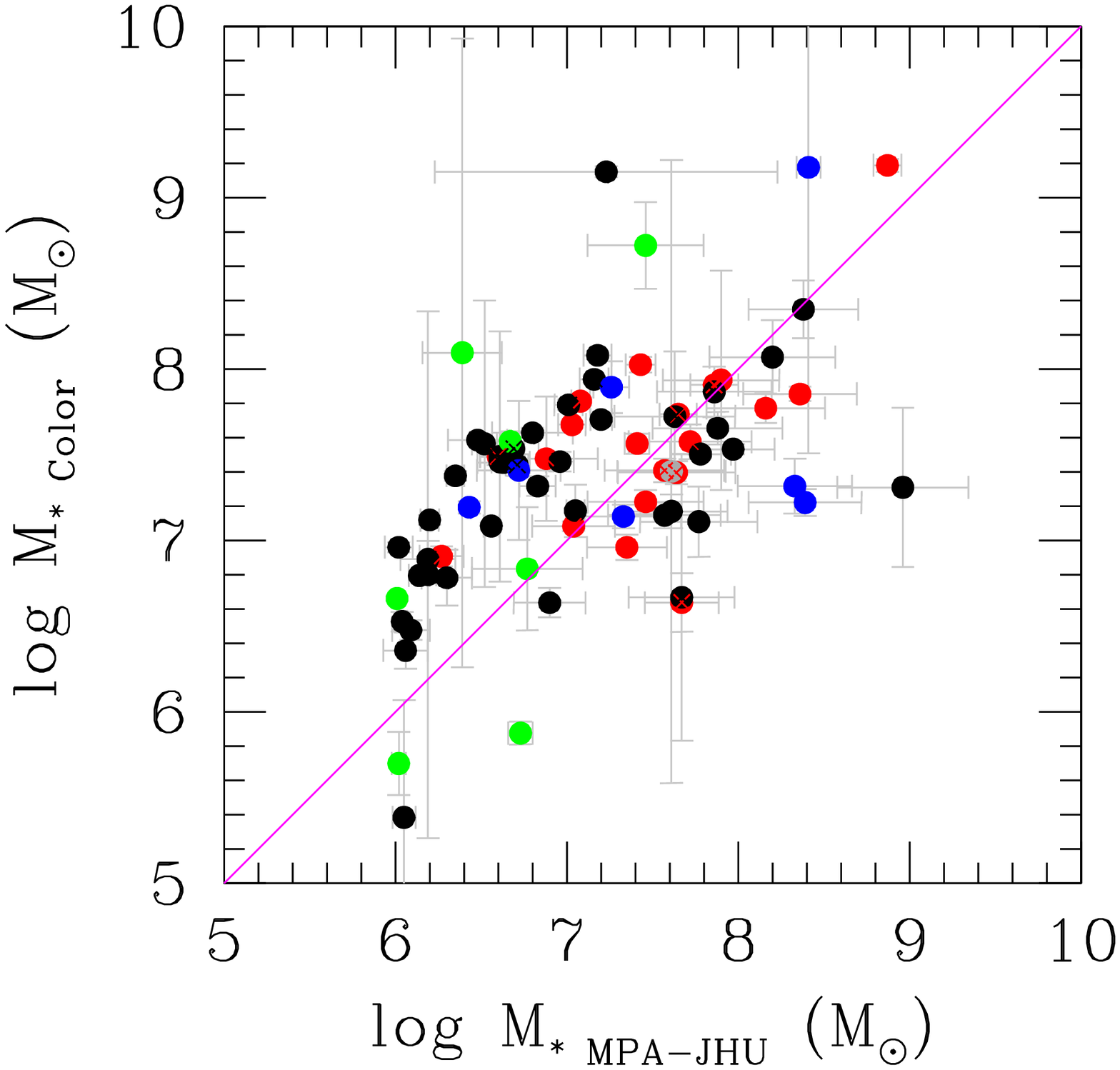}
\caption{The comparison between the MPA-JHU- and color-determined stellar masses. The color code parametrizes the morphology as follows: red = symmetric, black = cometary, blue = two-knot, green = multi-knot and grey = no morphological information. The magenta line is the one-to-one relation.}
\end{center}
\end{figure}


The colour-determined stellar masses are compared with the MPA-JHU-determined stellar masses in Fig.~3. There appears to be a break in the one-to-one relation at about log M$_{\star}$ = 7.5 M$_{\odot}$. MPA-JHU stellar masses below this value are underestimated according to the color-determined value, whereas stellar masses above this value are overestimated. The differences between these two methods of stellar mass determation is, however, rarely above 1~dex. 

By default, we use the MPA-JHU-determined stellar masses in the subsequent analysis. However, we also present color-determined stellar mass results when the outcome may depend on the stellar mass determination, in order to have an idea of potential uncertainties.


The total SFRs are computed using the nebular emission lines and galaxy photometry\footnote{http://www.mpa-garching.mpg.de/SDSS/DR7/sfrs.html}. The specific star formation rate (sSFR) is the SFR divided by the total stellar mass.



\setcounter{table}{3}

\begin{table*}

\footnotesize

\begin{center}

\begin{minipage}[c]{118mm}

\caption{The SDSS MPA-JHU data for the 140 XMP galaxies in the local Universe.
Col.~1: Source name.
Col.~2: Best-fit spectroscopic redshift, $z$ = v$_z$/c (heliocentric, optical convention).
Col.~3: Balmer emission-line velocity offset and error, measured relative to the best-fit radial velocity (v$_z$).
Col.~4: Forbidden emission-line velocity offset and error, measured relative to the best-fit radial velocity (v$_z$).
Col.~5: Logarithm of the stellar mass and 1$\sigma$ error.
Col.~6: Logarithm of the star-formation rate.}

\begin{tabular}{l c c c c c}

\hline

Name &   $z$  &   $\Delta$v$_{\rm Balmer}$  & $\Delta$v$_{\rm forbidden}$   & log M$_{\star}$ &   log SFR \\
     &        & km s$^{-1}$                & km s$^{-1}$               & M$_{\odot}$ &   M$_{\odot}$ yr$^{-1}$ \\
(1)  &  (2)   & (3)                 & (4)                        &   (5)                     &  (6)        \\

\hline
\hline

UGC12894   &   \ldots  &   \ldots  &   \ldots     &     \ldots   &   \ldots   \\ 
J0004+0025   &   0.0126  &    2.4  $\pm$    1.2  &    -11.0 $\pm$    1.9     &      7.72 $\pm$ 0.36   &   -2.96   \\ 
J0014-0044   &   0.0136  &   -5.0  $\pm$    0.4  &    -6.4 $\pm$    0.4     &      7.61 $\pm$ 0.31   &   -1.31   \\ 
J0015+0104   &   0.0069  &   -3.4  $\pm$    1.3  &    -3.1 $\pm$    2.3     &      7.46 $\pm$ 0.34   &   \ldots   \\ 
J0016+0108   &   0.0104  &   -0.5  $\pm$    1.0  &    -4.7 $\pm$    1.7     &      7.64 $\pm$ 0.34   &   \ldots   \\ 
HS0017+1055   &   \ldots  &   \ldots  &   \ldots     &     \ldots   &   \ldots   \\ 
J0029-0108   &   0.0132  &    1.3  $\pm$    1.9  &     8.0 $\pm$    4.0     &      7.88 $\pm$ 0.38   &   \ldots   \\ 
J0029-0025   &   0.0141  &    5.1  $\pm$    1.7  &    -3.4 $\pm$    2.5     &      6.88 $\pm$ 0.30   &   \ldots   \\ 
ESO473-G024   &   \ldots  &   \ldots  &   \ldots     &     \ldots   &   \ldots   \\ 
J0036+0052   &   0.0282  &   -1.2  $\pm$    0.3  &    -5.1 $\pm$    0.3     &      8.36 $\pm$ 0.33   &   -1.02   \\ 
AndromedaIV   &   \ldots  &   \ldots  &   \ldots     &     \ldots   &   \ldots   \\ 
J0057-0022   &   0.0094  &   -0.7  $\pm$    0.8  &    -4.1 $\pm$    1.0     &      7.46 $\pm$ 0.34   &   -2.45   \\ 
IC1613   &   \ldots  &   \ldots  &   \ldots     &     \ldots   &   \ldots   \\ 
J0107+0001   &   0.0181  &   -5.4  $\pm$    1.1  &    -5.0 $\pm$    1.3     &      7.86 $\pm$ 0.33   &   -2.35   \\ 
AM0106-382   &   \ldots  &   \ldots  &   \ldots     &     \ldots   &   \ldots   \\ 
J0113+0052   &   0.0038  &    1.1  $\pm$    0.8  &    -1.9 $\pm$    1.1     &      6.77 $\pm$ 0.32   &   \ldots   \\ 
J0119-0935   &   0.0064  &    2.9  $\pm$    0.3  &    -2.6 $\pm$    0.3     &      6.06 $\pm$ 0.13   &   -2.14   \\ 
HS0122+0743   &   0.0097  &    9.3  $\pm$    0.1  &     3.7 $\pm$    0.2     &      6.39 $\pm$ 0.23   &   -1.17   \\ 
J0126-0038   &   0.0065  &   -0.1  $\pm$    0.6  &     4.5 $\pm$    1.2     &      7.63 $\pm$ 0.35  &   -2.93   \\ 
J0133+1342   &   0.0087  &   -6.6  $\pm$    0.5  &     0.6 $\pm$    0.4     &      6.60 $\pm$ 0.10  &   -1.39   \\ 
J0135-0023   &   0.0169  &    2.8  $\pm$    1.1  &    -1.1 $\pm$    2.0     &      8.16 $\pm$ 0.35   &   \ldots   \\ 
UGCA20   &   \ldots  &   \ldots  &   \ldots     &     \ldots   &   \ldots   \\ 
UM133   &   \ldots  &   \ldots  &   \ldots     &     \ldots   &   \ldots   \\ 
J0158+0006   &   \ldots  &   \ldots  &   \ldots     &     \ldots   &   \ldots   \\ 
HKK97L14   &   \ldots  &   \ldots  &   \ldots     &     \ldots   &   \ldots   \\ 
J0204-1009   &   0.0063  &    1.9  $\pm$    1.1  &     5.0 $\pm$    2.1     &      7.20 $\pm$ 0.06   &   -1.54   \\ 
J0205-0949   &   \ldots  &   \ldots  &   \ldots     &     \ldots   &   \ldots   \\ 
J0216+0115   &   0.0093  &    1.5  $\pm$    1.0  &    -2.8 $\pm$    1.3     &      8.20 $\pm$ 0.37   &   -2.71   \\ 
096632   &   \ldots  &   \ldots  &   \ldots     &     \ldots   &   \ldots   \\ 
J0254+0035   &   0.0149  &    0.5  $\pm$    1.2  &     0.5 $\pm$    3.0     &      7.61 $\pm$ 0.33   &   \ldots   \\ 
J0301-0059   &   0.0383  &    3.5  $\pm$    1.0  &    -6.1 $\pm$    1.3     &      8.96 $\pm$ 0.38   &   -0.05   \\ 
J0301-0052   &   0.0073  &   -8.4  $\pm$    0.4  &    -10.2 $\pm$    0.4     &      7.57 $\pm$ 0.33   &   -1.99   \\ 
J0303-0109   &   0.0304  &    0.5  $\pm$    0.6  &     2.4 $\pm$    0.6     &      8.33 $\pm$ 0.33   &   \ldots   \\ 
J0313+0006   &   0.0292  &   -10.6  $\pm$    0.8  &    -15.5 $\pm$    0.4     &      7.67 $\pm$ 0.31   &   -1.02   \\ 
J0313+0010   &   0.0077  &    1.9  $\pm$    1.0  &    -7.7 $\pm$    1.8     &      7.57 $\pm$ 0.35   &   \ldots   \\ 
J0315-0024   &   0.0226  &   -1.3  $\pm$    1.1  &     0.2 $\pm$    2.0     &      7.77 $\pm$ 0.34   &   \ldots   \\ 
UGC2684   &   \ldots  &   \ldots  &   \ldots     &     \ldots   &   \ldots   \\ 
SBS0335-052W   &   \ldots  &   \ldots  &   \ldots     &     \ldots   &   \ldots   \\ 
SBS0335-052E   &   \ldots  &   \ldots  &   \ldots     &     \ldots   &   \ldots   \\ 
J0338+0013   &   0.0426  &    5.0  $\pm$    0.4  &     3.4 $\pm$    0.4     &     \ldots   &   \ldots   \\ 
J0341-0026   &   0.0306  &    1.4  $\pm$    0.7  &    -7.8 $\pm$    1.1     &      8.38 $\pm$ 0.32   &   -1.92   \\ 
ESO358-G060   &   \ldots  &   \ldots  &   \ldots     &     \ldots   &   \ldots   \\ 
G0405-3648   &   \ldots  &   \ldots  &   \ldots     &     \ldots   &   \ldots   \\ 
J0519+0007   &   \ldots  &   \ldots  &   \ldots     &     \ldots   &   \ldots   \\ 
Tol0618-402   &   \ldots  &   \ldots  &   \ldots     &     \ldots   &   \ldots   \\ 
ESO489-G56   &   \ldots  &   \ldots  &   \ldots     &     \ldots   &   \ldots   \\ 
J0808+1728   &   0.0442  &   -2.0  $\pm$    0.5  &    -3.0 $\pm$    0.5     &      7.41 $\pm$ 0.07   &   -0.79   \\ 
J0812+4836   &   0.0017  &    0.6  $\pm$    1.0  &    -4.4 $\pm$    2.0     &      6.35 $\pm$ 0.05   &   -2.64   \\ 
UGC4305   &   \ldots  &   \ldots  &   \ldots     &     \ldots   &   \ldots   \\ 
J0825+1846   &   0.0380  &   -9.6  $\pm$    0.3  &    -14.0 $\pm$    0.3     &      7.35 $\pm$ 0.23   &   -0.22   \\ 
HS0822+03542   &   0.0025  &   -31.4  $\pm$    0.3  &    -32.9 $\pm$    0.2     &      6.04 $\pm$ 0.03   &   -1.99   \\ 
DD053   &   \ldots  &   \ldots  &   \ldots     &     \ldots   &   \ldots   \\ 
UGC4483   &   \ldots  &   \ldots  &   \ldots     &     \ldots   &   \ldots   \\ 
HS0837+4717   &   0.0420  &   31.2  $\pm$    0.3  &    28.3 $\pm$    0.2     &      7.97 $\pm$ 0.24   &    0.55   \\ 
J0842+1033   &   0.0103  &   12.8  $\pm$    0.3  &    13.8 $\pm$    0.2     &      7.01 $\pm$ 0.08   &   -1.19   \\ 
HS0846+3522   &   \ldots  &   \ldots  &   \ldots     &     \ldots   &   \ldots   \\ 
J0859+3923   &   \ldots  &   \ldots  &   \ldots     &     \ldots   &   \ldots   \\ 
J0910+0711   &   \ldots  &   \ldots  &   \ldots     &     \ldots   &   \ldots   \\ 
J0911+3135   &   0.0025  &   -2.3  $\pm$    0.8  &    -2.7 $\pm$    1.5     &      6.14 $\pm$ 0.06   &   -2.73   \\ 
J0926+3343   &   0.0018  &    3.3  $\pm$    4.0  &    18.0 $\pm$   13.9     &      6.05 $\pm$ 0.07   &   -2.96   \\

\hline
 
\end{tabular}

\end{minipage}

\end{center}

\end{table*}



\setcounter{table}{3}

\begin{table*}

\footnotesize

\begin{center}

\begin{minipage}[c]{118mm}

\caption{The SDSS MPA-JHU data for the 140 XMP galaxies in the local Universe. Continued.}

\begin{tabular}{l c c c c c c}

\hline

Name &   $z$  &   $\Delta$v$_{\rm Balmer}$  & $\Delta$v$_{\rm forbidden}$   & log M$_{\star}$ &   log SFR \\
     &        & km s$^{-1}$                & km s$^{-1}$               & M$_{\odot}$ &   M$_{\odot}$ yr$^{-1}$ \\
(1)  &  (2)   & (3)                 & (4)                        &   (5)                     &  (6)           \\

\hline
\hline

IZw18   &   0.0024  &   12.9  $\pm$    0.4  &    10.9 $\pm$    0.7     &      6.43 $\pm$ 0.01   &   -1.74   \\ 
J0940+2935   &   0.0017  &   -0.6  $\pm$    0.4  &    -1.8 $\pm$    0.5     &      6.30 $\pm$ 0.15   &   -2.09   \\ 
J0942+3404   &   0.0225  &    5.0  $\pm$    0.3  &    -1.7 $\pm$    0.3     &      6.83 $\pm$ 0.11   &   -1.10   \\ 
CGCG007-025   &   0.0048  &    7.3  $\pm$    0.2  &    12.0 $\pm$    0.1     &      6.67 $\pm$ 0.16   &   -1.21   \\ 
SBS940+544   &   0.0054  &    0.3  $\pm$    1.0  &    -4.2 $\pm$    1.8     &      7.05 $\pm$ 0.05   &   -1.81   \\ 
CS0953-174   &   \ldots  &   \ldots  &   \ldots     &     \ldots   &   \ldots   \\ 
J0956+2849   &   0.0017  &   -7.8  $\pm$    1.3  &    -12.5 $\pm$    3.3     &      6.73 $\pm$ 0.07   &   -1.89   \\ 
LeoA   &   \ldots  &   \ldots  &   \ldots     &     \ldots   &   \ldots   \\ 
SextansB   &   \ldots  &   \ldots  &   \ldots     &     \ldots   &   \ldots   \\ 
J1003+4504   &   0.0092  &    2.5  $\pm$    0.3  &    -1.4 $\pm$    0.2     &      7.03 $\pm$ 0.08   &   -1.17   \\ 
SextansA   &   \ldots  &   \ldots  &   \ldots     &     \ldots   &   \ldots   \\ 
KUG1013+381   &   0.0039  &    7.2  $\pm$    0.3  &     9.4 $\pm$    0.2     &      6.58 $\pm$ 1.00   &   -1.46   \\ 
SDSSJ1025+1402   &   0.1004  &   110.4  $\pm$    3.0  &    67.4 $\pm$    0.5     &     10.22 $\pm$ 0.14   &    1.57   \\ 
UGCA211   &   0.0028  &    1.2  $\pm$    0.9  &     1.0 $\pm$    1.4     &      7.18 $\pm$ 0.08   &   -2.20   \\ 
J1031+0434   &   0.0039  &    0.3  $\pm$    0.5  &    -0.4 $\pm$    0.4     &      6.61 $\pm$ 0.05   &   -1.53   \\ 
HS1033+4757   &   0.0052  &   -2.7  $\pm$    0.4  &    -5.7 $\pm$    0.4     &      7.08 $\pm$ 0.05   &   -1.89   \\ 
J1044+0353   &   0.0129  &    2.8  $\pm$    0.2  &     3.7 $\pm$    0.2     &      6.80 $\pm$ 0.24   &   -0.85   \\ 
HS1059+3934   &   \ldots  &   \ldots  &   \ldots     &     \ldots   &   \ldots   \\ 
J1105+6022   &   0.0044  &   -1.5  $\pm$    0.3  &    -2.7 $\pm$    0.3     &      6.96 $\pm$ 0.08   &   -1.53   \\ 
J1119+5130   &   0.0045  &   -2.9  $\pm$    0.3  &    -5.2 $\pm$    0.3     &      6.48 $\pm$ 0.17   &   -1.75   \\ 
J1121+0324   &   0.0038  &   -0.3  $\pm$    0.4  &    -6.2 $\pm$    0.5     &      6.19 $\pm$ 0.16   &   -2.10   \\ 
UGC6456   &   \ldots  &   \ldots  &   \ldots     &     \ldots   &   \ldots   \\ 
SBS1129+576   &   \ldots  &   \ldots  &   \ldots     &     \ldots   &   \ldots   \\ 
J1145+5018   &   0.0056  &   -0.4  $\pm$    0.4  &    -6.0 $\pm$    0.3     &      6.71 $\pm$ 0.07   &   -1.90   \\ 
J1151-0222   &   0.0035  &    3.1  $\pm$    0.3  &    -2.7 $\pm$    0.2     &      6.65 $\pm$ 1.00   &   -1.88   \\ 
J1157+5638   &   0.0014  &   -6.8  $\pm$    0.3  &    -7.5 $\pm$    0.4     &     \ldots   &   -2.60   \\ 
J1201+0211   &   0.0032  &   -2.0  $\pm$    0.3  &    -2.0 $\pm$    0.3     &      6.09 $\pm$ 0.11   &   -1.99   \\ 
SBS1159+545   &   \ldots  &   \ldots  &   \ldots     &     \ldots   &   \ldots   \\ 
SBS1211+540   &   0.0031  &   -14.9  $\pm$    0.3  &    -14.6 $\pm$    0.3     &      6.02 $\pm$ 0.08   &   -1.94   \\ 
J1215+5223   &   0.0005  &   -6.0  $\pm$    0.4  &    -5.0 $\pm$    0.4     &      6.01 $\pm$ 0.01   &   -2.82   \\ 
Tol1214-277   &   \ldots  &   \ldots  &   \ldots     &     \ldots   &   \ldots   \\ 
VCC0428   &   0.0027  &   -0.9  $\pm$    0.4  &    -1.9 $\pm$    0.4     &      6.20 $\pm$ 0.05   &   -1.92   \\ 
HS1222+3741   &   0.0404  &    5.6  $\pm$    0.3  &     5.5 $\pm$    0.2     &      7.86 $\pm$ 0.14   &   -0.09   \\ 
Tol65   &   \ldots  &   \ldots  &   \ldots     &     \ldots   &   \ldots   \\ 
J1230+1202   &   0.0042  &    4.6  $\pm$    0.3  &     7.3 $\pm$    0.2     &      6.56 $\pm$ 0.05   &   -1.69   \\ 
KISSR85   &   \ldots  &   \ldots  &   \ldots     &     \ldots   &   \ldots   \\ 
UGCA292   &   \ldots  &   \ldots  &   \ldots     &     \ldots   &   \ldots   \\ 
HS1236+3937   &   \ldots  &   \ldots  &   \ldots     &     \ldots   &   \ldots   \\ 
J1239+1456   &   \ldots  &   \ldots  &   \ldots     &     \ldots   &   \ldots   \\ 
SBS1249+493   &   \ldots  &   \ldots  &   \ldots     &     \ldots   &   \ldots   \\ 
J1255-0213   &   0.0519  &   -8.2  $\pm$    0.4  &    -11.0 $\pm$    0.5     &      7.67 $\pm$ 0.22   &   -0.68   \\ 
GR8   &   0.0007  &    0.4  $\pm$    0.6  &    -7.2 $\pm$    1.1     &      6.02 $\pm$ 0.04   &   -3.16   \\ 
KISSR1490   &   \ldots  &   \ldots  &   \ldots     &     \ldots   &   \ldots   \\ 
DD0167   &   \ldots  &   \ldots  &   \ldots     &     \ldots   &   \ldots   \\ 
HS1319+3224   &   \ldots  &   \ldots  &   \ldots     &     \ldots   &   \ldots   \\ 
J1323-0132   &   0.0225  &   -2.9  $\pm$    0.2  &    -5.9 $\pm$    0.2     &      7.04 $\pm$ 0.24   &   -0.73   \\ 
J1327+4022   &   0.0105  &   -4.2  $\pm$    0.4  &    -6.6 $\pm$    0.4     &      6.27 $\pm$ 0.13   &   -1.70   \\ 
J1331+4151   &   0.0117  &   -4.9  $\pm$    0.2  &    -9.0 $\pm$    0.2     &      7.16 $\pm$ 0.08   &   -0.89   \\ 
ESO577-G27   &   \ldots  &   \ldots  &   \ldots     &     \ldots   &   \ldots   \\ 
J1355+4651   &   0.0281  &    3.8  $\pm$    1.0  &    -1.7 $\pm$    0.6     &      6.90 $\pm$ 0.21   &   -1.42   \\ 
J1414-0208   &   0.0052  &    0.4  $\pm$    0.7  &    -1.1 $\pm$    1.3     &      6.61 $\pm$ 0.12   &   -1.87   \\ 
SBS1415+437   &   \ldots  &   \ldots  &   \ldots     &     \ldots   &   \ldots   \\ 
J1418+2102   &   0.0085  &   15.8  $\pm$    0.2  &    19.8 $\pm$    0.2     &      6.63 $\pm$ 0.15   &   -1.16   \\ 
J1422+5145   &   \ldots  &   \ldots  &   \ldots     &     \ldots   &   \ldots   \\ 
J1423+2257   &   0.0328  &   -3.6  $\pm$    0.3  &    -2.3 $\pm$    0.3     &      7.65 $\pm$ 0.11   &   -0.19   \\ 
J1441+2914   &   \ldots  &   \ldots  &   \ldots     &     \ldots   &   \ldots   \\ 
HS1442+4250   &   0.0021  &   -0.8  $\pm$    0.6  &    -2.5 $\pm$    0.7     &      6.52 $\pm$ 0.07   &   -2.08   \\ 
J1509+3731   &   0.0325  &   10.4  $\pm$    0.3  &    14.2 $\pm$    0.2     &      7.78 $\pm$ 0.18   &    0.33   \\ 
KISSR666   &   \ldots  &   \ldots  &   \ldots     &     \ldots   &   \ldots   \\ 

\hline
 
\end{tabular}

\end{minipage}

\end{center}

\end{table*}



\setcounter{table}{3}

\begin{table*}

\footnotesize

\begin{center}

\begin{minipage}[c]{118mm}

\caption{The SDSS MPA-JHU data for the 140 XMP galaxies in the local Universe. Continued.}

\begin{tabular}{l c c c c c c c}

\hline

Name &   $z$  &   $\Delta$v$_{\rm Balmer}$  & $\Delta$v$_{\rm forbidden}$   & log M$_{\star}$ &   log SFR \\
     &        & km s$^{-1}$                & km s$^{-1}$               & M$_{\odot}$ &   M$_{\odot}$ yr$^{-1}$ \\
(1)  &  (2)   & (3)                 & (4)                        &   (5)                     &  (6)        \\

\hline
\hline

KISSR1013   &   0.0249  &    3.1  $\pm$    0.6  &     4.4 $\pm$    0.5     &      8.41 $\pm$ 0.07   &   -0.93   \\ 
J1644+2734   &   0.0232  &   -28.6  $\pm$    1.9  &    -140.9 $\pm$   14.0     &      8.87 $\pm$ 0.08   &   -0.94   \\ 
J1647+2105   &   \ldots  &   \ldots  &   \ldots     &     \ldots   &   \ldots   \\ 
W1702+18   &   \ldots  &   \ldots  &   \ldots     &     \ldots   &   \ldots   \\ 
HS1704+4332   &   \ldots  &   \ldots  &   \ldots     &     \ldots   &   \ldots   \\ 
SagDIG   &   \ldots  &   \ldots  &   \ldots     &     \ldots   &   \ldots   \\ 
J2053+0039   &   0.0131  &    3.7  $\pm$    1.9  &    -1.5 $\pm$    2.5     &      7.26 $\pm$ 0.10   &   -1.73   \\ 
J2104-0035   &   0.0046  &   -0.2  $\pm$    0.6  &     1.0 $\pm$    1.8     &      6.19 $\pm$ 0.05   &   -2.01   \\ 
J2105+0032   &   0.0142  &   -0.2  $\pm$    0.9  &    -7.0 $\pm$    1.2     &      7.23 $\pm$ 1.00   &   -1.63   \\ 
J2120-0058   &   0.0197  &   -0.4  $\pm$    0.3  &     1.2 $\pm$    0.4     &      7.43 $\pm$ 0.09   &   -1.15   \\ 
HS2134+0400   &   \ldots  &   \ldots  &   \ldots     &     \ldots   &   \ldots   \\ 
J2150+0033   &   0.0149  &    0.4  $\pm$    0.9  &    -2.1 $\pm$    1.5     &      7.90 $\pm$ 0.34   &   \ldots   \\ 
ESO146-G14   &   \ldots  &   \ldots  &   \ldots     &     \ldots   &   \ldots   \\ 
2dF171716   &   \ldots  &   \ldots  &   \ldots     &     \ldots   &   \ldots   \\ 
PHL293B   &   0.0053  &   16.0  $\pm$    0.3  &    15.2 $\pm$    0.2     &      6.69 $\pm$ 0.06   &   -1.52   \\ 
2dF115901   &   \ldots  &   \ldots  &   \ldots     &     \ldots   &   \ldots   \\ 
J2238+1400   &   0.0206  &    8.6  $\pm$    0.2  &     6.2 $\pm$    0.2     &      6.72 $\pm$ 0.24   &   -0.77   \\ 
J2250+0000   &   0.0808  &   -8.8  $\pm$    0.3  &    -11.9 $\pm$    0.3     &     \ldots   &    3.88   \\ 
J2259+1413   &   0.0939  &    8.5  $\pm$    5.8  &    -20.2 $\pm$   20.3     &     10.43 $\pm$ 1.00   &   -0.35   \\ 
J2302+0049   &   0.0331  &   -7.7  $\pm$    0.3  &    -8.8 $\pm$    0.3     &      8.39 $\pm$ 0.33   &   -0.56   \\ 
J2354-0005   &   0.0077  &    0.6  $\pm$    1.9  &    -7.0 $\pm$    3.9     &      7.33 $\pm$ 0.34   &   \ldots   \\

\hline
 
\end{tabular}

\end{minipage}

\end{center}

\end{table*}






\section{Multi-wavelength Analysis}

Combining the \ion{H}{i}, SDSS and other online data, we derive global galaxy parameters for the XMP galaxies, including bulk velocities and masses (Table~5). 

When multiple \ion{H}{i} entries are available from literature (12 sources; Table~1), we preferably chose interferometric values to provide better constraints on the \ion{H}{i} gas parameters. Sources HS0122+0743, J0133+1342, J1105+6022, J1121+0324, J1201+0211, J1215+5223 and HS1442+4250 are documented in Pustilnik \& Martin (2008) as belonging to merger systems or as having possible companions. This may produce possible \ion{H}{i} contamination and consequently (artifically) increase the value of the \ion{H}{i} mass.


We do not correct the Effelsberg \ion{H}{i} line widths (Table~2) for instrumental effects, turbulence, inclination and redshift stretching; these corrections are estimated to be small at low redshift (e.g., Springob et al. 2005). Indeed, even if line widths are small and the sources faint (Table~2), the errors in the line widths are dominated by the much larger errors generated in the estimation of line widths from low signal-to-noise spectra.

The Effelsberg \ion {H}{i} systemic radial velocities (Table~2) have been converted from the local to the barycentric standard-of-rest using the coordinates of each galaxy. The corresponding velocity corrections range from -30 to 13 km s$^{-1}$. The velocities have not been converted from the barycentric to the heliocentric standard-of-rest, since this correction has a maximum amplitude of 0.012 km s$^{-1}$, much smaller than the \ion{H}{i} gas velocity and velocity offset errors. We further converted the Effelsberg systemic radial velocities from the radio to the optical convention, according to the formula:

\begin{equation}
{\rm v}_{opt} = \frac{{\rm v}_{rad}}{1 - {\rm v}_{rad}/c}, 
\end{equation}

\noindent where v$_{opt}$ is the systemic radial velocity in the optical convention, v$_{rad}$ is the systemic radial velocity in the radio convention and $c$ is the speed of light. The difference between the radio and optical convention stems from the use of the frequency (radio) or the wavelength (optical) to infer the line shifts. 

The \ion{H}{i} velocity offset, $\Delta$v$_{\ion{H}{i}}$, has been defined as the displacement of the \ion{H}{i} systemic radial velocity (heliocentric, optical convention; Tables~1 and 2) with respect to the best-fit radial velocity, v$_{z}$ (heliocentric, optical convention; Sect.~3.3 and Table~4). The error in the \ion{H}{i} velocity offset contains the corresponding error in the \ion{H}{i} systemic radial velocity (Tables~1 and 2) and a 1 km s$^{-1}$ error in the best-fit radial velocity determination. When the former is not available, we adopt a value of 5 km s$^{-1}$ for the \ion{H}{i} systemic radial velocity error.

In order to compute dynamical masses, we require \ion{H}{i} galaxy sizes. From the \ion{H}{i} (Table~1) and optical (Table~3) radii, we estimate that, on average, the \ion{H}{i}-to-optical size ratio $\sim$3 for XMP galaxies in the local Universe, which is consistent with the results from literature (Thuan \& Martin 1981). We therefore adopt a value of r$_{\ion{H}{i}}$ = 3 $\times$ r$_{\rm opt}$. 

The dynamical mass, defined as the total baryonic plus non-baryonic mass contained within a certain radius, is estimated assuming a spherical system in dynamical equilibrium. We have determined the dynamical mass using the formula\footnote{http://www.cv.nrao.edu/course/astr534/HILine.html}:

\begin{equation}
\frac {\rm M_{dyn}}{\rm M_{\odot}} = 2.3 \times 10^5 \, \left( \frac{w_{50}}{2} \right)^2 \, {\rm r_{\ion{H}{i}}}, 
\end{equation}

\noindent where $w_{50}$ is the (uncorrected) \ion{H}{i} line width in km s$^{-1}$ (Tables~1 and 2) and r$_{\ion{H}{i}}$ is the \ion{H}{i} radius in kpc (see above). We stress that the dynamical mass calculated in this way may be an underestimation, since the \ion{H}{i} radius may not include the full extent of the galaxy.

The \ion{H}{i} mass estimate assumes the neutral atomic gas mass to be optically thin (Wild 1952). The formula\footnote{http://www.cv.nrao.edu/course/astr534/HILine.html} for deriving the \ion{H}{i} mass is:

\begin{equation}
\rm \frac{M_{\ion{H}{i}}}{M_{\odot}} = 2.36 \times 10^5 \, {D}^2 \, {S_{\ion{H}{i}}}, 
\end{equation}

\noindent where S$_{\ion{H}{i}}$ is the integrated flux density in Jy km s$^{-1}$ (Tables~1 and 2) and D is the Virgocentric infall-corrected Hubble flow distance in Mpc (Table~5).

The dark matter mass has been estimated as:

\begin{equation}
\rm M_{DM} = M_{dyn} - M_{\star} - M_{\ion{H}{i}} - M_{\rm He}, 
\end{equation}

\noindent where M$_{\rm He}$, the helium mass, is 0.34 $\times$ M$_{\ion{H}{i}}$, the value typical for standard Big Bang nucleosynthesis (Alpher et al. 1948; Coc et al. 2012).



The error determination in the dynamical mass includes a 20\% error in ${\rm r_{\ion{H}{i}}}$ and the corresponding error in the line width (Table~1). When the latter is not available, we have adopted a conservative error of 5 km s$^{-1}$. For the \ion{H}{i} gas mass, the error was estimated assuming a conservative error in integrated flux density (10\%; Sect.~2.2.2) and the error in the Virgocentric infall-corrected Hubble flow distance, as provided by NED. In the few cases when the latter is not available, we have attributed a conservative distance error of 10\%. The helium mass error determination follows from the propogation of the error in \ion{H}{i} gas mass. We adopt the 1$\sigma$ stellar mass errors (Sect.~3.3 and Table~4) which, typically, are of the order of 0.5 dex (Fig.~3). For the dark matter mass, the final error estimation includes the error in dynamical, \ion{H}{i} gas, helium and stellar mass.

All quoted velocities are in the heliocentric standard-of-rest and in the optical convention. Correlations found in literature for absolute $B$-band magnitudes have been converted to absolute $g$-band magnitudes using the standard SDSS conversion factors (e.g., Jester et al. 2005). 

Unless otherwise explicitly stated, metallicities refer to \ion{H}{ii} gas-phase metallicities (Table~3).

We call attention to the six sources with the lowest $g$-band luminosity -- UGC2684, DD053, LeoA, SextansB, UGCA292 and GR8. These are all nearby (D $<$ 6 Mpc), low-surface brightness galaxies, that also turn out to be XMPs.



\setcounter{table}{4}

\begin{table*}

\footnotesize

\begin{center}

\begin{minipage}[c]{184mm}

\caption{Global galaxy parameters for the 140 XMP galaxies in the local Universe.
Col.~1: Source name.
Col.~2: Best-fit radial velocity (v$_z$=c$z$; heliocentric, optical convention).
Col.~3: \ion{H}{i} gas radial velocity (heliocentric, optical convntion).
Col.~4: \ion{H}{i} gas velocity offset relative to the best-fit radial velocity.
Col.~5: Radial velocity for the Balmer emission lines (heliocentric, optical convention).
Col.~6: Radial velocity for the forbidden emission lines (heliocentric, optical convention).
Col.~7: Hubble flow distance corrected for Virgocentric infall, from NED.
Col.~8: Logarithm of the dynamical mass, estimated as M$_{\rm dyn}$/M$_{\odot}$ = 2.3 $\times$ 10$^5$ $\left( \frac{w_{50}}{2} \right)^2$ r$_{\ion{H}{i}}$, where $w_{50}$ is the (uncorrected) \ion{H}{i} line width (Tables~1 and 2) in km s$^{-1}$ and r$_{\ion{H}{i}}$ is the \ion{H}{i} radius (Table~1), assumed to be 3 $\times$ r$_{\rm opt}$ (Table~3) in pc. 
Col.~9: Logarithm of the \ion{H}{i} mass, estimated as M$_{\ion{H}{i}}$/M$_{\odot}$ = 2.36 $\times$ 10$^5$ D$^2$ S$_{\ion{H}{i}}$, where  S$_{\ion{H}{i}}$ is the integrated line flux density (Tables~1 and 2) in Jy km s$^{-1}$ and D is the Hubble flow distance corrected for Virgocentric infall in Mpc. 
Col.~10: Logarithm of the stellar mass (Table~4).
Col.~11: Logarithm of the stellar M/L ratio in the $g$-band.
}

\begin{tabular}{l c c c c c c c c c c}

\hline
 
Name    & v$_{z}$    & v$_{\ion{H}{i}}$  & $\Delta$v$_{\ion{H}{i}}$  & v$_{\rm Balmer}$  & v$_{\rm forbid}$ &  D  &  log M$_{\rm dyn}$         &  log M$_{\ion{H}{i}}$      &  log M$_{\star}$             & log M$_{\star}$/L$_{g}$ \\
        & km s$^{-1}$       & km s$^{-1}$         & km s$^{-1}$       & km s$^{-1}$    &   km s$^{-1}$    & Mpc           &  M$_{\odot}$         &  M$_{\odot}$          &  M$_{\odot}$               &  M$_{\odot}$/L$_{\odot}$      \\
(1)    & (2)          & (3)               & (4)                 & (5)               & (6)                 & (7)           & (8)                   & (9)                   & (10)                       & (11)     \\

\hline
\hline

UGC12894 & \ldots & 335.00 & \ldots & \ldots & \ldots &  8.36 &  8.34 &  8.05 &  \ldots & \ldots \\ 
J0004+0025 & 3786.00 &  \ldots & \ldots & 3783.58 & 3797.03 & 52.30 & \ldots & \ldots &   7.72 &  2.04 \\ 
J0014-0044 & 4077.00 &  \ldots & \ldots & 4081.96 & 4083.42 & 56.70 & \ldots & \ldots &   7.61 &  1.58 \\ 
J0015+0104 & 2058.00 & 2050.02 &  7.98 & 2061.42 & 2058.00 & 28.80 &  8.46 &  8.47 &   7.46 &  1.86 \\ 
J0016+0108 & 3111.00 &  \ldots & \ldots & 3111.00 & 3115.66 & 42.80 & \ldots & \ldots &   7.64 &  1.94 \\ 
HS0017+1055 & \ldots & 5630.00 & \ldots & \ldots & \ldots & 78.10 & \ldots & \ldots &  \ldots & \ldots \\ 
J0029-0108 & 3948.00 &  \ldots & \ldots & 3948.00 & 3939.95 & 53.90 & \ldots & \ldots &   7.88 &  2.10 \\ 
J0029-0025 & 4236.00 &  \ldots & \ldots & 4230.90 & 4236.00 & 59.10 & \ldots & \ldots &   6.88 &  1.50 \\ 
ESO473-G024 & \ldots & 542.00 & \ldots & \ldots & \ldots &  7.15 &  8.32 &  7.84 &  \ldots & \ldots \\ 
J0036+0052 & 8469.00 &  \ldots & \ldots & 8470.20 & 8474.05 & 118.50 & \ldots & \ldots &   8.36 &  1.73 \\ 
AndromedaIV & \ldots & 237.00 & \ldots & \ldots & \ldots &  6.30 &  9.17 &  8.26 &  \ldots & \ldots \\ 
J0057-0022 & 2811.00 &  \ldots & \ldots & 2811.00 & 2815.08 & 39.20 & \ldots & \ldots &   7.46 &  1.91 \\ 
IC1613     & \ldots  & 234.00  & \ldots & \ldots  & \ldots & 0.74   & 8.21   &  7.45  & \ldots & \ldots \\
J0107+0001 & 5439.00 &  \ldots & \ldots & 5444.38 & 5443.97 & 74.90 & \ldots & \ldots &   7.86 &  1.87 \\ 
AM0106-382 & \ldots &  \ldots & \ldots & \ldots & \ldots &  7.50 & \ldots & \ldots &  \ldots & \ldots \\ 
J0113+0052 & 1143.00 & 1155.61 & -12.61 & 1143.00 & 1143.00 & 15.80 &  7.71 &  8.53 &   6.77 &  2.41 \\ 
J0119-0935 & 1914.00 & 1932.00 & -18.00 & 1911.15 & 1916.58 & 24.80 & \ldots &  8.14 &   6.06 &  1.07 \\ 
HS0122+0743 & 2925.00 & 2926.00 & -1.00 & 2915.68 & 2921.30 & 40.30 &  8.98 &  9.33 &   6.39 & -0.54 \\ 
J0126-0038 & 1941.00 & 1904.70 & 36.30 & 1941.00 & 1936.53 & 25.80 &  8.24 &  8.63 &   7.63 &  2.17 \\ 
J0133+1342 & 2601.00 & 2580.00 & 21.00 & 2607.57 & 2601.00 & 36.10 &  8.27 &  7.49 &   6.60 &  0.72 \\ 
J0135-0023 & 5070.00 &  \ldots & \ldots & 5067.24 & 507\ldots & 69.50 & \ldots & \ldots &   8.16 &  2.04 \\ 
UGCA20 & \ldots & 498.00 & \ldots & \ldots & \ldots &  8.63 &  9.25 &  8.30 &  \ldots & \ldots \\ 
UM133 & \ldots & 1621.00 & \ldots & \ldots & \ldots & 22.40 & \ldots &  8.64 &  \ldots & \ldots \\ 
J0158+0006 & \ldots &  \ldots & \ldots & \ldots & \ldots & \ldots & \ldots & \ldots &  \ldots & \ldots \\ 
HKK97L14 & \ldots & \ldots & \ldots & \ldots & \ldots &  4.81 &  8.80 &  6.52 &  \ldots & \ldots \\ 
J0204-1009 & 1902.00 & 1907.72 & -5.72 & 1902.00 & 1897.01 & 25.20 & 9.43 &  9.17 &   7.20 &  1.24 \\ 
J0205-0949 & \ldots & 1885.00 & \ldots & \ldots & \ldots & 25.30 & 9.76 &  9.29 &  \ldots & \ldots \\ 
J0216+0115 & 2802.00 &  \ldots & \ldots & 2802.00 & 2804.84 & 38.70 & \ldots & \ldots &   8.20 &  1.98 \\ 
096632 & \ldots &  \ldots & \ldots & \ldots & \ldots & 12.90 & \ldots & \ldots &  \ldots & \ldots \\ 
J0254+0035 & 4458.00 &  \ldots & \ldots & 4458.00 & 4458.00 & 59.90 & \ldots & \ldots &   7.61 &  1.98 \\ 
J0301-0059 & 11484.00 &  \ldots & \ldots & 11480.46 & 11490.10 & 155.60 & \ldots & \ldots &   8.96 &  3.18 \\ 
J0301-0052 & 2196.00 & 2094.29 & 101.71 & 2204.43 & 2206.18 & 29.00 &  8.87 &  8.62 &   7.57 &  2.17 \\ 
J0303-0109 & 9120.00 &  \ldots & \ldots & 9120.00 & 9117.65 & 124.00 & \ldots & \ldots &   8.33 &  2.06 \\ 
J0313+0006 & 8748.00 &  \ldots & \ldots & 8758.57 & 8763.53 & 122.50 & \ldots & \ldots &   7.67 &  1.17 \\ 
J0313+0010 & 2322.00 &  \ldots & \ldots & 2322.00 & 2329.65 & 31.10 & \ldots & \ldots &   7.57 &  2.14 \\ 
J0315-0024 & 6774.00 & 6787.09 & -13.09 & 6774.00 & 6774.00 & 90.90 & 8.98 &  9.40 &   7.77 &  1.93 \\ 
UGC2684 & \ldots & 350.00 & \ldots & \ldots & \ldots &  5.95 &  7.88 &  7.94 &  \ldots & \ldots \\ 
SBS0335-052W & \ldots & 4014.70 & \ldots & \ldots & \ldots & 53.80 &  8.54 &  8.77 &  \ldots & \ldots \\ 
SBS0335-052E & \ldots & 4053.60 & \ldots & \ldots & \ldots & 54.00 &  8.88 &  8.62 &  \ldots & \ldots \\ 
J0338+0013 & 12795.00 &  \ldots & \ldots & 12790.03 & 12791.60 & 173.60 & \ldots & \ldots &  \ldots & \ldots \\ 
J0341-0026 & 9186.00 &  \ldots  & \ldots & 9186.00 & 9193.79 & 123.50 & \ldots & \ldots &   8.38 &  1.72 \\ 
ESO358-G060 & \ldots & 808.00 & \ldots & \ldots & \ldots &  8.90 &  9.10 &  8.30 &  \ldots & \ldots \\ 
G0405-3648 & \ldots &  \ldots & \ldots & \ldots & \ldots &  9.00 & \ldots & \ldots &  \ldots & \ldots \\ 
J0519+0007 & \ldots &  \ldots & \ldots & \ldots & \ldots & 180.40 & \ldots & \ldots &  \ldots & \ldots \\ 
Tol0618-402 & \ldots &  \ldots & \ldots & \ldots & \ldots & 140.0 & \ldots & \ldots &  \ldots & \ldots \\ 
ESO489-G56 & \ldots & 492.30 & \ldots & \ldots & \ldots &  4.23 &  7.65 &  6.95 &  \ldots & \ldots \\ 
J0808+1728 & 13263.00 &  \ldots & \ldots & 13265.01 & 13266.03 & 181.40 & \ldots & \ldots &   7.41 &  0.57 \\ 
J0812+4836 & 525.00 &  \ldots & \ldots & 525.00 & 529.39 &  9.44 & \ldots & \ldots &   6.35 &  0.80 \\ 
UGC4305 & \ldots & 157.10 & \ldots & \ldots & \ldots &  4.89 & 9.46 &  9.09 &  \ldots & \ldots \\ 
J0825+1846 & 11388.00 &  \ldots & \ldots & 11397.62 & 11401.95 & 156.10 & \ldots & \ldots &   7.35 &  0.56 \\ 
HS0822+03542 & 750.00 & 726.60 & 23.40 & 781.37 & 782.90 & 11.72 &  7.31 &  6.94 &   6.04 &  1.02 \\
DD053 & \ldots & 19.20 & \ldots & \ldots & \ldots &  2.42 &  7.87 &  7.47 &  \ldots & \ldots \\ 
UGC4483 & \ldots &  \ldots & \ldots & \ldots & \ldots &  5.00 & \ldots &  7.88 &  \ldots & \ldots \\ 
HS0837+4717 & 12588.00 &  \ldots & \ldots & 12556.84 & 12559.68 & 174.30 & \ldots & \ldots &   7.97 &  0.53 \\ 

\hline
					
\end{tabular}

\end{minipage}

\end{center}

\end{table*}



\setcounter{table}{4}

\begin{table*}

\footnotesize

\begin{center}

\begin{minipage}[c]{184mm}

\caption{Global galaxy parameters for the 140 XMP galaxies in the local Universe. Continued.}

\begin{tabular}{l c c c c c c c c c c}

\hline
 
Name    & v$_{z}$    & v$_{\ion{H}{i}}$  & $\Delta$v$_{\ion{H}{i}}$  & v$_{\rm Balmer}$  & v$_{\rm forbid}$ &  D &  log M$_{\rm dyn}$         &  log M$_{\ion{H}{i}}$      &  log M$_{\star}$             & log M$_{\star}$/L$_{g}$ \\
        & km s$^{-1}$       & km s$^{-1}$         & km s$^{-1}$       & km s$^{-1}$    &   km s$^{-1}$    & Mpc           &  M$_{\odot}$         &  M$_{\odot}$          &  M$_{\odot}$               &  M$_{\odot}$/L$_{\odot}$      \\
(1)    & (2)          & (3)               & (4)                 & (5)               & (6)                 & (7)           & (8)                   & (9)                   & (10)                       & (11)     \\

\hline 
\hline

J0842+1033 & 3096.00 &  \ldots & \ldots & 3083.20 & 3082.23 & 49.50 & \ldots & \ldots &   7.01 &  0.70 \\ 
HS0846+3522 & \ldots & 2169.00 & \ldots & \ldots & \ldots & 36.30 &  8.28 &  7.49 &  \ldots & \ldots \\ 
J0859+3923 & \ldots &  \ldots & \ldots & \ldots & \ldots &  9.50 & \ldots & \ldots &  \ldots & \ldots \\ 
J0910+0711 & \ldots &  \ldots & \ldots & \ldots & \ldots & 22.00 & \ldots & \ldots &  \ldots & \ldots \\ 
J0911+3135 & 756.00 &  \ldots & \ldots & 758.26 & 756.00 & 11.60 & \ldots & \ldots &   6.14 &  1.13 \\ 
J0926+3343 & 528.00 &  \ldots & \ldots & 528.00 & 528.00 &  8.25 & \ldots & \ldots &   6.05 &  1.34 \\ 
IZw18 & 726.00 & 745.00 & -19.00 & 713.06 & 715.13 & 13.90 &  7.76 &  8.12 &   6.43 &  0.59 \\ 
J0940+2935 & 504.00 & 505.00 & -1.00 & 504.00 & 505.80 &  7.23 &  8.77 &  7.40 &   6.30 &  1.18 \\ 
J0942+3404 & 6747.00 &  \ldots & \ldots & 6741.99 & 6748.69 & 94.10 & \ldots & \ldots &   6.83 &  0.52 \\ 
CGCG007-025 & 1434.00 &  \ldots & \ldots & 1426.68 & 1421.95 & 20.80 & \ldots & \ldots &   6.67 &  0.43 \\ 
SBS940+544 & 1623.00 &  \ldots & \ldots & 1623.00 & 1627.23 & 26.60 & \ldots & \ldots &   7.05 &  1.84 \\ 
CS0953-174 & \ldots &  \ldots & \ldots & \ldots & \ldots & \ldots & \ldots & \ldots &  \ldots & \ldots \\ 
J0956+2849 & 504.00 &  \ldots & \ldots & 511.75 & 516.53 &  5.85 & \ldots & \ldots &   6.73 &  1.56 \\ 
LeoA & \ldots & 21.70 & \ldots & \ldots & \ldots &  1.54 &  7.75 &  7.37 &  \ldots & \ldots \\ 
SextansB & \ldots & 300.50 & \ldots & \ldots & \ldots &  1.63 &  8.41 &  7.66 &  \ldots & \ldots \\ 
J1003+4504 & 2766.00 &  \ldots & \ldots & 2763.54 & 2767.39 & 41.10 & \ldots & \ldots &   7.03 &  0.80 \\ 
SextansA & \ldots & 324.00 & \ldots & \ldots & \ldots &  1.43 &  8.60 &  7.81 &  \ldots & \ldots \\ 
KUG1013+381 & 1164.00 & 1169.00 & -5.00 & 1156.84 & 1154.56 & 19.90 &  8.30 &  8.15 &   6.58 &  0.34 \\ 
SDSSJ1025+1402 & 30129.00 &  \ldots & \ldots & 30018.59 & 30061.57 & 413.90 & \ldots & \ldots &  10.22 &  3.15 \\ 
UGCA211 & 837.00 &  \ldots & \ldots & 837.00 & 837.00 & 15.50 & \ldots &  8.23 &   7.18 &  1.28 \\ 
J1031+0434 & 1170.00 &  \ldots & \ldots & 1170.00 & 1170.00 & 18.10 & \ldots & \ldots &   6.61 &  0.57 \\ 
HS1033+4757 & 1560.00 & 1541.00 & 19.00 & 1562.70 & 1565.65 & 25.60 &  9.04 &  8.31 &   7.08 &  1.26 \\ 
J1044+0353 & 3861.00 &  \ldots & \ldots & 3858.21 & 3857.27 & 53.50 & \ldots & \ldots &   6.80 &  0.34 \\ 
HS1059+3934 & \ldots & 3019.00 & \ldots & \ldots & \ldots & 48.10 &  9.05 &  8.88 &  \ldots & \ldots \\ 
J1105+6022 & 1326.00 & 1333.00 & -7.00 & 1327.49 & 1328.69 & 23.30 &  8.89 &  8.50 &   6.96 &  0.79 \\ 
J1119+5130 & 1338.00 &  \ldots & \ldots & 1340.89 & 1343.24 & 23.30 & \ldots & \ldots &   6.48 &  0.51 \\ 
J1121+0324 & 1149.00 & 1171.00 & -22.00 & 1149.00 & 1155.21 & \ldots &  9.11 &  8.40 &   6.19 &  0.83 \\ 
UGC6456 & \ldots & -93.69 & \ldots & \ldots & \ldots &  1.42 &  7.74 &  6.68 &  \ldots & \ldots \\ 
SBS1129+576 & \ldots & 1506.00 & \ldots & \ldots & \ldots & 26.40 &  8.82 &  8.81 &  \ldots & \ldots \\ 
J1145+5018 & 1674.00 &  \ldots & \ldots & 1674.00 & 1680.04 & 27.70 & \ldots & \ldots &   6.71 &  0.95 \\ 
J1151-0222 & 1056.00 &  \ldots & \ldots & 1052.91 & 1058.67 & 13.10 & \ldots & \ldots &   6.65 &  1.14 \\ 
J1157+5638 & 417.00 &  \ldots & \ldots & 423.81 & 424.46 &  8.56 & \ldots & \ldots &  \ldots & \ldots \\ 
J1201+0211 & 975.00 & 974.00 &  1.00 & 977.00 & 976.96 &  8.60 &  7.80 &  7.22 &   6.09 &  1.26 \\ 
SBS1159+545 & \ldots & 3560.00 & \ldots & \ldots & \ldots & 53.30 & \ldots & \ldots &  \ldots & \ldots \\ 
SBS1211+540 & 918.00 & 894.00 & 24.00 & 932.92 & 932.58 & 17.20 &  8.26 &  7.64 &   6.02 &  0.51 \\ 
J1215+5223 & 153.00 & 159.00 & -6.00 & 158.98 & 158.03 &  3.33 &  7.69 &  7.09 &   6.01 &  1.05 \\ 
Tol1214-277 & \ldots & 7785.00 & \ldots & \ldots & \ldots & 105.70 & \ldots & \ldots &  \ldots & \ldots \\ 
VCC0428 & 801.00 & 794.00 &  7.00 & 801.00 & 802.93 & 13.10 & 8.58 &  7.42 &   6.20 &  0.77 \\ 
HS1222+3741 & 12114.00 &  \ldots & \ldots & 12108.43 & 12108.51 & 170.90 & \ldots & \ldots &   7.86 &  0.55 \\ 
Tol65 & \ldots & 2790.00 & \ldots & \ldots & \ldots & 37.90 &  8.49 &  8.86 &  \ldots & \ldots \\ 
J1230+1202 & 1254.00 & 1227.00 & 27.00 & 1249.42 & 1246.75 & 13.10 &  7.93 &  7.62 &   6.56 &  1.01 \\ 
KISSR85 & \ldots &  \ldots & \ldots & \ldots & \ldots & 104.60 & \ldots & \ldots &  \ldots & \ldots \\ 
UGCA292 & \ldots & 308.30 & \ldots & \ldots & \ldots &  3.41 & 6.50  &  7.59 &  \ldots & \ldots \\ 
HS1236+3937 & \ldots &  \ldots & \ldots & \ldots & \ldots & 79.00 & \ldots & \ldots &  \ldots & \ldots \\ 
J1239+1456 & \ldots &  \ldots & \ldots & \ldots & \ldots & 296.70 & \ldots & \ldots &  \ldots & \ldots \\ 
SBS1249+493 & \ldots &  \ldots & \ldots & \ldots & \ldots & 103.60 & \ldots & \ldots &  \ldots & \ldots \\ 
J1255-0213 & 15564.00 &  \ldots & \ldots & 15572.15 & 15575.00 & 213.70 & \ldots & \ldots &   7.67 &  0.65 \\ 
GR8 & 219.00 & 217.00 &  2.00 & 219.00 & 226.21 &  1.43 &  7.28 &  6.64 &   6.02 &  2.87 \\ 
KISSR1490 & \ldots &  \ldots & \ldots & \ldots & \ldots & 53.00 & \ldots & \ldots &  \ldots & \ldots \\ 
DD0167 & \ldots & 150.24 & \ldots & \ldots & \ldots &  3.19 &  7.26 &  6.95 &  \ldots & \ldots \\ 
HS1319+3224 & \ldots &  \ldots & \ldots & \ldots & \ldots & 78.10 & \ldots & \ldots &  \ldots & \ldots \\ 
J1323-0132 & 6738.00 &  \ldots & \ldots & 6740.87 & 6743.93 & 92.50 & \ldots & \ldots &   7.04 &  0.39 \\ 
J1327+4022 & 3150.00 &  \ldots & \ldots & 3154.24 & 3156.55 & 48.00 & \ldots & \ldots &   6.27 &  0.51 \\ 
J1331+4151 & 3510.00 &  \ldots & \ldots & 3514.90 & 3519.01 & 52.60 & \ldots & \ldots &   7.16 &  0.56 \\ 
ESO577-G27 & \ldots &  \ldots & \ldots & \ldots & \ldots & 21.30 & \ldots & \ldots &  \ldots & \ldots \\ 
J1355+4651 & 8433.00 &  \ldots & \ldots & 8429.22 & 8434.65 & 118.30 & \ldots & \ldots &   6.90 &  0.47 \\ 

\hline
					
\end{tabular}

\end{minipage}

\end{center}

\end{table*}
 


\setcounter{table}{4}

\begin{table*}

\footnotesize

\begin{center}

\begin{minipage}[c]{184mm}

\caption{Global galaxy parameters for the 140 XMP galaxies in the local Universe. Continued.}

\begin{tabular}{l c c c c c c c c c c}

\hline
 
Name    & v$_{z}$    & v$_{\ion{H}{i}}$  & $\Delta$v$_{\ion{H}{i}}$  & v$_{\rm Balmer}$  & v$_{\rm forbid}$ &  D &  log M$_{\rm dyn}$         &  log M$_{\ion{H}{i}}$      &  log M$_{\star}$             & log M$_{\star}$/L$_{g}$ \\
        & km s$^{-1}$       & km s$^{-1}$         & km s$^{-1}$       & km s$^{-1}$    &   km s$^{-1}$    & Mpc           &  M$_{\odot}$         &  M$_{\odot}$          &  M$_{\odot}$               &  M$_{\odot}$/L$_{\odot}$      \\
(1)    & (2)          & (3)               & (4)                 & (5)               & (6)                 & (7)           & (8)                   & (9)                   & (10)                       & (11)     \\

\hline
\hline

J1414-0208 & 1548.00 &  \ldots & \ldots & 1548.00 & 1548.00 & 24.90 & \ldots & \ldots &   6.61 &  1.02 \\ 
SBS1415+437 & \ldots & 605.00 & \ldots & \ldots & \ldots & 11.66 &  7.52 &  8.24 &  \ldots & \ldots \\ 
J1418+2102 & 2565.00 &  \ldots & \ldots & 2549.16 & 2545.19 & 35.30 & \ldots & \ldots &   6.63 &  0.57 \\ 
J1422+5145 & \ldots &  \ldots & \ldots & \ldots & \ldots & 163.70 & \ldots & \ldots &  \ldots & \ldots \\ 
J1423+2257 & 9855.00 &  \ldots & \ldots & \ldots & 9857.29 & 138.00 & \ldots & \ldots &   7.65 &  0.53 \\ 
J1441+2914 & \ldots &  \ldots & \ldots & \ldots & \ldots & 191.60 & \ldots & \ldots &  \ldots & \ldots \\ 
HS1442+4250 & 639.00 & 647.00 & -8.00 & 639.00 & 641.47 & 12.58 &  9.39 &  8.42 &   6.52 &  0.68 \\ 
J1509+3731 & 9765.00 &  \ldots & \ldots & 9754.56 & 9750.79 & 137.90 & \ldots & \ldots &   7.78 &  0.42 \\ 
KISSR666 & \ldots &  \ldots & \ldots & \ldots & \ldots & 139.90 & \ldots & \ldots &  \ldots & \ldots \\ 
KISSR1013 & 7467.00 &  \ldots & \ldots & 7463.94 & 7462.61 & 106.30 & \ldots & \ldots &   8.41 &  1.64 \\ 
J1644+2734 & 6966.00 &  \ldots & \ldots & 6994.60 & 7106.91 & 98.20 & \ldots & \ldots &   8.87 &  1.97 \\ 
J1647+2105 & \ldots &  \ldots & \ldots & \ldots & \ldots & 41.90 & \ldots & \ldots &  \ldots & \ldots \\ 
W1702+18 & \ldots &  \ldots & \ldots & \ldots & \ldots & \ldots & \ldots & \ldots &  \ldots & \ldots \\ 
HS1704+4332 & \ldots & 2082.00 & \ldots & \ldots & \ldots & 33.60 &  8.03 &  7.81 &  \ldots & \ldots \\ 
SagDIG & \ldots & -78.50 & \ldots & \ldots & \ldots &  1.02 & \ldots &  6.75 &  \ldots & \ldots \\ 
J2053+0039 & 3936.00 & 3971.79 & -35.79 & 3936.00 & 3936.00 & 56.40 &  8.93 &  9.08 &   7.26 &  1.52 \\ 
J2104-0035 & 1395.00 & 1422.81 & -27.81 & 1395.00 & 1395.00 & 20.30 & \ldots &  8.16 &   6.19 &  0.74 \\ 
J2105+0032 & 4266.00 &  \ldots & \ldots & 4266.00 & 4272.96 & 60.50 & \ldots & \ldots &   7.23 &  1.27 \\ 
J2120-0058 & 5904.00 &  \ldots & \ldots & 5904.00 & 5902.77 & 82.70 & \ldots & \ldots &   7.43 &  1.11 \\ 
HS2134+0400 & \ldots & 5090.00 & \ldots & \ldots & \ldots & 71.10 & \ldots &  8.16 &  \ldots & \ldots \\ 
J2150+0033 & 4482.00 & 4457.28 & 24.72 & 4482.00 & 4482.00 & 63.30 &  9.13 &  9.02 &   7.90 &  2.02 \\ 
ESO146-G14 & \ldots & 1691.10 & \ldots & \ldots & \ldots & 21.30 & 10.27 &  8.95 &  \ldots & \ldots \\ 
2dF171716 & \ldots &  \ldots & \ldots & \ldots & \ldots & 44.30 & \ldots & \ldots &  \ldots & \ldots \\ 
PHL293B & 1581.00 & 1606.38 & -25.38 & 1565.04 & 1565.81 & 22.70 &  7.99 &  7.93 &   6.69 &  0.86 \\ 
2dF115901 & \ldots &  \ldots & \ldots & \ldots & \ldots & 162.90 & \ldots & \ldots &  \ldots & \ldots \\ 
J2238+1400 & 6183.00 & 6160.00 & 23.00 & 6174.38 & 6176.85 & 86.10 &  8.73 & \ldots &   6.72 &  0.45 \\ 
J2250+0000 & 24237.00 &  \ldots & \ldots & 24245.75 & 24248.88 & 332.20 & \ldots & \ldots &  \ldots & \ldots \\ 
J2259+1413 & 28158.00 &  \ldots & \ldots & 28158.00 & 28158.00 & 123.60 & \ldots & \ldots &  10.43 &  3.89 \\ 
J2302+0049 & 9936.00 &  \ldots & \ldots & 9943.73 & 9944.84 & 136.20 & \ldots & \ldots &   8.39 &  1.64 \\ 
J2354-0005 & 2322.00 &  \ldots & \ldots & 2322.00 & 2322.00 & 32.20 & \ldots & \ldots &   7.33 &  1.79 \\ 

\hline
					
\end{tabular}

\end{minipage}

\end{center}

\end{table*}
 

\section{Results}

In this section, and in Fig.~4 -- 12, we present the properties of the XMP galaxy class, particularly the \ion{H}{i} content. All plots use a common color code to parametrize the morphology. The morphological color code is red for symmetric, black for cometary, blue for two-knot, green for multi-knot and grey for sources with no morphological information. 


\subsection{\ion{H}{i} Line Profile -- Optical Morphology Relation}

If we include the three marginal detections (and exclude J0014-0044), \ion{H}{i} Effelsberg integrated flux densities range from 1 to 15 Jy km s$^{-1}$ and (uncorrected) line widths range from 20 to 120 km s$^{-1}$ (Sect.~2.2 and Table~2). The line profile shapes are varied (Fig.~1 and Table~2) -- two sources show asymmetric double-horn profiles, one source shows a symmetric single-peak profile and the remaining seven have asymmetric single-peak profiles. Double-horn profiles are typical of disk rotation, while single-peak profiles may arise from non-rotation, face-on disk rotation or preponderance of random motions (relative to ordered motions) in the gas. An asymmetry in the line profile suggests an asymmetry in the kinematics or possible companions (Fig.~1, Tables~2 and 3). We have excluded the latter hypothesis in the case of the Effelsberg observations (Sect.~2.2.2). In the two double-horn sources, the highest intensity and/or widest peak occurs on the (spectral) blue side (Fig.~1).

The optical morphology of the Effelsberg targets shows the varied nature of the optical structures (Fig.~1 and Table~3) and an association with the \ion{H}{i} line profile (Table~2). The two asymmetric double-horn profile sources show cometary or multiple star-formation knots in the SDSS images. The remaining asymmetric single-peak sources are cometary (five sources), multi-knot (one source) or two-knot (one source), while the only single-peak symmetric source has a symmetric optical morphology. We find that an asymmetry in the \ion{H}{i} kinematics, as suggested by the \ion{H}{i} line shape, is systematically associated with an asymmetry in the optical morphology.

\subsection{Mass -- Luminosity Relation}

Figure~4 contains the relation between the mass of the different XMP galaxy constituents and the absolute $g$-band magnitude. 

Though the scatter is large, a trend is observed between the (dynamical, \ion{H}{i} and stellar) mass and the luminosity, reflecting that the more massive galaxies are more luminous. The largest deviations occur at low luminosities, in the low-surface brightness, nearby galaxies UGC2684, DD053, LeoA, SextansB, UGCA292 and GR8. At a given stellar mass, these galaxies are significantly offset to lower luminosities compared with most XMPs.


\setcounter{figure}{3}


\begin{figure} 
\begin{center}
\includegraphics[width=6cm]{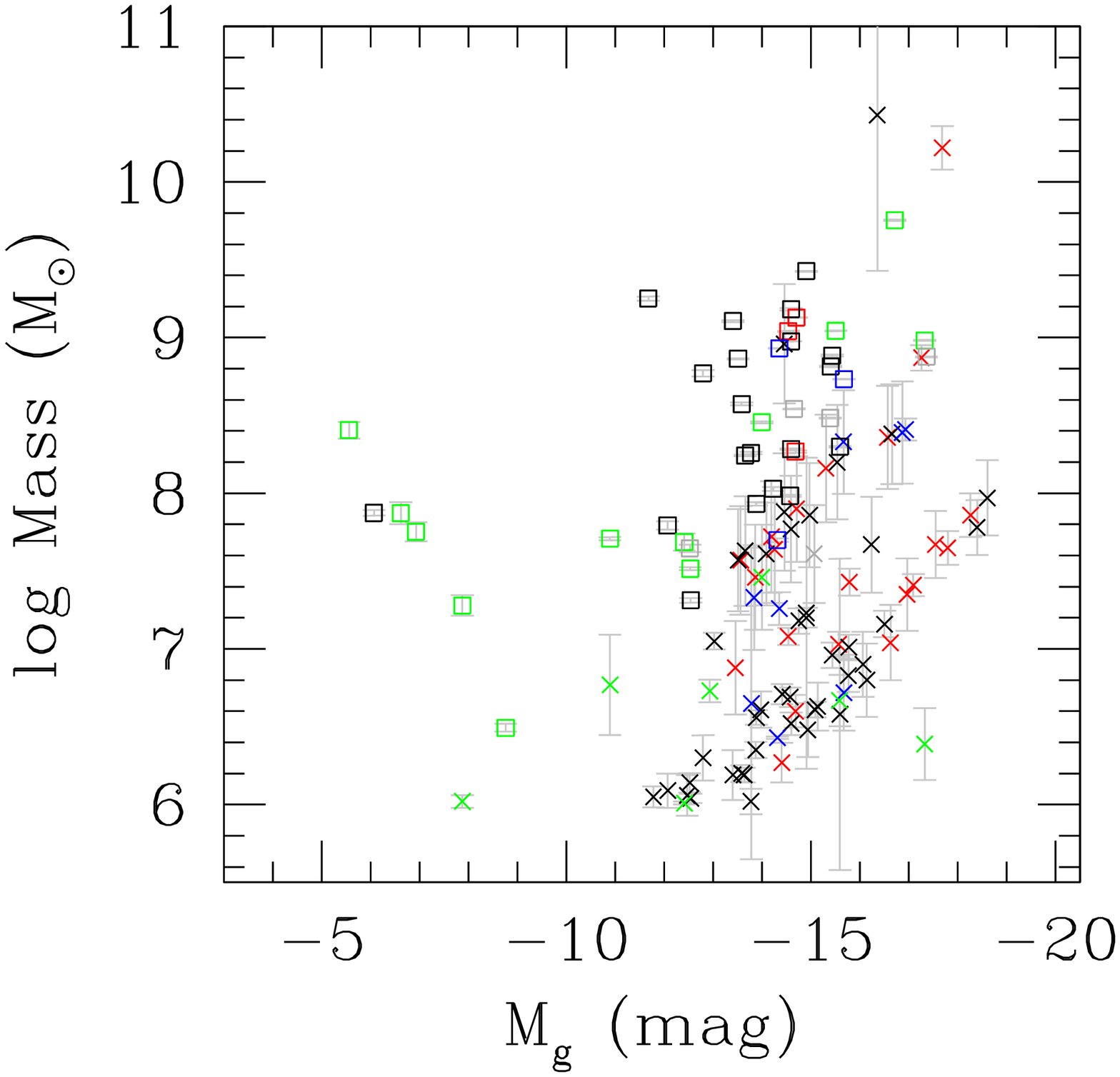}
\includegraphics[width=6cm]{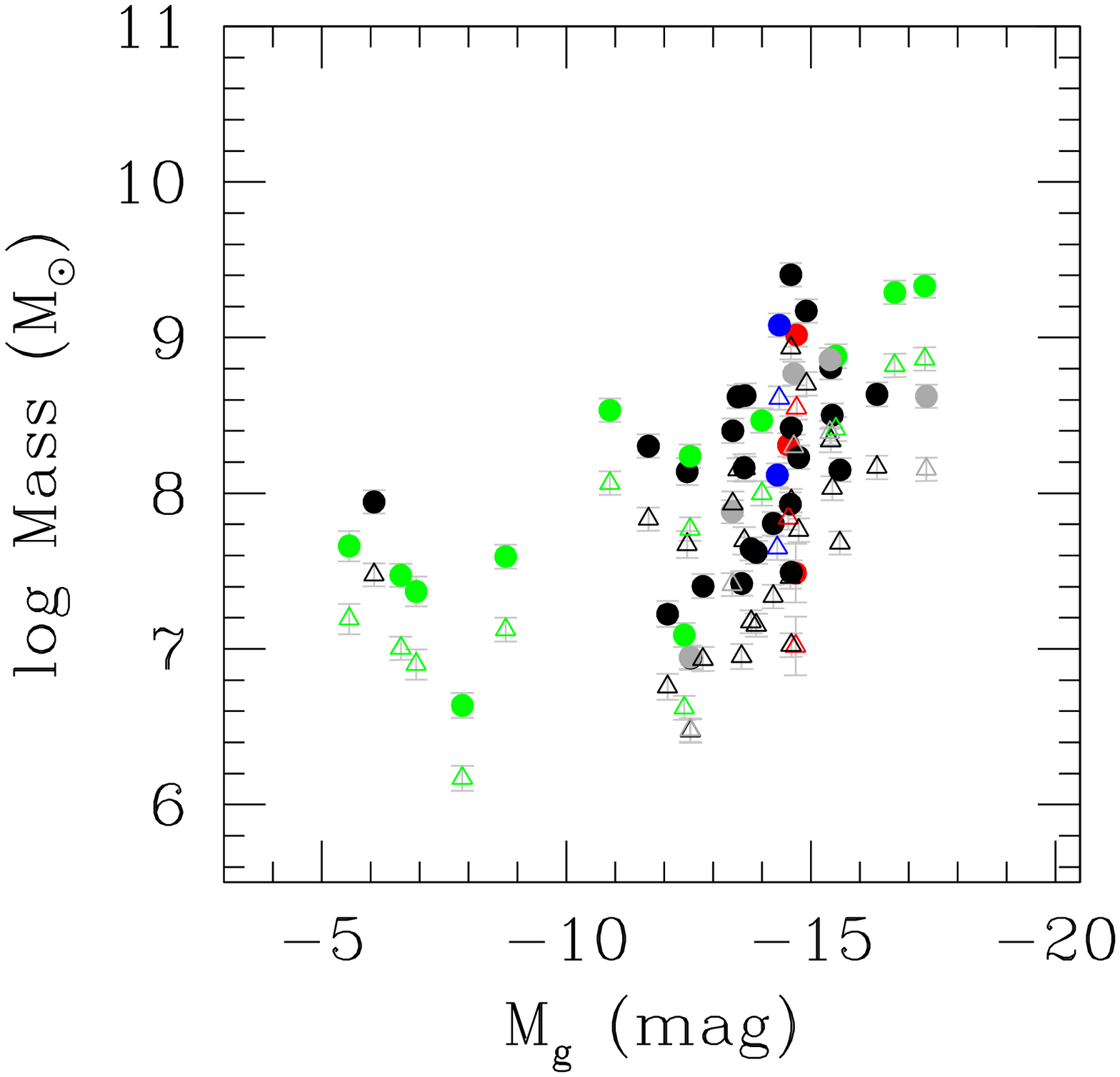}
\caption{The mass -- luminosity relation of the XMP galaxies. The color code parametrizes the morphology as follows: red = symmetric, black = cometary, blue = two-knot, green = multi-knot and grey = no morphological information. Top: the symbol code is $\square$ = M$_{\rm dyn}$ and $\times$ = M$_{\star}$. Bottom:  the symbol code is $\bullet$ = M$_{\ion{H}{i}}$ and $\triangle$ = M$_{\rm He}$.}
\end{center}
\end{figure}


\subsection{Mass-to-Light Ratio}

The XMP (\ion{H}{ii} gas-phase) metallicity and absolute $g$-band magnitude are plotted as a function of the stellar (M$_{\star}$/L$_g$) and the \ion{H}{i} gas (M$_{\ion{H}{i}}$/L$_g$) M/L ($g$-band) ratio in Fig.~5, where we have excluded the dynamical M/L ratio for clarity.

The M$_{\ion{H}{i}}$/L$_g$ and M$_{\star}$/L$_g$ ratios measured for the XMPs are typically $\lesssim$10 and $\sim$0.1, respectively. The lowest luminosity XMPs -- UCG2684, DD053, LeoA, SextansB, UGCA292 and GR8 -- have M$_{\ion{H}{i}}$/L$_g$ $>$ 10. The estimated M$_{\star}$/L$_g$ ratios likely reflect the M/L ratios of the galaxy templates used as input in the photometry fitting-based stellar mass determinations (Sect.~3.3). Papaderos et al. (1996b) have determined the average M/L ratio of the star-forming component in BCDs to be $\sim$0.5 (orange line; top; Fig.~5). This ratio (and its large scatter) has been confirmed in several subsequent studies (e.g., Cair\'os et al. 2001; Gil de Paz \& Madore 2005; Amor\'\i n et al. 2007, 2009). We recall that nearby XMPs are commonly found to be BCDs (Sect.~1). These values can be compared to the typical stellar M/L ratios observed in star-forming galaxies ($<$ 1; cyan line; top; Fig.~5), spirals (4 -- 6; magenta line; top;  Fig.~5), lenticulars ($\sim$ 10; cyan dashed line; top; Fig.~5), dwarf irregulars (10 -- 15; orange dashed line; top; Fig.~5) and elliptical ($\sim$ 20; magenta dashed line; top; Fig.~5) galaxies (Faber \& Gallagher 1979).



Regarding the absolute $g$-band magnitude versus the M$_{\ion{H}{i}}$/L$_g$ ratio (bottom; Fig.~5), we confirm previous published results found for a sample of dwarfs (cyan line; bottom; Fig.~5; Staveley-Smith, Davies \& Kinman 1992). We find an anti-correlation between the M$_{\ion{H}{i}}$/L$_g$ ratio and the luminosity, whereby fainter XMPs are more gas-rich than brighter XMPs. The result suggests that the less luminous sources have converted a smaller fraction of their \ion{H}{i} gas into stars. However, our sources deviate from the observed correlation found for dwarf galaxies (cyan line; bottom; Fig.~5; Staveley-Smith, Davies \& Kinman 1992) particularly at low luminosities, which correspond to the low-surface brightness, nearby galaxies UGC2684, DD053, LeoA, SextansB, UGCA292 and GR8. In this low luminosity regime, the XMP galaxies appear to be 10 to 100 times more gas-rich than typical dwarf galaxies. 
 



\begin{figure} 
\begin{center}
\includegraphics[width=6cm]{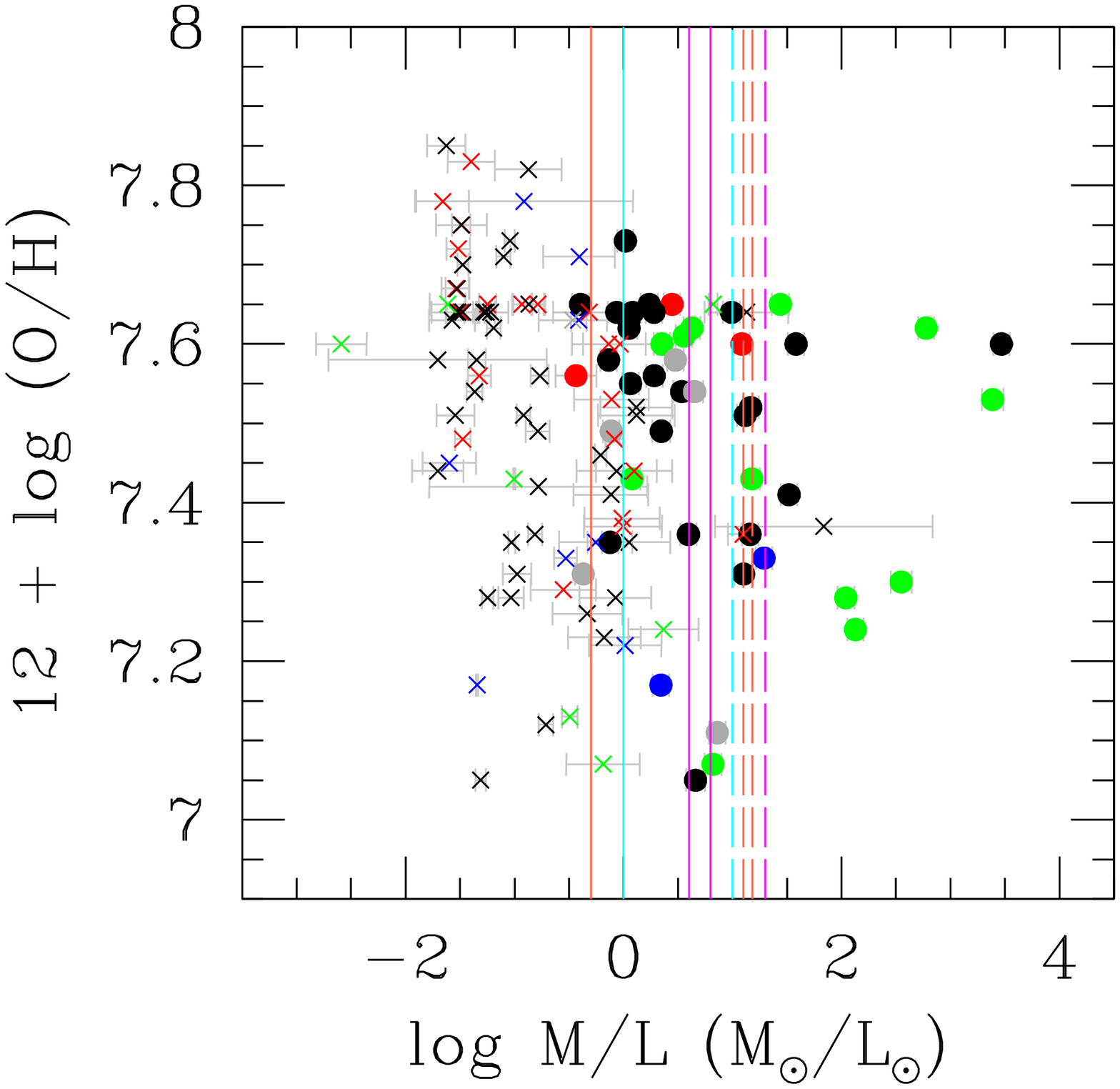}
\includegraphics[width=6cm]{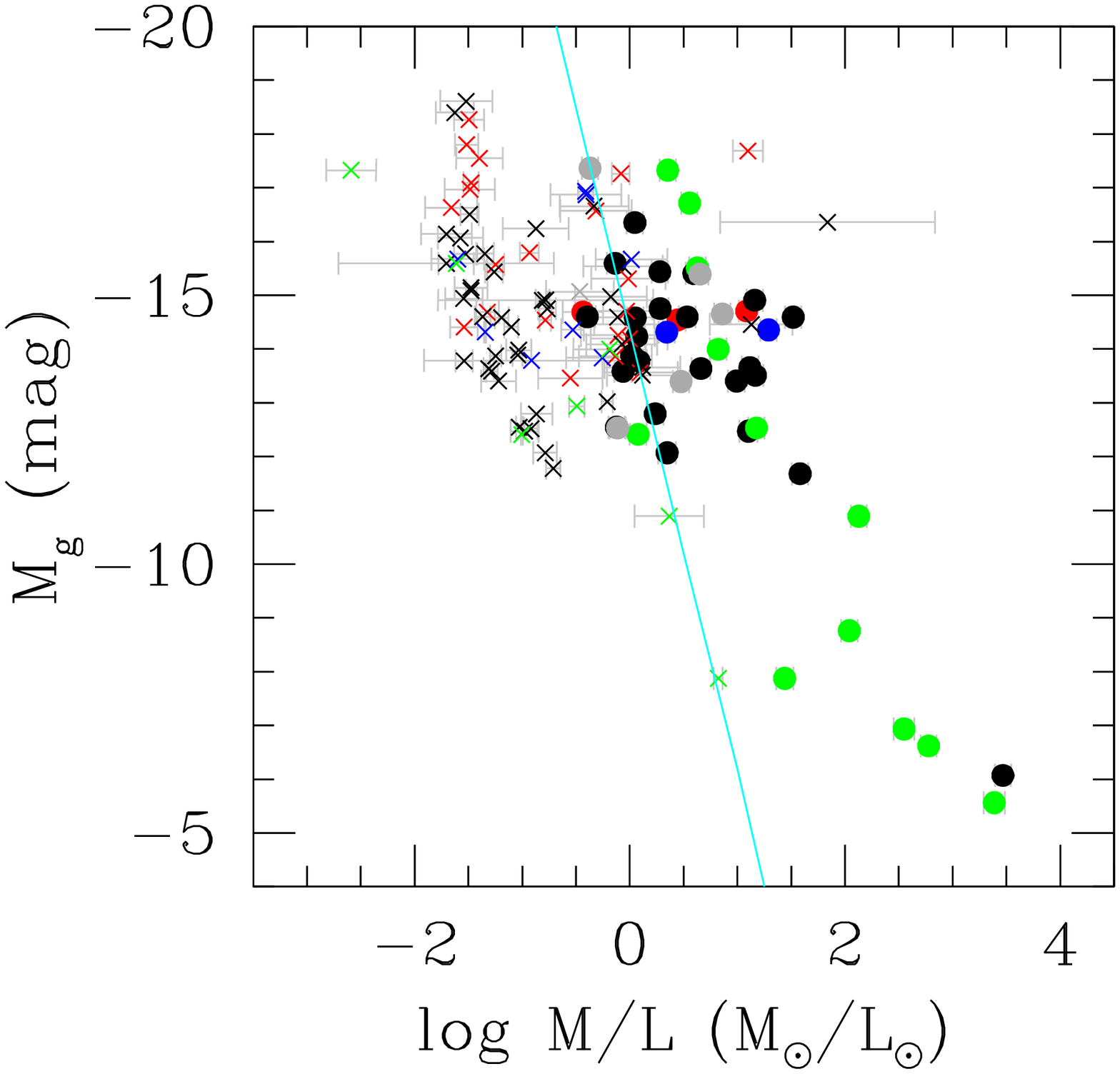}
\caption{The \ion{H}{ii} gas-phase metallicity and $g$-band magnitude of the XMP galaxies as a function of the stellar and \ion{H}{i} gas M/L ratio. The color code parametrizes the morphology as follows: red = symmetric, black = cometary, blue = two-knot, green = multi-knot and grey = no morphological information. The symbol code is $\bullet$ = M$_{\ion{H}{i}}$/L$_g$ and $\times$ = M$_{\star}$/L$_g$. Top: the plot includes a range of stellar mass-to-light ratios found in spirals (magenta line), lenticulares (cyan dashed line), dwarf irregulars (orange dashed line) and  ellipticals (magenta dashed line), an upper limit found in star-forming (cyan line) galaxies (Faber \& Gallagher 1979) and an average value (orange line) found for BCDs (Papaderos et al 1996a). Bottom: we plot in cyan the \ion{H}{i} gas M/L ratio-to-luminosity correlation found for dwarf galaxies (Staveley-Smith, Davies \& Kinman 1992). }
\end{center}
\end{figure}


\subsection{Mass Fraction and Dark Matter Content}

Figures~6 and 7 show the relation between the mass and the mass fraction of the different XMP galaxy constituents. In the figures, we mark the one-to-one (magenta line; top; Fig.~6; Fig.~7) and 0.1 $\times$ Mass (cyan line; top; Fig.~6) relation for reference. We recall that the dynamical mass estimation is a lower limit (Eq.~3; Sect.~4) and that the dark matter content is the fraction of dynamical mass which is not stellar, \ion{H}{i} or helium mass (Eq.~4; Sect.~4). In 12 of the sources with \ion{H}{i} and dynamical mass determinations, M$_{\ion{H}{i}}$ $>$ M$_{\rm dyn}$ by less than 0.5 dex (top; Fig.~6); these sources shall be excluded from the following discussion.

Typically, the stellar component constitutes less than 5\% of the total (baryonic and non-baryonic) mass in the XMP systems (middle; Fig.~6). The \ion{H}{i} gas mass fraction, relative to the dynamical mass, falls typically between 20 and 60\% (middle and bottom; Fig.~6), higher than the values found in late-type systems (4 -- 25\%; Young \& Scoville 1991). Moreover, the \ion{H}{i} gas mass is 10 to 20 times larger than the stellar mass (middle; Fig.~6; Fig.~7), denoting that XMPs are extremely gas-rich. As expected, this ratio depends on the stellar mass estimate. If we use the color-determined stellar masses (Sect.~3.3 and Fig.3), the \ion{H}{i}-to-stellar mass ratios fall, typically, to about 5 -- 10 (Fig.~7), which are still large values. As a comparison, \ion{H}{i} gas-to-stellar mass ratios in excess of 100 are observed in some of the lowest metallicity BCDs (Pustilnik et al. 2001), while ratios of less than 4 are commonly observed in spirals and irregulars (Swaters 1999; Dalcanton 2007).



\begin{figure} 
\begin{center}
\includegraphics[width=6cm]{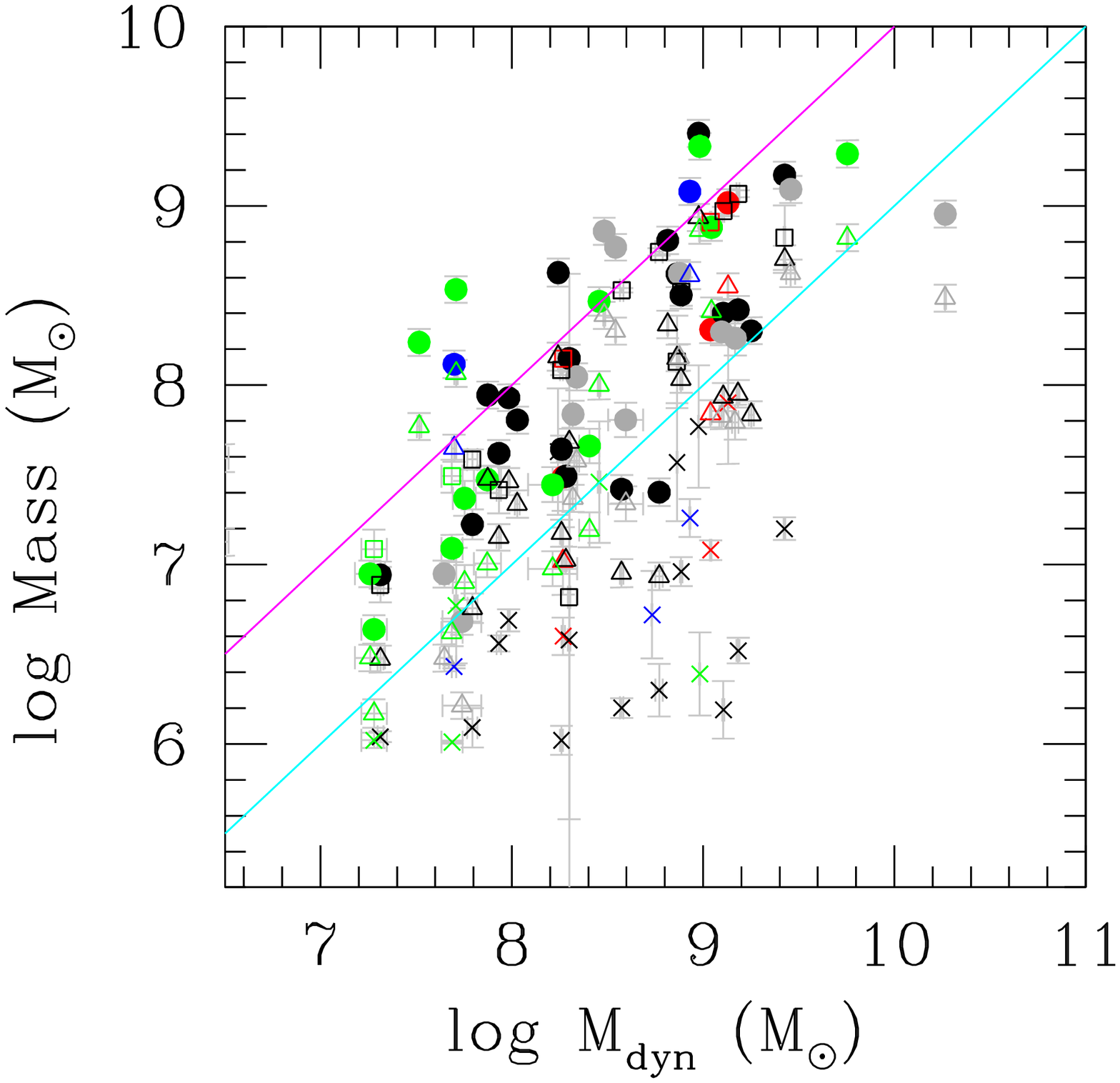} 
\includegraphics[width=6cm]{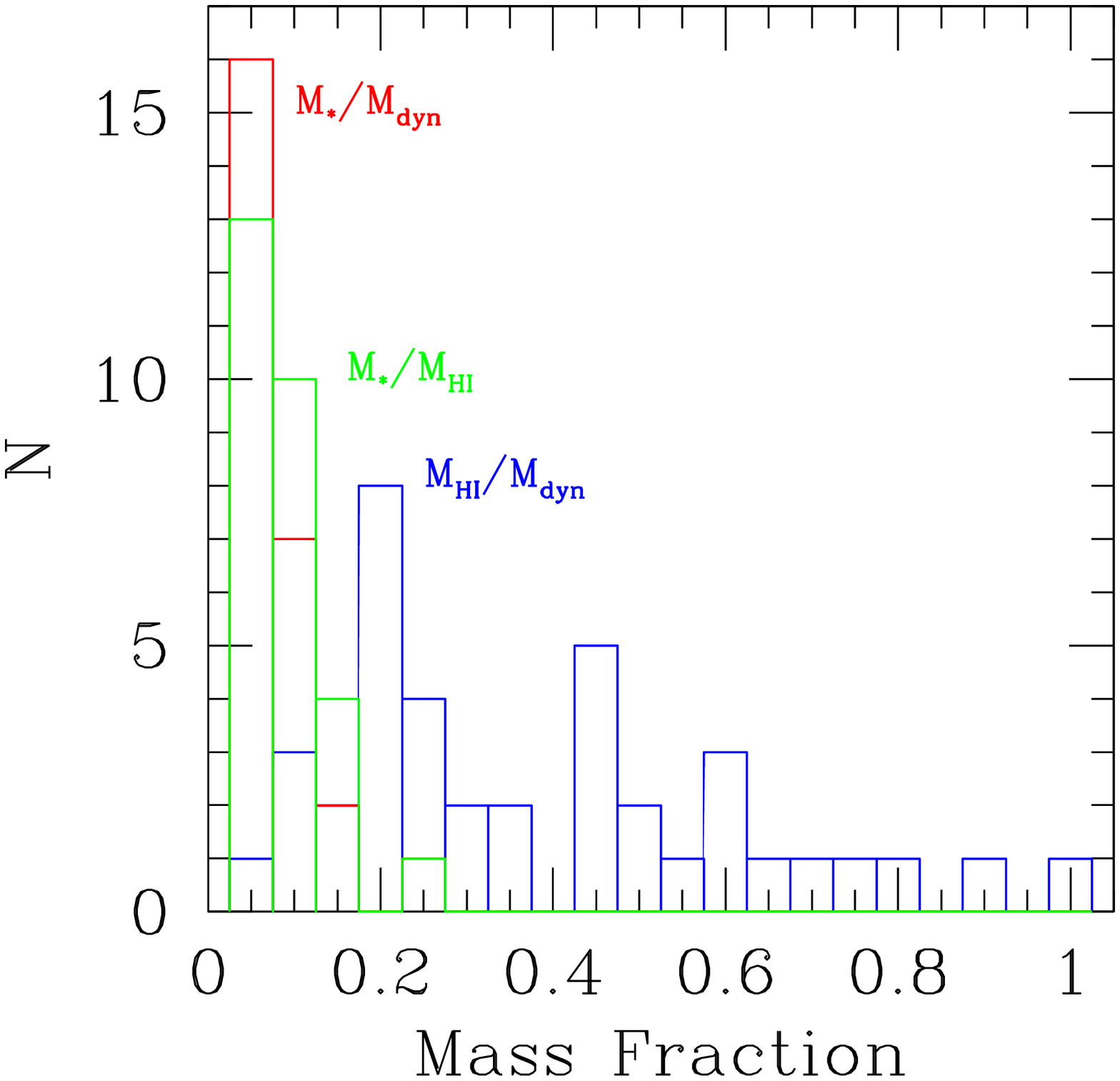}
\includegraphics[width=6cm]{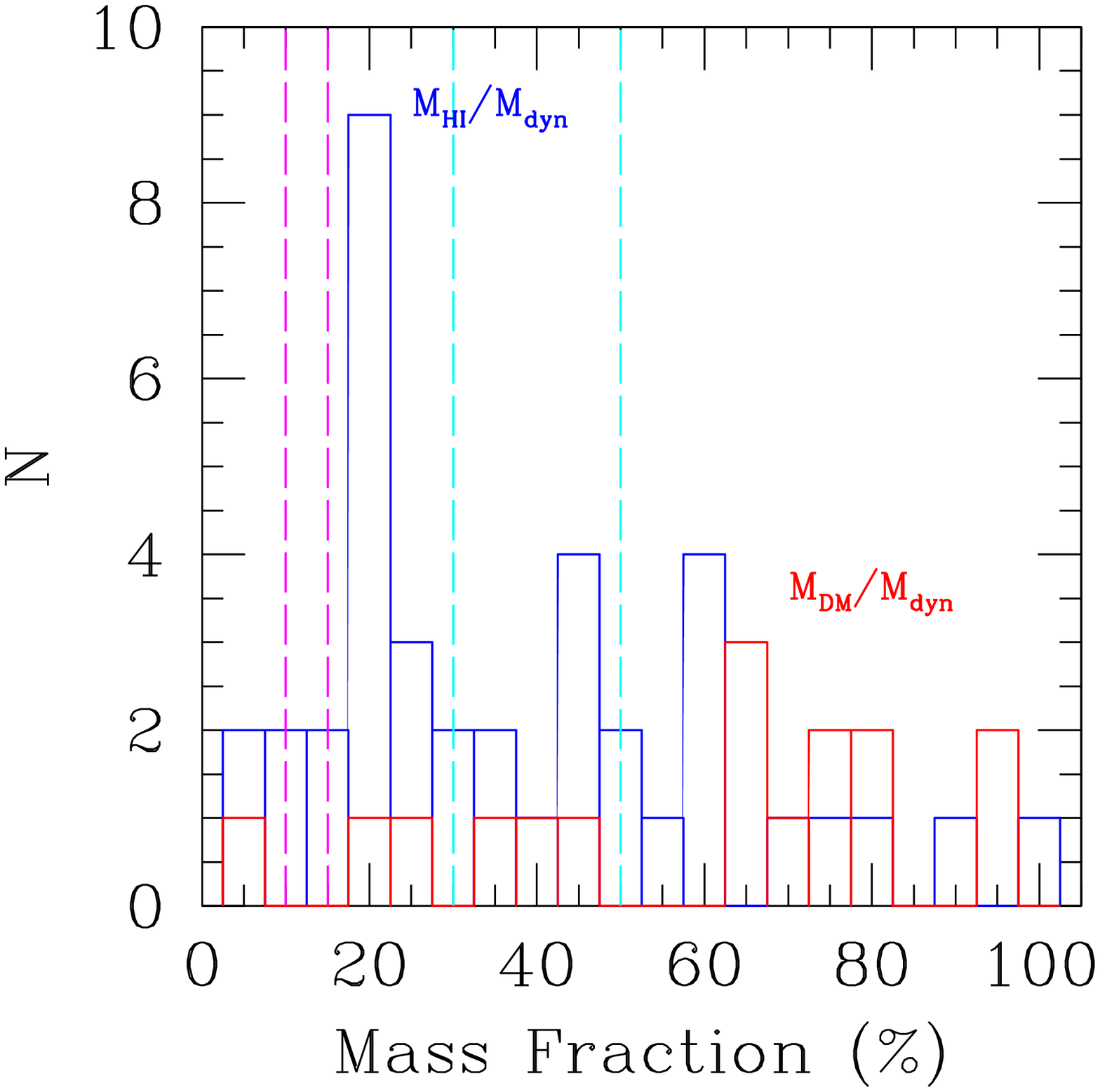}
\caption{The mass fraction of the various XMP galaxy constituents. Top: the color code parametrizes the morphology as follows: red = symmetric, black = cometary, blue = two-knot, green = multi-knot and grey = no morphological information. The symbol code is $\square$ = M$_{\rm DM}$, $\bullet$ = M$_{\ion{H}{i}}$, $\triangle$ = M$_{\rm He}$ and $\times$ = M$_{\star}$. The magenta and cyan line are, respectively, the one-to-one and 0.1 $\times$ Mass relation. Middle: the distribution of the mass fractions relative to the dynamical or \ion{H}{i} mass. The color code is blue = $\frac{\rm M_{\ion{H}{i}}}{\rm M_{dyn}}$, red = $\frac{\rm M_{\star}}{\rm M_{dyn}}$ and green = $\frac{\rm M_{\star}}{\rm M_{\ion{H}{i}}}$. Bottom: the distribution of the dark matter and \ion{H}{i} mass fractions relative to the dynamical mass, in percentage. The color code is blue = $\frac{\rm M_{\ion{H}{i}}}{\rm M_{dyn}}$ and red = $\frac{\rm M_{\rm DM}}{\rm M_{dyn}}$. The magenta and cyan dashed lines are, respectively, the range for dark matter mass fractions in spiral and elliptical galaxies.}
\end{center}
\end{figure}


\setcounter{figure}{6}


\begin{figure} 
\begin{center}
\includegraphics[width=6cm]{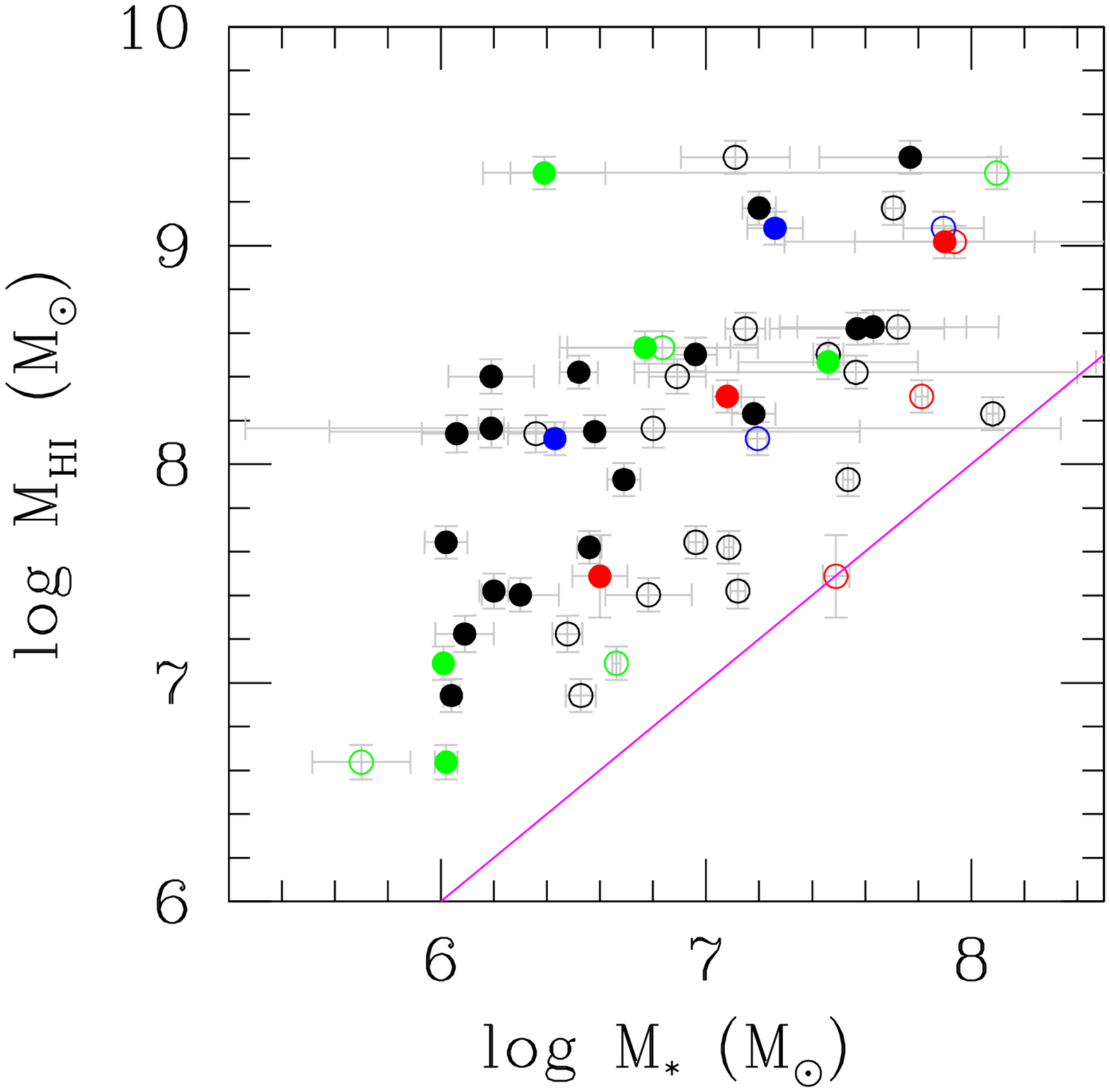} 
\caption{The \ion{H}{i}-to-stellar mass relation of the XMP galaxies. The color code parametrizes the morphology as follows: red = symmetric, black = cometary, blue = two-knot, green = multi-knot and grey = no morphological information. The symbol code is $\bullet$ = MPA-JHU-determined stellar mass values and $\circ$ = color-determined stellar mass values. The magenta line is the one-to-one relation.}
\end{center}
\end{figure}


Although our statistics are poor (16 sources), the values for the putative dark matter content in XMPs show a wide range, with a peak at 65\% of the dynamical mass, and skewed to higher values (bottom; Fig.~6). These latter estimates are typically larger than those observed in the inner parts of spirals (10 -- 15\%; magenta dashed line; bottom; Fig.~6) and elliptical (30 -- 50\%; cyan dashed line; bottom; Fig.~6) galaxies (Swaters 1999; Borriello, Salucci \& Danese 2003; Cappellari et al. 2006; Thomas et al. 2007; Williams et al. 2009). 






\subsection{Effective Yield}

Supernova-driven outflows or accretion of external gas may be partially responsible for the observed low metallicities in XMP galaxies (e.g. Kunth \& \"Ostlin 2000) and in low-metallicity starbursts (Amor\'\i n, P\'erez-Montero \& V\'\i lchez 2010; Amor\'\i n et al. 2012a). In a closed box model (no inflow or outflow), as the gas is converted into stars, the gas mass fraction decreases and the gas metallicity increases. Deviations from this model, signaling outflow or inflow, are investigated using the effective yield (e.g., Dalcanton 2007):

\begin{equation}
\rm y_{eff} = \frac{Z_{gas}}{ln(1/f_{gas})},
\end{equation}

\noindent where Z$_{\rm gas}$ is the fraction of the gas mass in metals and f$_{\rm gas}$ is the gas mass fraction, defined as the \ion{H}{i}, helium and metal mass divided by the total (\ion{H}{i}, helium, stellar and metal) mass.

For a galaxy that evolves as a closed box, the effective yield must coincide with the theoretical yield, y, derived from stellar evolution models (e.g., Edmunds 1990). Therefore, differences between y and y$_{\rm eff}$ provide a direct tool to diagnose gas inflows and outflows.


Figure~8 (top) shows the effective yields of the XMPs as a function of the gas fraction (as defined above), for MPA-JHU- and color-determined stellar masses (Sect.~3.3 and Fig.~3). The yields were computed assuming the metallicity from the optical emission lines (Table~3) to be the metallicity of the \ion{H}{i} gas, i.e., Z$_{\rm gas}\simeq12\times$O/H (e.g., Pilyugin, V\'\i lchez \& Contini 2004). Figure~8 (top) also shows, for reference, the effective yields for the set of spirals and dwarf irregular galaxies compiled by Pilyugin, V\'\i lchez \& Contini (2004). The gas fraction from Pilyugin, V\'\i lchez \& Contini (2004) includes molecular hydrogen, but its contribution is negligible.

The theoretical oxygen yield is $\log$ y $\simeq$ -2.4 (magenta line; top; Fig.~8; Dalcanton et al. 2007; Meynet \& Maeder 2002), which coincides with the average effective yield of the gas-poor galaxies in the reference set, as is expected to happen in chemical evolution models, when the gas is exhausted (K\"oppen \& Hensler 2005; Dalcanton 2007). However, many XMPs have yields which largely exceed the theoretical limit. There are two ways in which this excess could be explained. Either the theoretical yield of our XMPs is significantly larger than that of regular spirals, or the metallicity of the \ion{H}{i} gas is much lower than the metallicity of the ionized \ion{H}{ii} gas.

Regarding the former explanation, a dependence of the IMF on the metallicity may play a role (e.g., Bromm \& Larson 2004). In particular, a top-heavy IMF will increase the oxygen yields by significant amounts (e.g., Meynet \& Maeder 2002). The second possibility, which we explain below, is the extremely low metallicity of the \ion{H}{i} gas, that must be much lower than the already low metallicity measured in the \ion{H}{ii} regions of the XMPs. 


When the gas mass fraction is large, as is the case in XMPs, one can easily show the effective yield to be the ratio between the mass of metals in the gas, ${\rm Z_{gas} \, M_{gas}}$, and the stellar mass:

\begin{equation}
\rm y_{eff} \simeq \frac{Z_{gas}\,M_{gas}}{M_{\star}},
\end{equation}

\noindent where M$_{\rm gas}$ stands for the mass of the gas. This expression shows how difficult it is to increase the effective yield beyond the theoretical yield because, taken at face value, ${\rm y_{eff} > y}$ implies creating more metals than those allowed by the stellar evolution that produced the mass in stars. However, if the gas metallicity has been overestimated by a significant amount, this would artificially increase the effective yields (Eq.~7). In other words, the large effective yields that we infer may be easily explained if the \ion{H}{i} gas is mostly pristine (Thuan, Lecavelier des Etangs \& Izotov 2005), with a metallicity much lower than the already low metallicity measured in the gas forming stars at present (\ion{H}{ii} gas-phase metallicity; Table~3). 

As a sanity check, we have also computed ${\rm y_{eff}}$ using the color-determined stellar masses (Sect.~3.3 and Fig.~3; top; Fig.~8). Equation~6 shows how an increase in stellar mass decreases ${\rm y_{eff}}$. However, it does not suffice to explain the observed ${\rm y_{eff} > y}$, in most cases.


The \ion{H}{i} gas metallicity reproducing both the theory and observations can be estimated assuming that the mass of metals in the \ion{H}{i} gas coincides with the mass of metals produced by the stars, i.e., y = Z$_{\rm gas}$ M$_{\rm gas}$/M$_{\star}$. The ratio between the effective yield and the true yield is then given by the ratio between the metallicity measured in the \ion{H}{ii} regions, Z$_{\ion{H}{ii}}$, used to compute y$_{\rm eff}$, and the true \ion{H}{i} gas metallicity, Z$_{\ion{H}{i}}$ (bottom; Fig.~8). Explicitly:

\begin{equation}
{\rm Z_{\ion{H}{ii}}/Z_{\ion{H}{i}} \simeq y_{eff}/y}.
\end{equation}

\noindent The results show that the ratio ${\rm Z_{\ion{H}{ii}}/Z_{\ion{H}{i}}}$ falls, typically, between 1 and 10 (bottom; Fig.~8). This is in agreement with the findings of Lebouteiller et al. (2013) on the metallicity of the \ion{H}{i} gas in the XMP prototype, IZw18. The \ion{H}{i} region abundances were found to be lower by a factor of $\sim$ 2 compared to the \ion{H}{ii} regions, and it may contain pockets of pristine gas, with an essentially null metallicity.




\setcounter{figure}{7}

\begin{figure} 
\begin{center}
\includegraphics[width=6cm]{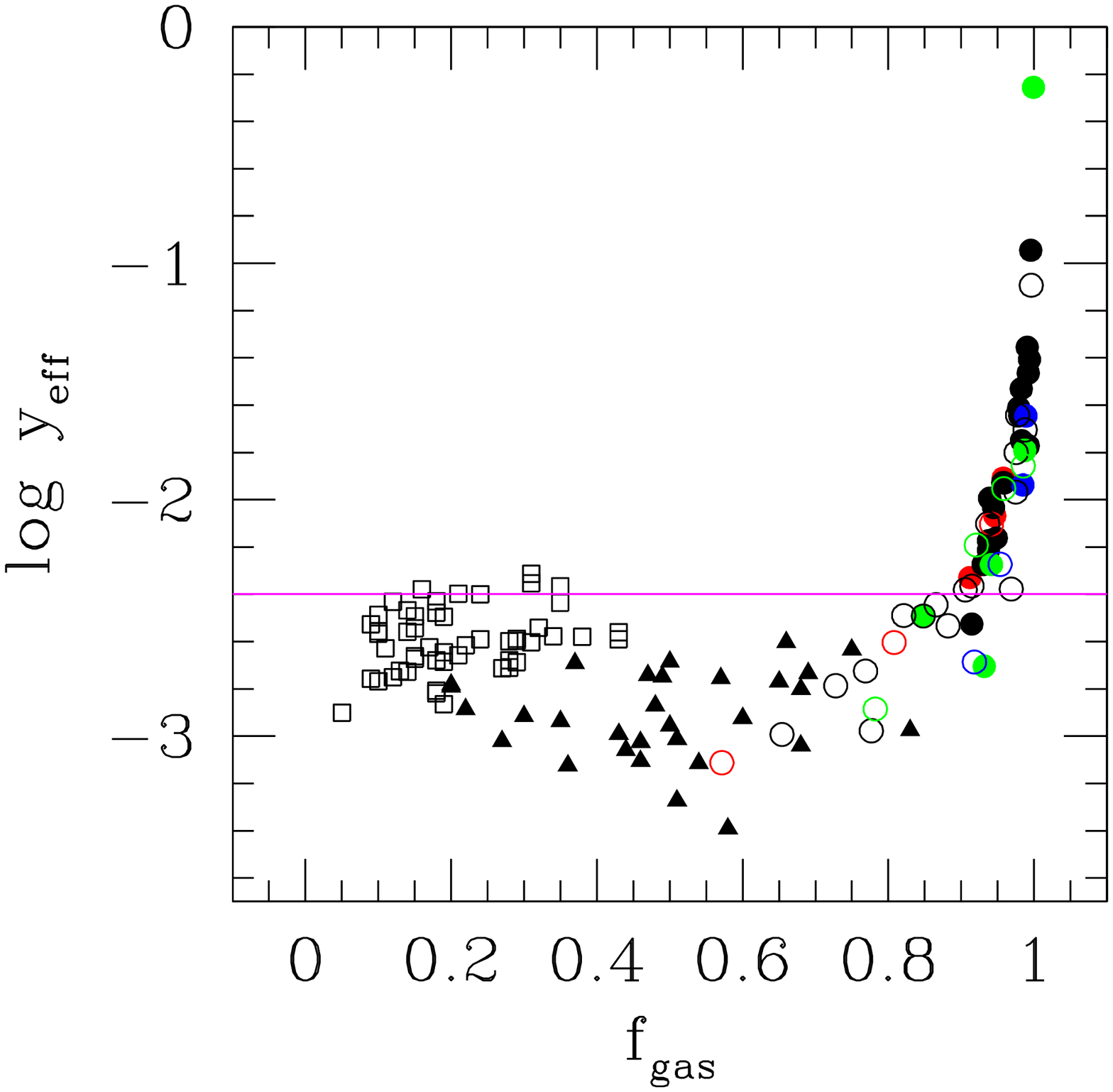} 
\includegraphics[width=6cm]{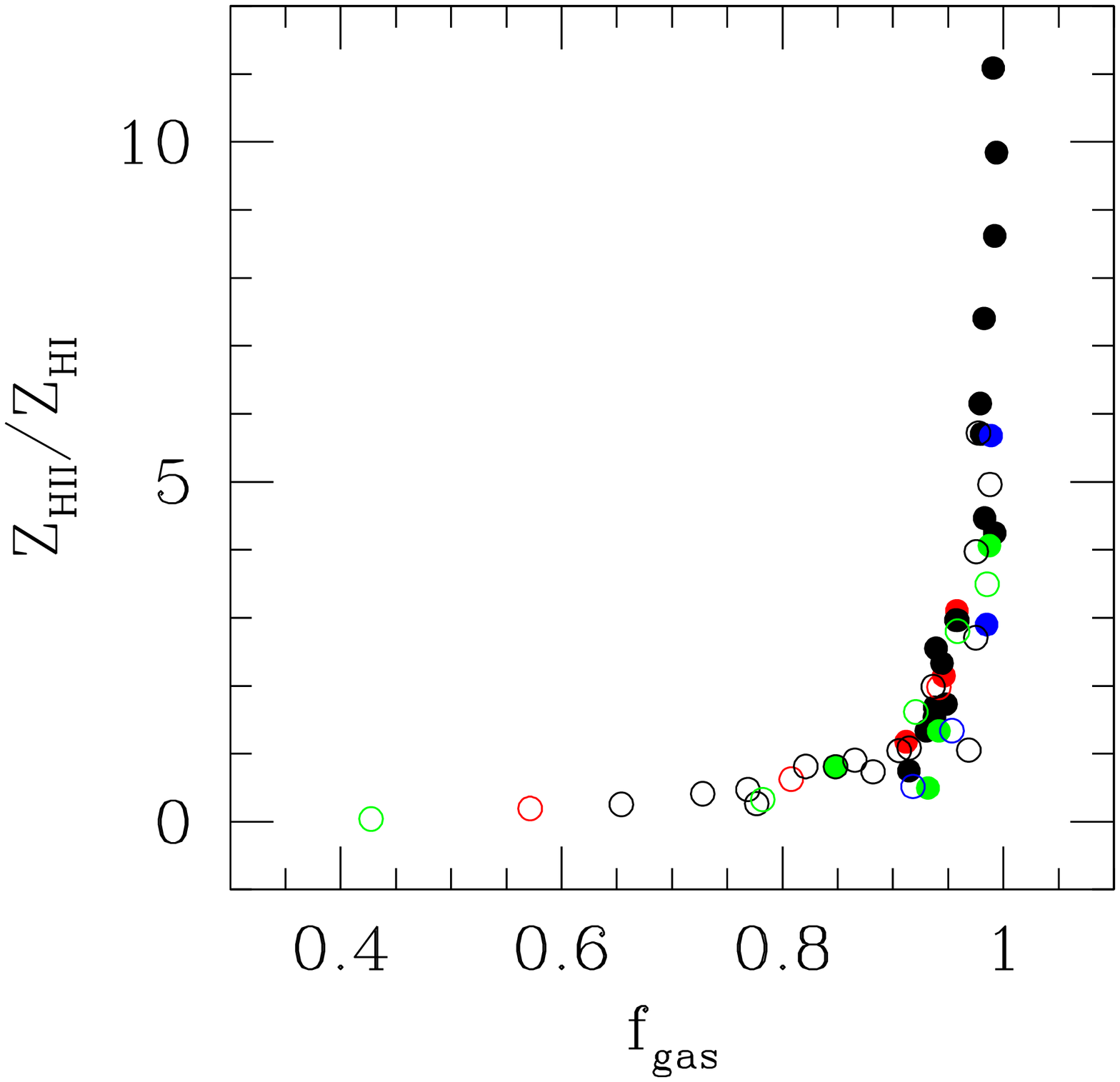} 
\caption{The effective yield and the \ion{H}{ii}-to-\ion{H}{i} metallicity ratio of the XMP galaxies as a function of the gas mass fraction. The color code parametrizes the morphology as follows: red = symmetric, black = cometary, blue = two-knot, green = multi-knot and grey = no morphological information. Top: the symbol code is $\bullet$ = XMPs with MPA-JHU-determined stellar masses, $\circ$ = XMPs with color-determined stellar masses, $\square$ = spiral and $\blacktriangle$ = irregular galaxies. The spiral and irregular galaxy data were obtained from Pilyugin, V\'\i lchez \& Contini (2004). The magenta line represents the theoretical yield limit for a closed-box model. Bottom: the symbol code is $\bullet$ = XMPs with MPA-JHU-determined stellar masses and $\circ$ = XMPs with color-determined stellar masses.}
\end{center}
\end{figure}


\subsection{Mass -- and Luminosity -- Metallicity Relation}

The mass -- metallicity relation, or equivalently the luminosity -- metallicity relation (for similar M/L ratios; Lequeux et al. 1979), reflects the fundamental role that the mass plays in galaxy chemical evolution. We recall that the metallicity pertains to the \ion{H}{ii} gas-phase metallicity (Table~3).

Besides the mass -- metallicity relation for the XMPs, Fig.~9 (top) includes the empirical relations for extreme starburst galaxies called {\em green peas} (magenta line; top; Fig.~9; Amor\'\i n, P\'erez-Montero \& V\'\i chez 2010) and for extremely metal-poor BCDs (cyan line; top; Fig.~9; Papaderos et al. 2008). The latter is determined from the luminosity of the underlying host, after the subtraction of the starburst contamination using surface photometric techniques, and therefore, refers to the stellar mass.


The XMP galaxies show a large scatter in the mass -- luminosity relation, with a significant fraction of the stellar mass points falling to the left of the correlations (top; Fig.~9). This is similar to what is found in more distant low-metallicity galaxies (e.g. $z$ $>$ 0.1; Kakazu, Cowie \& Hu 2007; Amor\'\i n, P\'erez-Monteiro \& V\'\i lchez 2010). Furthermore, for a given dynamical mass, the measured metallicity is low, suggesting the presence of pristine material.

In the bottom figure (Fig.~9) we include the luminosity -- metallicity relation for the XMPs, the correlation for a sample of dwarf irregular galaxies (magenta line; bottom; Fig.~9; Skillman, Kennicutt \& Hodge 1989) and for a sample of SDSS star-forming galaxies (cyan line; bottom; Fig.~9; Guseva et al. 2009), linearly extrapolated to fainter magnitudes and lower metallicities. 

We observe that most XMPs follow a similar slope to that observed for dwarf irregular and SDSS star-forming galaxies (magenta and cyan line; bottom; Fig.~9; Skillman, Kennicutt \& Hodge 1989; Guseva et al. 2009), but are too metal-poor for their luminosity. The largest deviations from the correlation occur at low luminosities, corresponding to the low-surface brightness, nearby galaxies UGC2684, DD053, LeoA, SextansB, UGCA292 and GR8. In this low luminosity regime, the inverse situation occurs; the XMPs are underluminous for their metallicity.



\begin{figure} 
\begin{center}
\includegraphics[width=6cm]{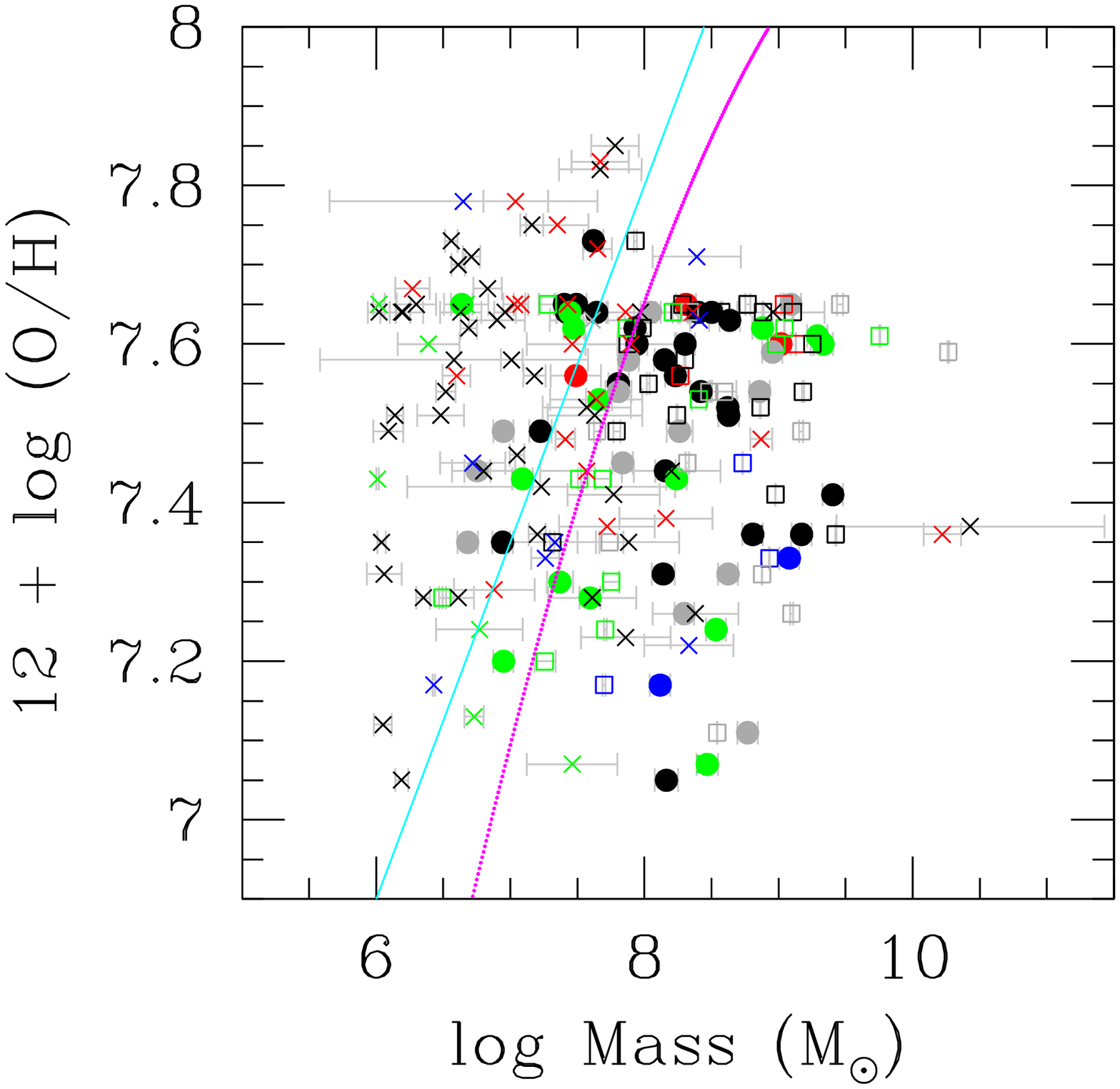}
\includegraphics[width=6cm]{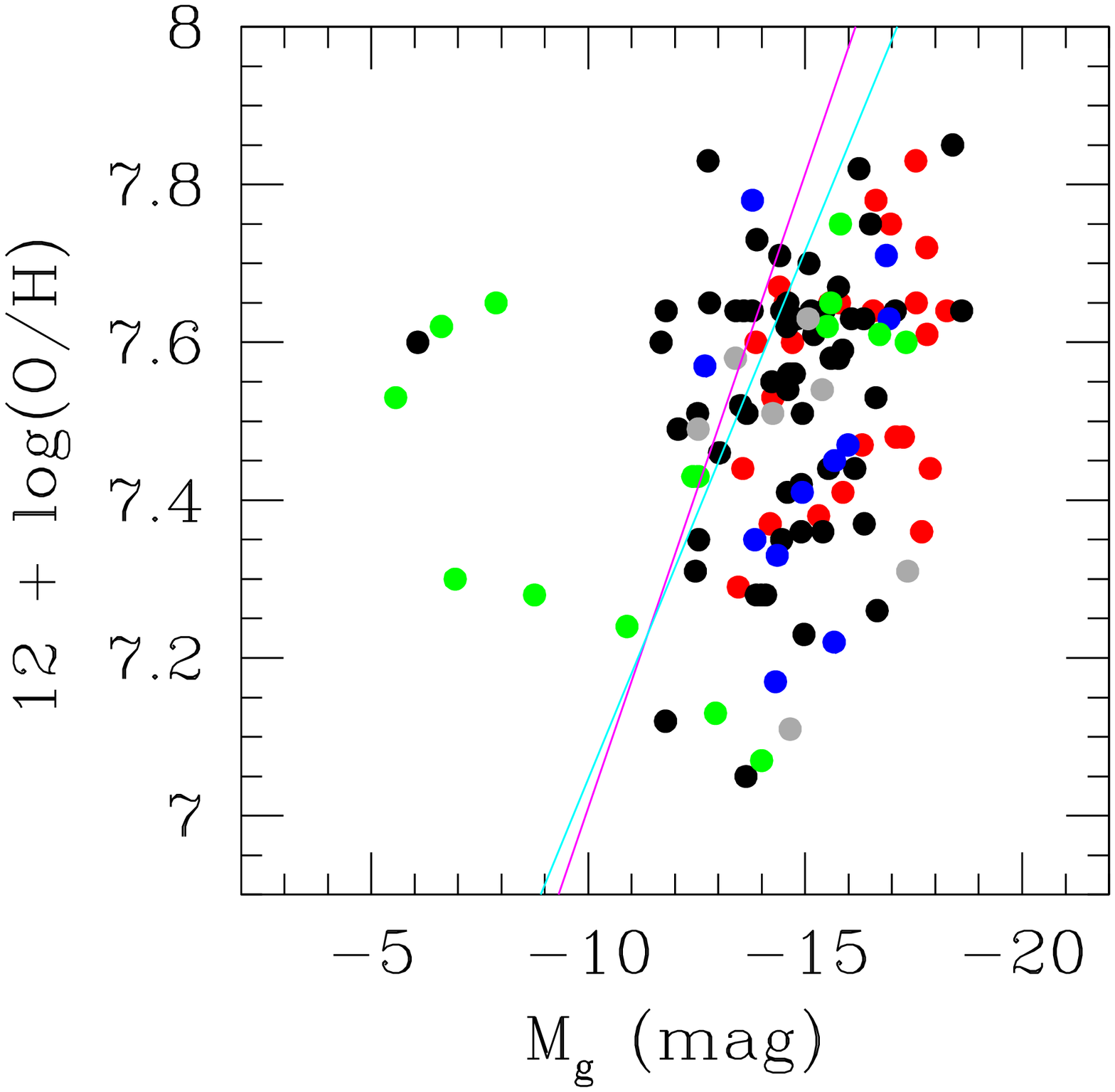}
\caption{The mass -- and luminosity -- \ion{H}{ii} gas-phase metallicity relation of the XMP galaxies. The color code parametrizes the morphology as follows: red = symmetric, black = cometary, blue = two-knot, green = multi-knot and grey = no morphological information. Top: the symbol code is $\square$ = M$_{\rm dyn}$, $\bullet$ = M$_{\ion{H}{i}}$ and $\times$ = M$_{\star}$. The figure includes the empirical correlation established for BCDs by Papaderos (2008; cyan line) and for {\em green pea} starburst galaxies (magenta line; Amor\'\i n, P\'erez-Montero \& V\'\i lchez 2010). Bottom: the figure shows the luminosity -- metallicity relation found for a sample of dwarf irregular galaxies (magenta line; Skillman, Kennicutt \& Hodge 1989) and for a sample of SDSS star-forming galaxies (cyan line; Guseva et al. 2009).}
\end{center}
\end{figure}



\subsection{Mass -- and Luminosity -- Tully -- Fisher Relation}

The Tully -- Fisher (TF) relation (Tully \& Fisher 1977) is an empirical correlation found for spiral galaxies, reflecting the fact that the bigger or brighter the galaxy, the faster it rotates. Indeed, all galaxies are found to follow the same baryonic mass -- TF relation, including low-surface brightness (LSB) galaxies, that contain a large amount of \ion{H}{i} gas (e.g., van de Kruit \& Freeman 2011).

As pointed out in Kannappan et al. (2002), the dispersion in the TF relation, particularly regarding the dwarf galaxy population, is mainly related to recent perturbations in their evolution, which is consistent with Kassin et al. (2007) and de Rossi, Tissera \& Pedrosa (2012), who argue that the dispersion and residuals correlate with the morphology and kinematical indicators.

Figure~10 (top) contains the luminosity -- TF relation for the XMPs, together with the empirical relation found for early- and late-type dwarf and giant galaxies (cyan and magenta line, respectively; top; Fig.~10; de Rijcke et al. 2007) and their 1$\sigma$ errors (cyan and magenta dashed line, respectively; top; Fig.~10; de Rijcke et al. 2007), extrapolated to fainter and slower rotating galaxies. We note that the 1$\sigma$ lower limit for the late-type dwarf and giant luminosity -- TF relation (magenta dashed line; top; Fig.~10; de Rijcke et al. 2007) falls almost on top of the early-type dwarf and giant luminosity -- TF relation (cyan line; top; Fig.~10; de Rijcke et al. 2007). We have converted the circular velocity used in de Rijcke et al. (2007) to $w_{50}$, assuming a Gaussian shape for the line profile, such that v$_{circ}$ = 1.52 $\times$ $w_{50}$/2 (e.g., Gurovich et al. 2010).

The majority of the XMP galaxies follow the luminosity -- TF relation (top; Fig.~10) within 1$\sigma$, with a wide spread in luminosity. However, differences are particularly large in the lowest luminosity targets, which correspond to the low-surface brightness, nearby galaxies UGC2684, DD053, LeoA, SextansB, UGCA292 and GR8. Indeed, for a given stellar luminosity, the line widths are too broad compared to late- and early-type giant and dwarf galaxies. In these sources, the main support against gravity comes from random motions rather than from rotation (e.g., Carignan, Beaulieu \& Freeman 1990).



Figure~10 (bottom) contains the mass -- TF relation for the XMPs, together with the mass -- TF relation for the dwarf and giant early- and late-type galaxy samples by de Rijcke et al. (2007), in terms of stellar mass (magenta line; bottom; Fig.~10) and gas plus stellar (baryonic) mass (cyan line; bottom; Fig.~10), and their 1$\sigma$ errors (magenta and cyan dashed line, respectively; bottom; Fig.~10), extrapolated to lower masses and line widths. We define the baryonic mass, M$_{\rm baryon}$, as the total stellar, \ion{H}{i} and helium mass. 


The stellar mass points fall generally below the (stellar) correlation, with a large scatter. The majority of these points are not within 1$\sigma$ of the (stellar) correlation; their stellar masses are unusually low for their potential wells. This is comparable to what has been observed for BCDs (Amor\'\i n et al. 2009). 
In simulations performed by de Rossi, Tissera \& Pedrosa (2010), it is shown that the tendency for slowly rotating galaxies (low-mass systems) to lie below the stellar mass -- TF relation can be explained by the action of supernova feedback and stellar winds.

The XMPs do follow the baryonic mass -- TF relation; a correlation is present between the line width and the \ion{H}{i} gas and the baryonic mass, as expected for extremely gas-rich galaxies (e.g., McGaugh et al. 2000; Gurovich et al. 2010). These results suggest that the \ion{H}{i} gas is largely virialized and may be partially rotationally supported.





\begin{figure} 
\begin{center}
\includegraphics[width=6cm]{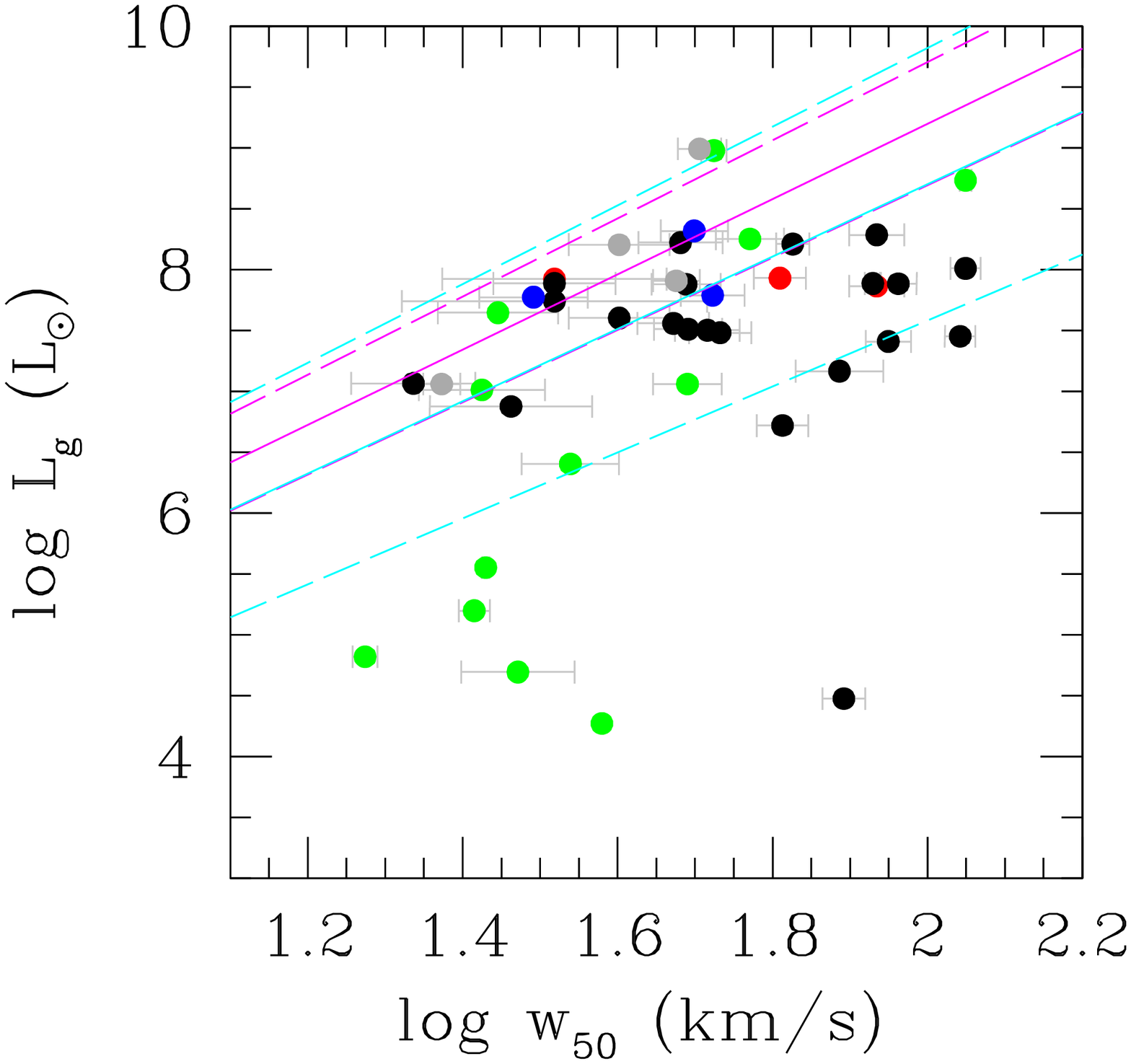}
\includegraphics[width=6cm]{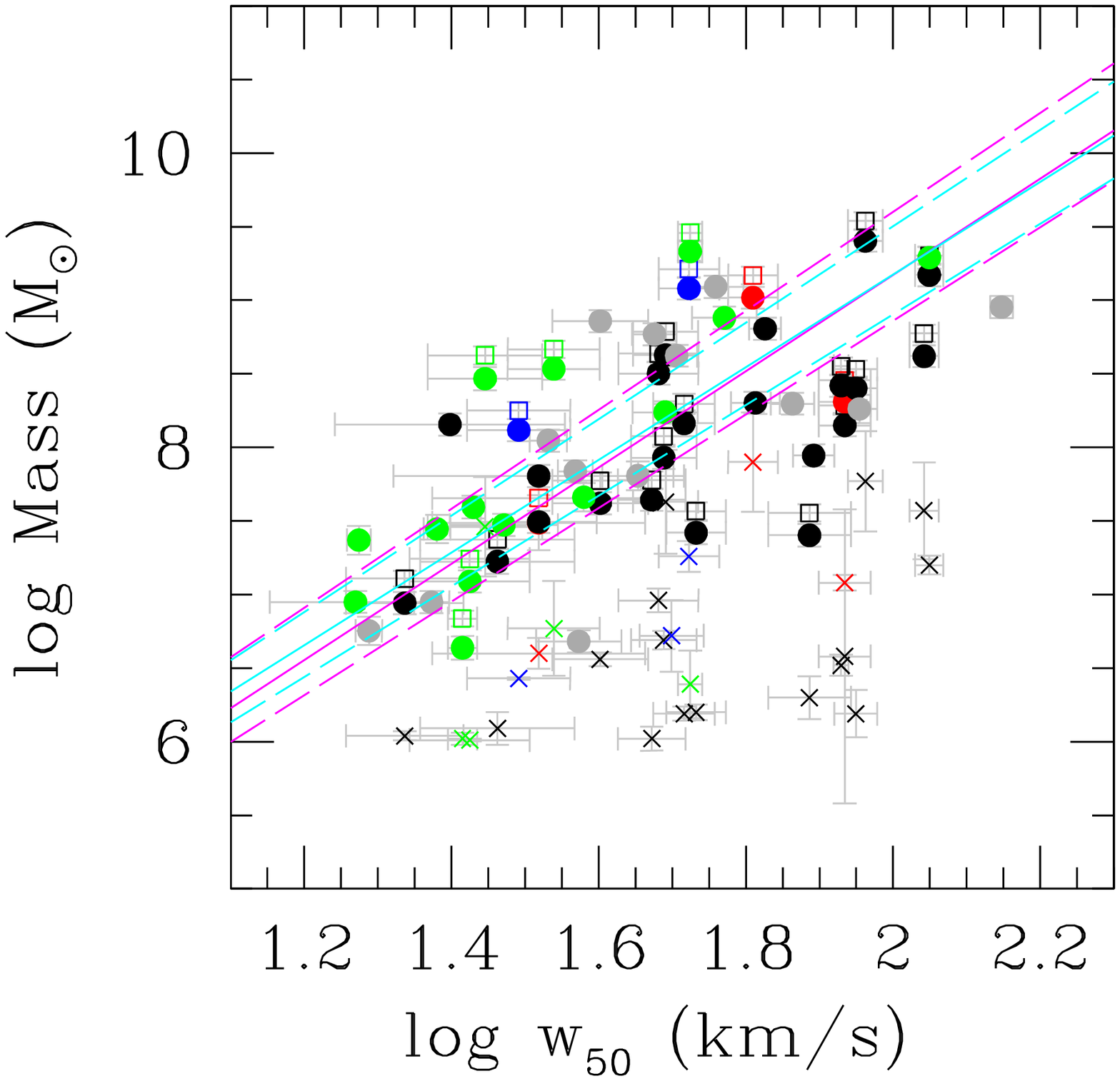}
\caption{The luminosity and mass -- TF relation of the XMP galaxies. The color code parametrizes the morphology as follows: red = symmetric, black = cometary, blue = two-knot, green = multi-knot and grey = no morphological information. Top: the figure contains the luminosity -- TF relation found for early-type (cyan line; de Rijcke et al. 2007) and late-type (magenta line; de Rijcke et al. 2007) dwarf and giant galaxies and 1$\sigma$ errors (cyan and magenta dashed line, respectively; de Rijcke et al. 2007). We note that the 1$\sigma$ lower limit for the late-type dwarf and giant luminosity -- TF relation (magenta dashed line; de Rijcke et al. 2007) falls almost on top of the early-type dwarf and giant luminosity -- TF relation (cyan line; de Rijcke et al. 2007). Bottom: the symbol code is $\square$ = M$_{\rm baryon}$, $\bullet$ = M$_{\ion{H}{i}}$ and $\times$ = M$_{\star}$. The magenta and cyan lines show the mass -- TF correlation for a sample of dwarf and giant early- and late-type galaxies (de Rijcke et al. 2007), in terms of stellar mass and \ion{H}{i} gas plus stellar mass, respectively, and their 1$\sigma$ errors (magenta and cyan dashed line, respectively; de Rijcke et al. 2007).}
\end{center}
\end{figure}


\subsection{Velocity Offset}

In Fig.~11 we investigate the presence of velocity differences (offsets) between the \ion{H}{i} gas, Balmer emission lines and forbidden emission lines, measured relative to the best-fit radial velocity (Sect.~3.3 and Tables~4 and 5). We plot only non-zero velocity offsets. Balmer line, forbidden line and \ion{H}{i} velocity offsets are only considered reliable if the offset is larger than 3 times the respective error (Table~4). The error in the \ion{H}{i} gas offset is dominated by the error in the systemic velocity estimation, which is typically 5 km s$^{-1}$. However, \ion{H}{i} offsets lower than the intrinsic \ion{H}{i} gas velocity dispersion in a dwarf galaxy (typically $\sim$10 km s$^{-1}$) were considered unreliable. Best-fit (spectroscopic) redshift errors for these sources are quite low (e.g. Maddox et al. 2013), resulting in best-fit radial velocity errors of about 1 km s$^{-1}$. Furthermore, because of the typical velocity dispersion of the warm gas (20 -- 30 km s$^{-1}$), we have disregarded Balmer and forbidden line offsets below 10 km s$^{-1}$. 

The majority (80 -- 90\%) of the XMPs with measured Balmer and forbidden emission-line velocities show no shift (Table~4). However, in 60\% of the sources with \ion{H}{i} gas and optical data, we find a small offset (10 -- 40 km s$^{-1}$) between the systemic velocity of the \ion{H}{i} gas and the best-fit radial velocity (top; Fig.~11; Table~5; e.g. Maddox et al. 2013). We have verified the SDSS DR9 images to find that the majority of these sources are cometary or knotted (Table~3) and that the SDSS fiber position for the spectra is generally located at the head or at the brightest star-forming knot, not at the center of the XMP system. We recall that the spatial offset we defined (Sect.~3.2 and Table~3) is a measurement of this displacement. Therefore, this could, in principle, explain the small velocity offsets observed between the \ion{H}{i} gas and the nebular/stellar emission. However, we find no evidence for a correlation between the spatial offset and the \ion{H}{i} gas velocity offset (bottom; Fig.~11). The result suggests that the \ion{H}{i} gas and the nebular/stellar component are not tightly coupled in these XMPs.




In the three XMP sources where the \ion{H}{i} gas, Balmer line or forbidden line offsets are larger ($>$ 50 km s$^{-1}$), we searched the SDSS DR9 and radio continuum (NVSS and FIRST; Condon et al. 1998; Becker, White \& Helfand 1995) fields (6\arcmin~radius) for potential galaxy contaminants. We discuss these below.

$\bullet$ {\bf J0301-0052} -- This cometary XMP (Table~3) has been observed with Effelsberg (Table~2) and shows a large \ion{H}{i} gas offset ($\sim$100 km s$^{-1}$; Table~5). We had previously discarded \ion{H}{i} contamination (Sect.~2.2.2) by checking the SDSS spectroscopic or photometric redshifts of the neighbours and found no sources within 200 km s$^{-1}$ of the target. In addition, the nearest radio continuum source (NVSS) present in the field is about 1\arcmin~away and has a flux of 3.2~mJy. The SDSS has classified this source as a starburst at a redshift of 0.0073 (Table~4). Indeed, the SDSS spectra of the source shows narrow emission lines, with weak [\ion{O}{i}] and [\ion{S}{ii}], typical of a star-forming galaxy. This is consistent with the observed [\ion{N}{ii}]$\lambda$6584 \AA/H$\alpha$ and [\ion{O}{iii}]$\lambda$5007 \AA/H$\beta$ ratios and their location on the so-called BPT diagram (after Baldwin, Phillips \& Terlevich 1981). The fact that the SDSS spectra has been taken at the head of the cometary structure could explain the large velocity displacement relative to the \ion{H}{i} gas.

$\bullet$ {\bf SDSSJ1025+1402} -- This red symmetric (Table~3) source shows Balmer and forbidden line offsets in excess of 60 km s$^{-1}$ (Table~4). The neighbouring sources in the SDSS field show typically lower redshifts, which discards possible \ion{H}{i} contamination, and the nearest radio continuum source (NVSS) lies 5\arcmin~away. The SDSS classifies this source as a broad-line quasar at a redshift of 0.1004 (Table~4). The [\ion{N}{ii}]$\lambda$6584 \AA/H$\alpha$ and [\ion{O}{iii}]$\lambda$5007 \AA/H$\beta$ ratios put this source in the realm of star-forming galaxies in the BPT diagram. However, the red color, broad H$\alpha$ line, large redshift and large instrinsic luminosity indicate that the source may host an Active Galactic Nuclei (AGN; Izotov, Thuan \& Guseva 2007). The interaction of the AGN jet with the Narrow Line Region (NLR) may explain the Balmer and forbidden line offsets.


$\bullet$ {\bf J1644+2734}  -- This disk-like symmetric (Table~3) source shows a large forbidden line offset (-140 km s$^{-1}$; Table~4). Sources in the SDSS field show higher radial velocities, which discards possible \ion{H}{i} contamination, and the nearest radio continnum source (NVSS) is at a distance of 3\arcmin. The SDSS classifies this source as a quasar at a redshift of 0.0232 (Table~4). The [\ion{N}{ii}]$\lambda$6584 \AA/H$\alpha$ and [\ion{O}{iii}]$\lambda$5007 \AA/H$\beta$ ratios put this source in the realm of composite galaxies in the BPT diagram. The disk-like morphology, associated with the presence of broad H$\alpha$ emission indicate that the source may host an AGN (Izotov, Thuan \& Guseva 2007). The large forbidden line offset may be the result of cloud motion in the NLR, as a result of the interaction with the AGN jet.

See Izotov, Thuan \& Guseva (2007) for a discussion on the presence of broad-line emission and AGN in XMP galaxies.



\begin{figure} 
\begin{center}
\includegraphics[width=6cm]{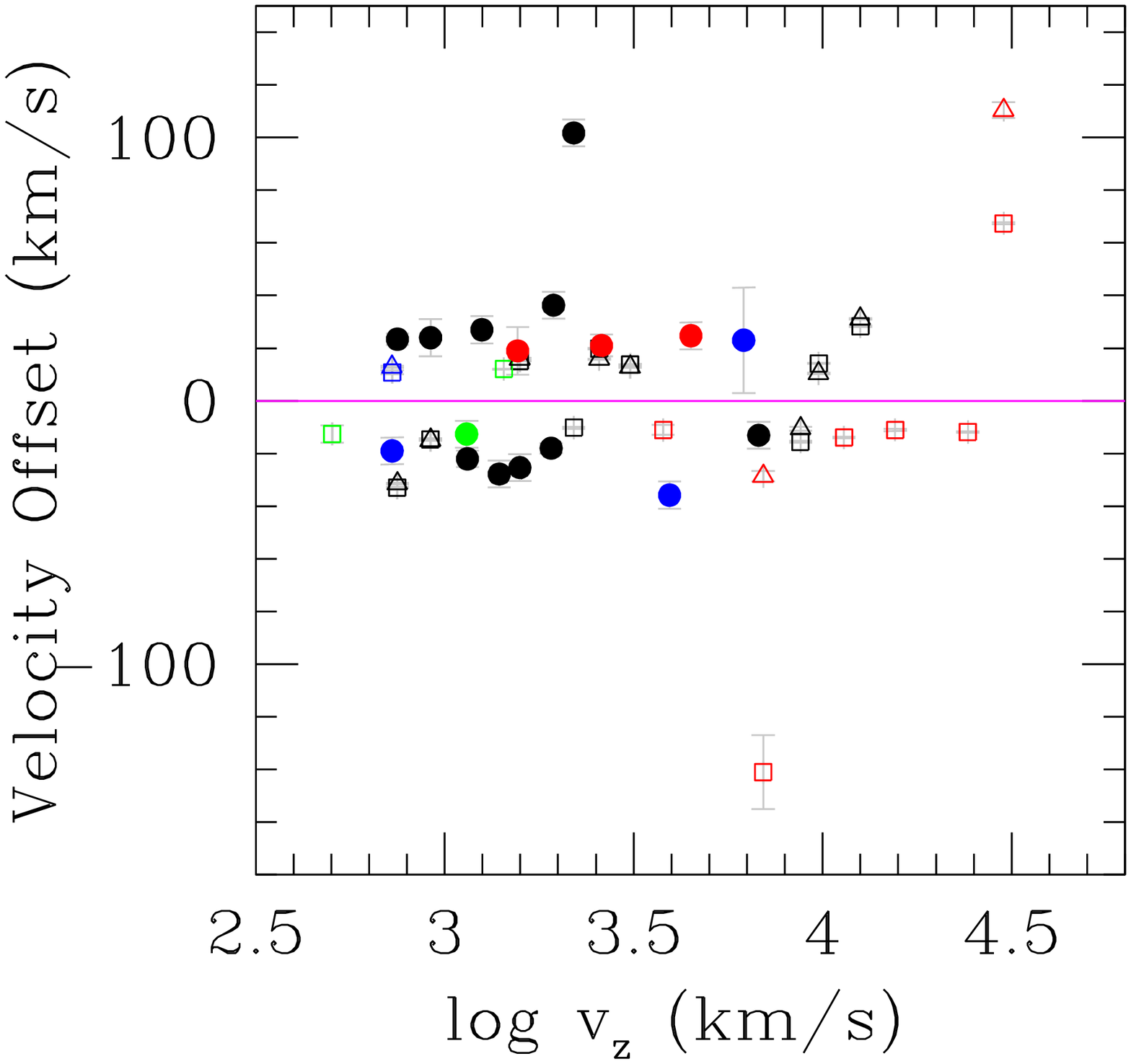} 
\includegraphics[width=6cm]{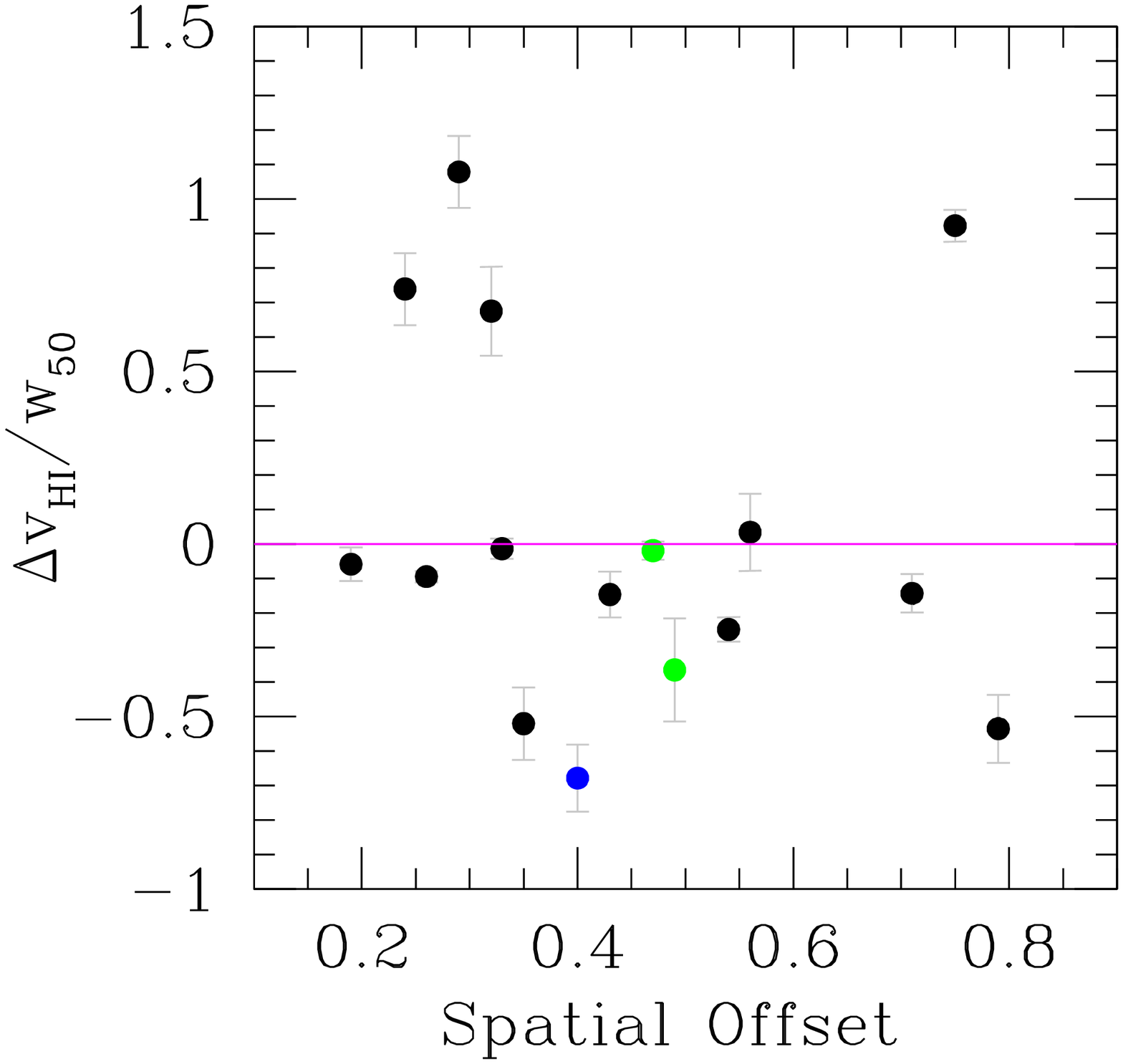}
\caption{The velocity offset of the XMP galaxies as a function of the best-fit radial velocity and spatial offset. The \ion{H}{i}, Balmer emission-line and forbidden emission-line velocity offsets, $\Delta$v$_{\ion{H}{i}}$, $\Delta$v$_{\rm Balmer}$ and $\Delta$v$_{\rm forbidden}$, respectively, are the displacements of the \ion{H}{i} systemic radial velocity,  Balmer lines and forbidden lines with respect to the best-fit radial velocity, v$_{z}$ (Sect.~3.3 and Table~5). Only non-zero values are plotted. The color code parametrizes the morphology as follows: red = symmetric, black = cometary, blue = two-knot, green = multi-knot and grey = no morphological information. The symbol code is $\bullet$ = $\Delta$v$_{\ion{H}{i}}$, $\square$ = $\Delta$v$_{\rm forbidden}$ and $\triangle$ = $\Delta$v$_{\rm Balmer}$. The magenta line defines the null velocity offset.}
\end{center}
\end{figure}


\section{Complementary Results From Optical Data}

\subsection{Optical Morphology}

Our optical morphological classification scheme (Sect.~3.2; Table~3; Fig.~1 and 2) has yielded 27 symmetric, 60 cometary, 11 two-knot and 18 multi-knot galaxies. Of the 116 out of 140 sources with SDSS DR9 information, $\sim$80\% show asymmetric (cometary, two-knot or multi-knot) optical morphology. This result puts on a firm statistical base the association between XMP galaxies and cometary morphology, noted by Papaderos et al. (2008) and quantified by ML11.

From Fig.~4 -- 12, we observe that the multi-knot sources show the broadest range in galaxy properties, with the lowest/highest dynamical, \ion{H}{i} and stellar masses, (\ion{H}{i} gas and stellar) mass fractions and (\ion{H}{i} gas and stellar) mass-to-light ratios, faintest/brightest magnitudes and the narrowest/widest line widths of the XMP galaxies.







\subsection{Star-Formation and Specific Star-Formation Rate}

Figure~12 contains the SFR and sSFR as a function of the stellar mass.  

The XMP SFRs (top; Fig.~12), which measure the global star-formation rate, are similar to those typically observed in BCDs (log~SFR $\sim$ -3.6 -- 0.4, for log~M$_{\star}$ = 6 -- 10~M$_{\odot}$; S\'anchez Almeida et al. 2009). However, more luminous BCDs can show SFRs up to 60 M$_{\odot}$~yr$^{-1}$ and sSFRs up to 10$^{-7}$~yr$^{-1}$ (Cardamone et al. 2009; Amor\'\i n et al. 2012b; also Izotov, Guseva \& Thuan 2011). The XMPs SFR values are lower than the range of SFRs that are typically observed in the disks of large spiral galaxies ($<$~20~M$_{\odot}$~yr$^{-1}$; magenta line; top; Fig.~12; Kennicutt 1998). Given the positive correlation between the SFR and the stellar mass (top; Fig. 12), this is expected, given the low masses of the XMPs. 

Regarding the sSFR (bottom; Fig.~12), we observe a wide range of values. We find some high sSFRs compared to local SDSS samples (Tremonti et al. 2004; also Peeples, Pogge \& Stanek 2008, 2009). The timescale to double their stellar mass, at the present SFR, 1/sSFR, is typically lower than 1 Gyr, smaller than the age of the galaxies, if we assume that they are about the age of the Universe ($\sim$ 14~Gyr; e.g., Caon et al. 2005; Amor\'\i n et al. 2007). This implies that XMPs are now undergoing a major starburst episode, which can not be sustained for very long.




\begin{figure} 
\begin{center}
\includegraphics[width=6cm]{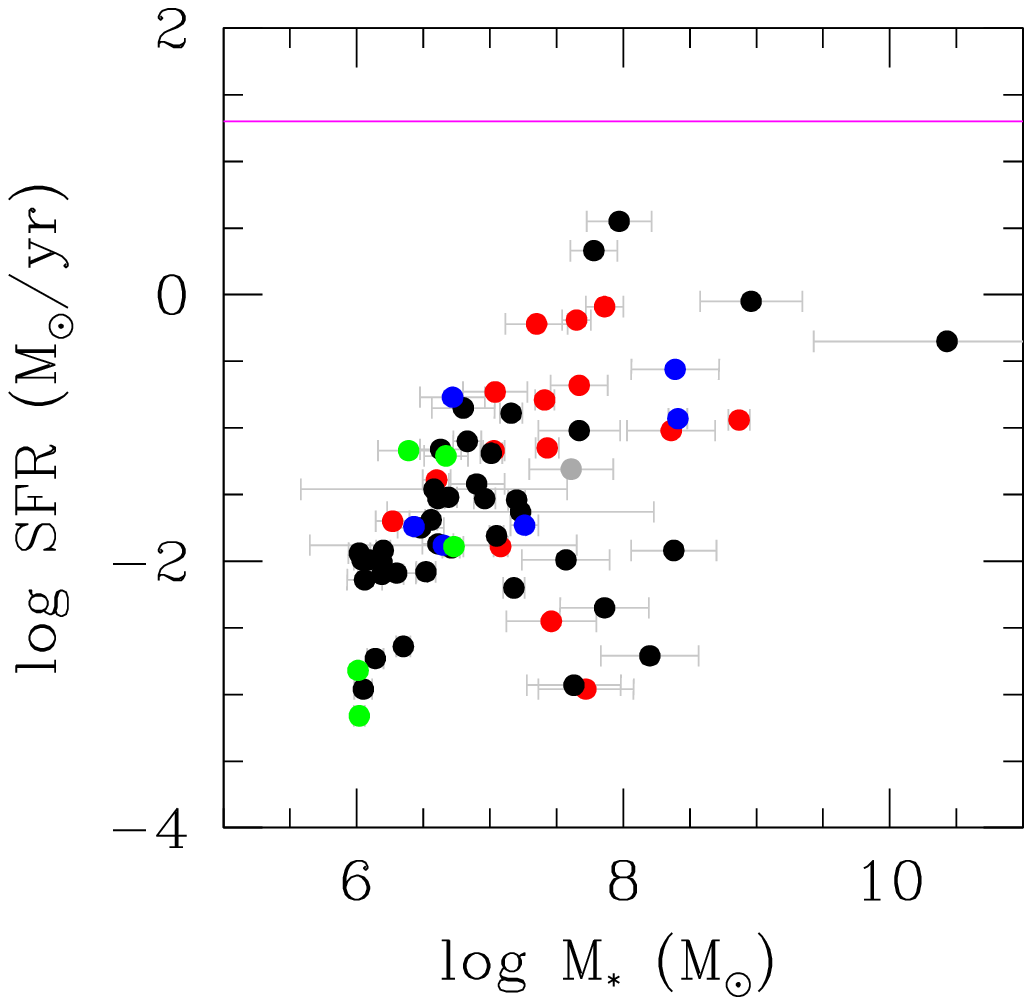}
\includegraphics[width=6cm]{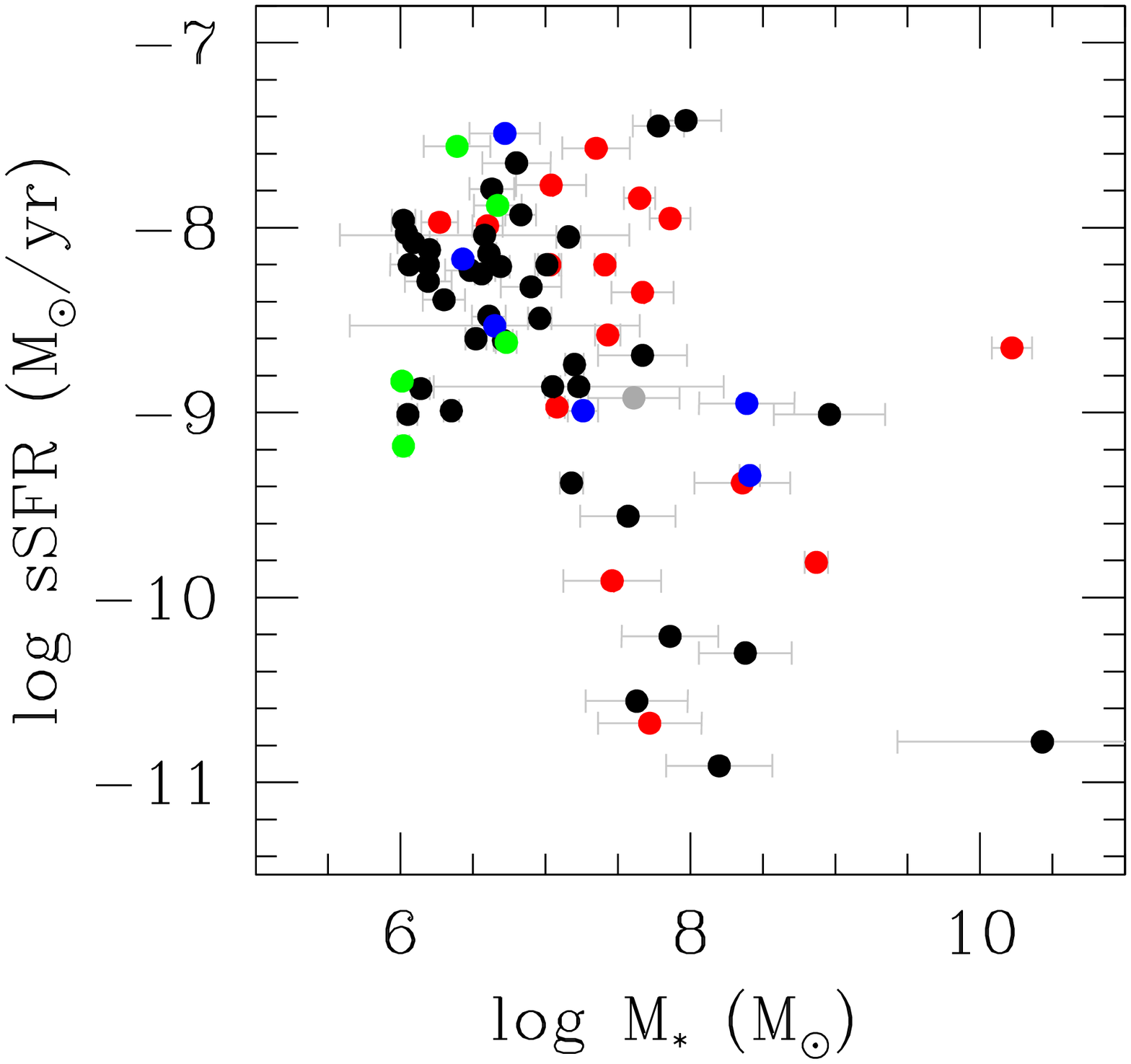} 
\caption{The SFR and sSFR of the XMP galaxies as a function of stellar mass. The color code parametrizes the morphology as follows: red = symmetric, black = cometary, blue = two-knot, green = multi-knot and grey = no morphological information. Top: the magenta line provides the upper limit to the SFRs typically observed in the disks of local spiral galaxies.}
\end{center}
\end{figure}



\section{Conclusions}

We have investigated the \ion{H}{i} content of local XMP galaxies as a class, using the 140 sources compiled by ML11. These sources were selected either as showing negliglible [\ion{N}{ii}] lines in the SDSS DR7 (Abazajian et al. 2009) spectra, or \ion{H}{ii} gas-phase metallicities smaller than a tenth the solar value. 

53 of these galaxies have published \ion{H}{i} data, providing integrated flux densities and line widths in the range 0.1 -- 200 Jy km s$^{-1}$ and 15 -- 150 km s$^{-1}$, respectively (Sect.~2.1 and Table~1). We have also obtained new Effelsberg observations for 29 sources with no published 21 cm data, yielding 10 new detections (Sect.~2.2 and Table~2). Effelsberg integrated flux densities are in the range 1 -- 15 Jy km s$^{-1}$ and line widths cover a wide spectrum (20 -- 120 km s$^{-1}$). The Effelsberg line profiles are varied; one source shows a symmetric single-peak profile, seven show skewed single-peak profiles, demonstrating an asymmetry in the kinematics of the \ion{H}{i} gas, and two show asymmetric double-horn profiles (Fig.~1 and Table~2). Combining the optical morphology with the \ion{H}{i} line profile shapes (Fig.1, Tables~2 and 3), we find that the double-horn sources are associated with multi-knot or cometary optical morphology. Single-peak asymmetric profiles are associated with either cometary, two-knot or multi-knot morphology, while the single-peak symmetric profile is associated with a symmetric source. We conclude that an asymmetry in the \ion{H}{i} gas line profile is systematically associated with an asymmetry in the optical morphology (Sect.~5.1 and Fig.~1). 

When the new Effelsberg and literature data are taken together, typically, the estimated dynamical, \ion{H}{i} and stellar masses are in the range 10$^{6.5 - 10.5}$, 10$^{6.5 - 9.5}$ and 10$^{6 - 9}$~M$_{\odot}$, respectively (Sect.~5, Fig.~6 and Table~5). \ion{H}{i} gas-to-stellar mass ratios are about 10 -- 20 (Sect.~5.4, Fig.~6 and 7). We find that brighter XMPs have converted a larger fraction of their \ion{H}{i} gas into stars (Sect.~5.3 and Fig.~5). Moreover, M$_{\star}$/L$_g$ ratios are found to be on average 0.1, whereas M$_{\ion{H}{i}}$/L$_g$ ratios may be up to 100 times larger (Sect.~5.3 and Fig.~5). Therefore, we conclude that local XMP galaxies are extremely gas-rich. The \ion{H}{i} gas and stellar mass constitute 20 -- 60\% and $<$5\% of the dynamical mass, respectively (Sect.~5.4 and Fig.~6). Furthermore, dark matter mass content (Sect.~5.4 and Fig.~6) spans a wide range of values for XMP systems, but in some cases it accounts for over 65\% of the dynamical mass, higher than the values determined for spirals and ellipticals (10 -- 50\%; Swaters 1999; Borriello, Salucci \& Danese 2003; Cappellari et al. 2006; Thomas et al. 2007; Williams et al. 2009).

The global SFRs (Sect.~6.2, Table~4 and Fig.~12) in XMPs are found to similar to those found in typical BCDs (S\'anchez Almeida et al. 2009). The apparent low SFRs are due to the lower stellar masses of the XMPs, because the sSFRs (SFR per unit mass) are high and are, on average, higher than those observed for local galaxies (Tremonti et al. 2004), with timescales to double their stellar mass, at the current rate, of typically less than 1~Gyr.

XMPs are found to fall off of the mass -- and luminosity -- metallicity relations (Sect.~5.6 and Fig.~9) found for BCDs (Papaderos et al. 2008), extreme starburst galaxies (Amor\'\i n, P\'erez-Montero \& V\'\i lchez 2010), SDSS star-forming galaxies (Guseva et al. 2009) and dwarf irregulars (Skillman, Kennicutt \& Hodge 1989), signaling the presence of pristine material. XMPs generally uphold the baryonic mass -- and luminosity -- TF relation (Sect.~5.7 and Fig.~10) found for samples of late- and early-type dwarf and giant galaxies (de Rijcke et al. 2007). The results suggest that the \ion{H}{i} gas is partly virialized and may contain some rotational support. However, for the lowest luminosity XMPs, most of the gravitational support comes from random motions.	


The effective yield of oxygen in XMPs is often larger than the theoretical yield (Sect.~5.5 and Fig.~8). This is unusual and suggests that either the theoretical yields are underestimating the production of oxygen in these low-metallicity environments, or the \ion{H}{i} gas metallicity is 0.1 -- 1 times that measured in the \ion{H}{ii} regions. This second possibility is in agreement with the recent work of Leboutellier et al. (2013) on the metallicity of the \ion{H}{i} gas of the XMP prototype, IZw18 (also Thuan, Lecavelier des Etangs \& Izotov 2005). 

We have also completed and/or revised the optical morphological classification presented in ML11, bringing up to 83\% the percentage of classified sources (Sect.~3.2, Sect.~6.1, Fig.~2 and Table~3). Of the 116 sources with SDSS imaging, $\sim$80\% present an asymmetric optical morphology, signature of asymmetric star-formation. 27 galaxies show a symmetric spherical, elliptical or disk-like structure, 60 present a clear head-tail cometary morphology, 11 show a two-knot structure, and 18 present a multiple knot structure due to the presence of multiple star-forming regions. 

Velocity offsets between the \ion{H}{i} gas and the nebular/stellar component have also been investigated (Sect.~5.8, Table~5 and Fig.~11). We find that in 60\% of the XMPs with \ion{H}{i} and optical data, small displacements (10 -- 40 km s$^{-1}$) occur and these do not correlate with the morphology. This result suggests that, in these sources, the \ion{H}{i} gas is not highly coupled to the nebular/stellar component.

We conclude that XMP galaxies are extremely gas-rich, with evidence that the \ion{H}{i} component is kinematically disturbed and relatively metal-free.


\begin{acknowledgements}

We would like to thank J. Brinchmann for his help regarding the MPA-JHU data and the anonymous referee for their suggestions and comments.

M. E. F. is supported by a Post-Doctoral grant SFRH/BPD/36141/2007, by the Funda\c c\~ao para a Ci\^encia e Tecnologia (FCT, Portugal). J. M. G. is supported by a Post-Doctoral grant SFRH/BPD/66958/2009 by the FCT (Portugal). P. L. is supported by a Post-Doctoral grant SFRH/72308/2010, funded by the FCT (Portugal). P. P. is supported by a Ci\^encia 2008 Contract, funded by the FCT/MCTES and POPH/FSE (EC). A. H. is supported by a Marie Curie Fellowship, cofunded by the FCT and the FP7. Y. A. receives financial support from project AYA2010-21887-C04-03, from the former Ministerio de Ciencia e Innovaci\'on (MICINN, Spain), as well as the Ram\'{o}n y Cajal programme (RyC-2011-09461), now managed by the Ministerio de Econom\'{i}a y Competitividad (fiercely cutting back on the Spanish scientific infrastructure). P. P., J. M. G., P. L. and A. H. acknowledge support by the FCT, under project FCOMP-01-0124-FEDER-029170 (Reference FCT PTDC/FIS-AST/3214/2012), funded by the FCT-MEC (PIDDAC) and FEDER (COMPETE). This work has been partly funded by the Spanish Ministery for Science, through project AYA 2010-21887-C04-04. The research leading to these results has received funding from the European Commission Seventh Framework Programme (FP/2007-2013) under grant agreement No. 283393 (RadioNet3).

This work is based on observations obtained with the 100-m telescope of the MPIfR (Max-Planck-Institut f\"ur Radioastronomie) at Effelsberg. 

Funding for SDSS-III has been provided by the Alfred P. Sloan Foundation, the Participating Institutions, the National Science Foundation, and the U.S. Department of Energy Office of Science. The SDSS-III web site is http://www.sdss3.org/. SDSS-III is managed by the Astrophysical Research Consortium for the Participating Institutions of the SDSS-III Collaboration including the University of Arizona, the Brazilian Participation Group, Brookhaven National Laboratory, University of Cambridge, Carnegie Mellon University, University of Florida, the French Participation Group, the German Participation Group, Harvard University, the Instituto de Astrofisica de Canarias, the Michigan State/Notre Dame/JINA Participation Group, Johns Hopkins University, Lawrence Berkeley National Laboratory, Max Planck Institute for Astrophysics, Max Planck Institute for Extraterrestrial Physics, New Mexico State University, New York University, Ohio State University, Pennsylvania State University, University of Portsmouth, Princeton University, the Spanish Participation Group, University of Tokyo, University of Utah, Vanderbilt University, University of Virginia, University of Washington, and Yale University. 

\end{acknowledgements}

{}

\end{document}